\documentclass[natbib]{svjour3}             
\smartqed                  
\usepackage{graphicx}
\def\gapprox{\;\rlap{\lower 2.5pt            
 \hbox{$\sim$}}\raise 1.5pt\hbox{$>$}\;}       
\def\lapprox{\;\rlap{\lower 2.5pt            
 \hbox{$\sim$}}\raise 1.5pt\hbox{$<$}\;} 
\begin{document}

\title{Magnetic Fields in Massive Stars, their Winds, and their Nebulae 
}


\titlerunning{Magnetic Fields in Massive Stars}  

\author{Rolf Walder         \and
        Doris Folini        \and
        Georges Meynet  
}


\institute{R. Walder \at
              CRAL: Centre de Recherche Astrophysique de Lyon \\
              ENS-Lyon, 46, all\'ee d'Italie                    \\
              69364 Lyon Cedex 07, France                     \\
              UMR CNRS 5574, Universit\'e de Lyon             \\
              \email{Rolf.Walder@ens-lyon.fr}           
           \and
           D. Folini \at
              CRAL: Centre de Recherche Astrophysique de Lyon \\
              ENS-Lyon, 46, all\'ee d'Italie                    \\
              69364 Lyon Cedex 07, France                     \\
              UMR CNRS 5574, Universit\'e de Lyon             \\
              \email{Doris.Folini@ens-lyon.fr}           
          \and
           G. Meynet \at
              Observatoire de Gen\`eve                  \\
              Facult\'e des Sciences                     \\
              Chemin des Maillettes 51                 \\
              1290 Versoix, Suisse                     \\
              \email{Georges.Meynet@unige.ch}           
}

\date{Received: date / Accepted: date}

\maketitle

\begin{abstract}
Massive stars are crucial building blocks of galaxies and the
universe, as production sites of heavy elements and as stirring agents
and energy providers through stellar winds and supernovae. The field
of {\it magnetic} massive stars has seen tremendous progress in recent
years. Different perspectives -- ranging from direct field
measurements over dynamo theory and stellar evolution to colliding
winds and the stellar environment -- fruitfully combine into a most
interesting and still evolving overall picture, which we attempt to
review here. Zeeman signatures leave no doubt that at least some O-
and early B-type stars have a surface magnetic field. Indirect
evidence, especially non-thermal radio emission from colliding winds,
suggests many more. The emerging picture for massive stars shows
similarities with results from intermediate mass stars, for which much
more data are available. Observations are often compatible with a
dipole or low order multi-pole field of about 1 kG (O-stars) or 300 G
to 30 kG (Ap / Bp stars). Weak and unordered fields have been detected
in the O-star $\zeta$ Ori A and in Vega, the first normal A-type star
with a magnetic field. Theory offers essentially two explanations for
the origin of the observed surface fields: fossil fields, particularly
for strong and ordered fields, or different dynamo mechanisms,
preferentially for less ordered fields. Numerical simulations yield
the first concrete stable (fossil) field configuration, but give
contradictory results as to whether dynamo action in the radiative
envelope of massive main sequence stars is possible. Internal magnetic
fields, which may not even show up at the stellar surface, affect
stellar evolution as they lead to a more uniform rotation, with more
slowly rotating cores and faster surface rotation. Surface
metallicities may become enhanced, thus affecting the mass-loss rates.
%
%
%
\end{abstract}
\section{Introduction}
\label{Sec:Intro}
One may wonder why a review on magnetic fields in massive stars enters
a book dealing with large scale magnetic fields in the
universe. Several reasons may be given. 

A first line of argument may stress the relevance of massive stars as
cosmic engines for the chemical and dynamical evolution of galaxies
and the universe, their role as production sites of heavy elements and
as stirring agents and energy providers in the form of stellar winds,
supernova explosions, gamma ray bursts, and galactic
super-bubbles. Also, if the stars are magnetic, their wind driven
structures will be magnetized as well. One may further argue with the
magnetic dipole field observed at the surface of some massive stars,
whose large scale ordering not trivial to achieve or maintain. Or with
the conditions under which magnetic fields in massive stars exist or
form, which represent relative exotic locations of the parameter
space. Massive stars may be viewed as laboratories to study and
observe magnetic fields under such conditions. Better understanding of
the role of magnetic fields in massive stars then will not only
help to better understand the role of massive stars in the
universe but will also broaden our knowledge on the existence and
generation of magnetic fields in the universe. Like the field of
magnetism in massive stars has profited from combining different
points of view, from direct field measurements to modeling of
colliding winds, the reader of this book may profit from the presented
largely different perspectives on magnetism in the universe.
 
The term massive star is typically used for stars with initial masses
above 8 M$_{\odot}$, which will ultimately end their lives as
supernovae. Early type massive stars refer to massive stars of
spectral type O or B, thus stars on or close to the main sequence but
not, for example, massive stars having entered the red super-giant
phase. The first definite magnetic field detections in massive stars, by
means of spectropolarimetry, date back only about a decade, to around
the year 2000. Since then, observational techniques have been further
refined and field detections, although still difficult for physical
reasons, have multiplied. We stress that all these detections refer
only to {\it surface magnetic fields}, and that any {\it magnetic
fields in the interior} of stars need not even show up at the
surface. A 'snapshot' overview of the rapidly evolving observational
results is given within the frame of this review,
Sect.~\ref{Sec:Observations}.

Crucial and interesting with regard to magnetic fields in massive
stars is the radiative envelope of these stars, which prevents the
existence of a solar type dynamo. Any magnetic field thus must be due
to another cause. Theories on the origin of magnetic fields in early
type massive stars may be roughly divided into two categories: dynamos
working differently than the solar dynamo or fossil fields of debated
origin that are present before the star reaches the main sequence. The
dynamo aspects may link to another review in this book on turbulent
dynamos. Similarly, the fossil fields may link to the review by Beck
et al., which touches on magnetic fields in the ISM. Part of this
review, Sect.~\ref{Sec:Models}, is devoted to a more detailed
presentation of these theories, their analytical argumentation as
well as numerical simulations which allow to access the non-linear and
time-dependent regime.

Equally important as the cause of magnetic fields in massive stars are
the consequences of such fields. The issue may again be roughly
subdivided into two parts, consequences for the stellar evolution and
consequences for the ambient medium. Even weak magnetic fields in the
stellar interior, which may not even show up as surface fields, are
expected to affect the stellar evolution from 'cradle to graveyard',
by modifying the internal transport of angular momentum. Subsequent
changes in the transport of chemical yields, surface abundances, or
mass loss, may even influence the ultimate fate of the star as a
neutron star, gamma-ray burst, or black hole. Sufficiently strong
surface fields can, by contrast, convincingly explain a number of
observational signatures originating at the surface of or at some
distance from the star. Termination shocks of magnetized winds or wind
collision in binaries or open clusters can contain indirect evidence
on the presence of a stellar magnetic field. Moreover, such
structures could potentially serve as a laboratory to test non-thermal
shock models, super-enhancement of magnetic fields in shocks, or
particle acceleration, aspects which may link to two other
contributions in this volume, one on supernova remnants and pulsar
nebulae, the other on particle acceleration. The consequences of
stellar magnetic fields add another two parts to this review,
Sect.~\ref{Sec:Implications} and~\ref{Sec:Environment}.

The presented material and the rapid speed at which this knowledge,
observational and theoretical, currently grows leads us to believe
that some of the above questions will have a more definite answer in
the not too far future. Some more room to these considerations will be
given in the closing section of this review,
Sect.~\ref{Sec:Conclusions}.
\section{Observations of Surface Fields in Intermediate and High Mass Stars}
\label{Sec:Observations}
Stellar magnetism is ubiquitous in most parts of the HR diagram, along
the main sequence as well as during pre- and post-main-sequence
evolution~\citep{2009IAUS..259..323B, 2009ARA&A..47..333D}. With the
possible exception of a narrow mass range between 1.5 and 1.6
M$_{\odot}$~\citep{2009ARA&A..47..333D} stellar magnetism has been
found throughout the main sequence. Widely different are, however, the
concrete manifestations of stellar magnetism, such as the strength of
the field, the field topology, or just the percentage of stars of a
given spectral type which have a detectable magnetic field. Our
knowledge on these aspects has been rapidly evolving in recent years,
as better instruments became available and data analysis techniques
were refined. This development is still ongoing and, consequently,
some of the current conclusions may need revision in the future, when
both the quality and quantity of the observational data will have
increased even further. We include a brief overview of current
observational techniques in Sect.~\ref{sec:obs_techniques} in the hope
that this will help the reader to better rate the reviewed
observational results.

Current observational data suggests the following rules of thumb,
which may be subject to change as new data becomes available. Magnetic
fields are common among low-mass stars (M $<$ 1.5 M$_{\odot}$), but
only a minority of intermediate- and high-mass stars (M $>$ 1.5
M$_{\odot}$) show a magnetic field. The field topology in
intermediate- and high-mass stars is frequently dominated by a simple
dipole, while low-mass stars typically display a much richer
topology. For intermediate- and high-mass stars, the presence of a
magnetic field often goes hand in hand with a comparatively slow
rotation of the star and with a chemically peculiar photosphere.

These rules of thumb demand for physical explanation, thus inspiring a
lively and ongoing debate on the detailed physical origins and
consequences of magnetic fields in intermediate- and high-mass
stars. Some aspects of this debate deserve mentioning here as they
directly feed back in new observational campaigns. Fossil fields (see
Sect.~\ref{SubSec:Models_Fossil}), for example, are the currently
favored explanation for the dipole magnetic fields in intermediate-
and high-mass stars, in contrast to convection in low-mass stars. This
explanation promoted recent observational surveys to search for
magnetic fields in intermediate and high-mass pre-main-sequence
stars. Magnetic braking as the proposed agent to explain the on
average slower rotation of magnetic stars inspired observational
efforts to find a possible age - magnetic field strength relation
ship. The argument that magnetic dipole fields are stable only above a
certain threshold field strength led to an intense search for stars
with field strengths below this threshold value (see
Section~\ref{sec:obs_ab}).


%
%
\subsection{Measuring Stellar Magnetic Fields}
\label{sec:obs_techniques}
\citet{1947ApJ...105..105B} showed that the circular polarization
observed in some absorption lines of the spectrum of 78 Vir can be
interpreted in terms of the Zeeman effect and thus give access to the
measurement of magnetic fields in stars. Since then, observational
techniques have been greatly refined~\citep{1989FCPh...13..143M,
  2006ASPC..358..345K, 2009ARA&A..47..333D}. A recent review on
current instruments and techniques, given at the IAU symposium 272 on 
active OB stars, can be found in~\citet{2010arXiv1010.2248P}.

The basic idea behind Zeeman-measurements is simple. The component of
a magnetic field parallel to the line of sight of an observer,
$B_{l}$, manifests itself in a split of spectral lines into two
components of opposite circular
polarization~\citep{2009EAS....39....1L}. For relatively weak magnetic
fields the resulting Zeeman shift $\Delta
\lambda$ between the two components is proportional to $B_{l}$, the
magnetic splitting sensitivity or effective Land\'{e} factor
$g_{eff}$, and the wave length squared,
\begin{equation}
\Delta \lambda \propto B_{l} g_{eff} \lambda^{2}.
\end{equation}
The Zeeman shift thus is particularly pronounced for strong fields
(large $B_{l}$) and large wave lengths (infra red).

In real observations, complications of this simple picture arise from
mainly two sides. First, thermal and stellar rotation broadening are
superimposed on the the $\Delta \lambda$ of the Zeeman shift. The
Zeeman effect thus may only result in some additional line broadening
and not in any clear line splitting. Second, the observed spectral
signal is an average signal from the entire stellar disk, thus not due
to one particular value of $B_{l}$ but rather due to a composite of
magnetic field strengths and orientations. If polarity of the field
changes over the stellar disk, significant cancellation of the
observable signal results. Taken together these effects demand for a
measurement accuracy on the order of $10^{-3}$ to $10^{-4}$ if stellar
magnetic fields are to be detected~\citep{2009ARA&A..47..333D,
2009IAUS..259..323B}. Only for strong dipole fields of slowly rotating
stars (as in some Ap stars) the Zeeman-effect may be directly visible
as a splitting of spectral lines in intensity spectra (Stokes I).  In
these cases, or if at least a substantial line broadening exists, one
may determine the mean value of the field strength averaged over the
stellar disk, $<|B|>$, also known as the mean field modulus or mean
surface field $<B>$ without detailed line profile
modeling~\citep{1995A&A...293..746M, 2006A&A...453..699M,
2009EAS....39....1L}.

The magnetic field observations reviewed in this article address these
difficulties in mainly two ways. First, on the 'hardware' side, by
measuring circular polarization profiles (Stokes V) with high
frequency resolution. Second, on the 'software' side, by combining
Zeeman signatures from several spectral lines to increase the signal
to noise ratio and by combining observations taken at different times,
thus different rotational phases of the star, to learn more about the
3D field topology.

The currently most advanced instruments for magnetic field
observations in massive stars are two high-resolution
spectropolarimeters. One of them is
ESPaDOnS~\citep{2003ASPC..307...41D}, installed in 2004 on the
Canada-France-Hawaii Telescope on Mauna Kea, Hawaii. The other is,
NARVAL~\citep{2003EAS.....9..105A}, a clone of ESPaDOnS, installed at
the T\'{e}lescope Bernard Lyot on Pic du Midi, France. Both
instruments are regularly used by the MiMeS (Magnetism in Massive
Stars) project~\citep{2009IAUS..259..333W,2010arXiv1012.2925W}. Two other instruments that
were frequently used in recent years for magnetic field observations
in massive stars are the low-resolution spectropolarimeter FORS1 on
the ESO Very Large Telescope (VLT) on Cerro Paranal, Chile, and
MUSICOS~\citep{1999A&AS..134..149D}, also on Pic du Midi, France.

\begin{figure}[tp]
\centerline{\includegraphics[width=10cm]{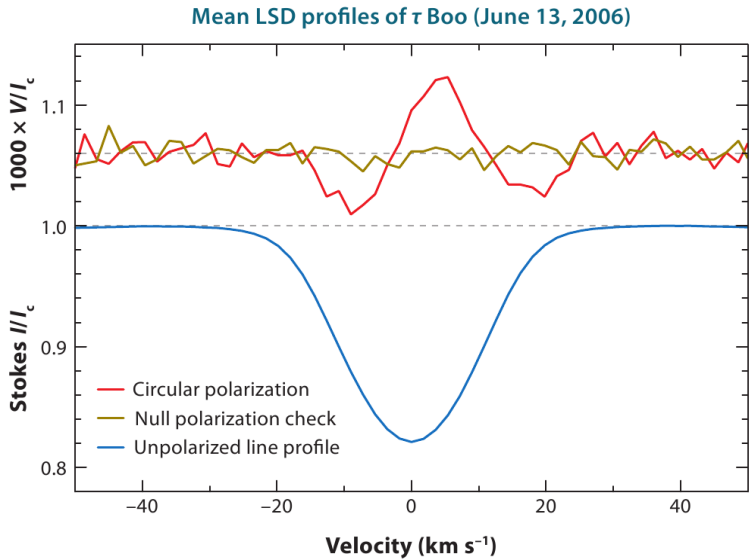}}
\caption{LSD circular polarization (Stokes V) Zeeman signature (red
  line), null polarization check (dark yellow line, both expanded by
  1000 and shifted vertically by 1.06 for graphical purposes) and
  unpolarized (Stokes I) profile (blue line) from the photospheric
  lines of $\tau$ Boo, as derived from ESPaDOnS data.  A clear Zeeman
  signature is detected. The Figure is taken
  from~\citet{2009ARA&A..47..333D}, their Figure~1. }
\label{fig:don_land_fig1}
\end{figure}

The basic idea for increasing the signal to noise ratio is to combine
the information from numerous spectral lines instead of analyzing only
a single line. Different approaches exist to achieve this combination
of multiple lines and~\citet{2009A&A...504.1003S} recently suggested
to summarize all these different methods under the term multi-line
Zeeman signature (MZS).  One very basic approach is known as line
addition technique~\citep{1989A&A...225..456S, 1996SoPh..164..417S}.
Here, several observed lines are just added up. Another technique is
the Least Squares Deconvolution (LSD) introduced
by~\citet{1997MNRAS.291..658D} and illustrated in
Figure~\ref{fig:don_land_fig1}. Comparing the two techniques by
applying them to a synthetic test, case~\citet{2009A&A...504.1003S}
obtain very similar results. A detailed investigation of the
capabilities and limitations of LSD can be found
in~\citet{2010arXiv1008.5115K}. More recently, yet another technique
was proposed for de-noising the Zeeman signatures, based on principle
component analysis (PCA)~\citep{2006ASPC..358..355S,
  2008A&A...486..637M}.

\begin{figure}[tbp]
\centerline{\includegraphics[height=11.cm]{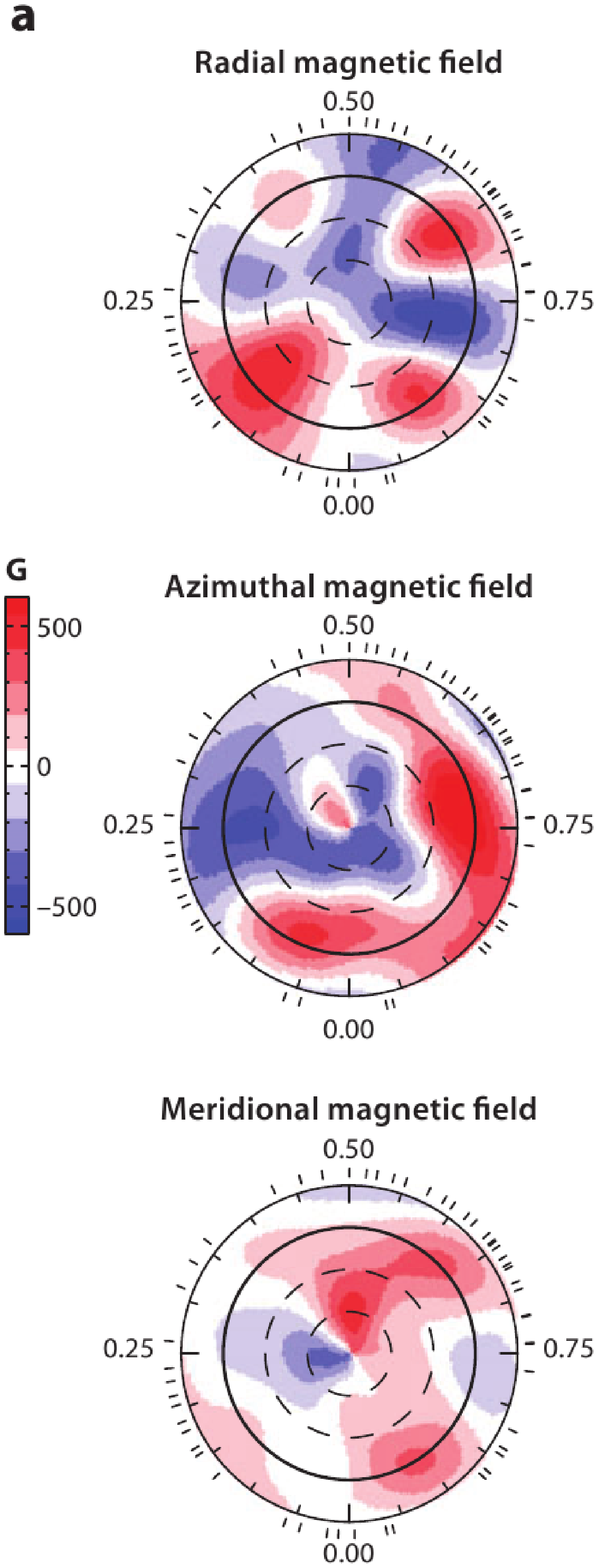}
            \includegraphics[height=11.cm]{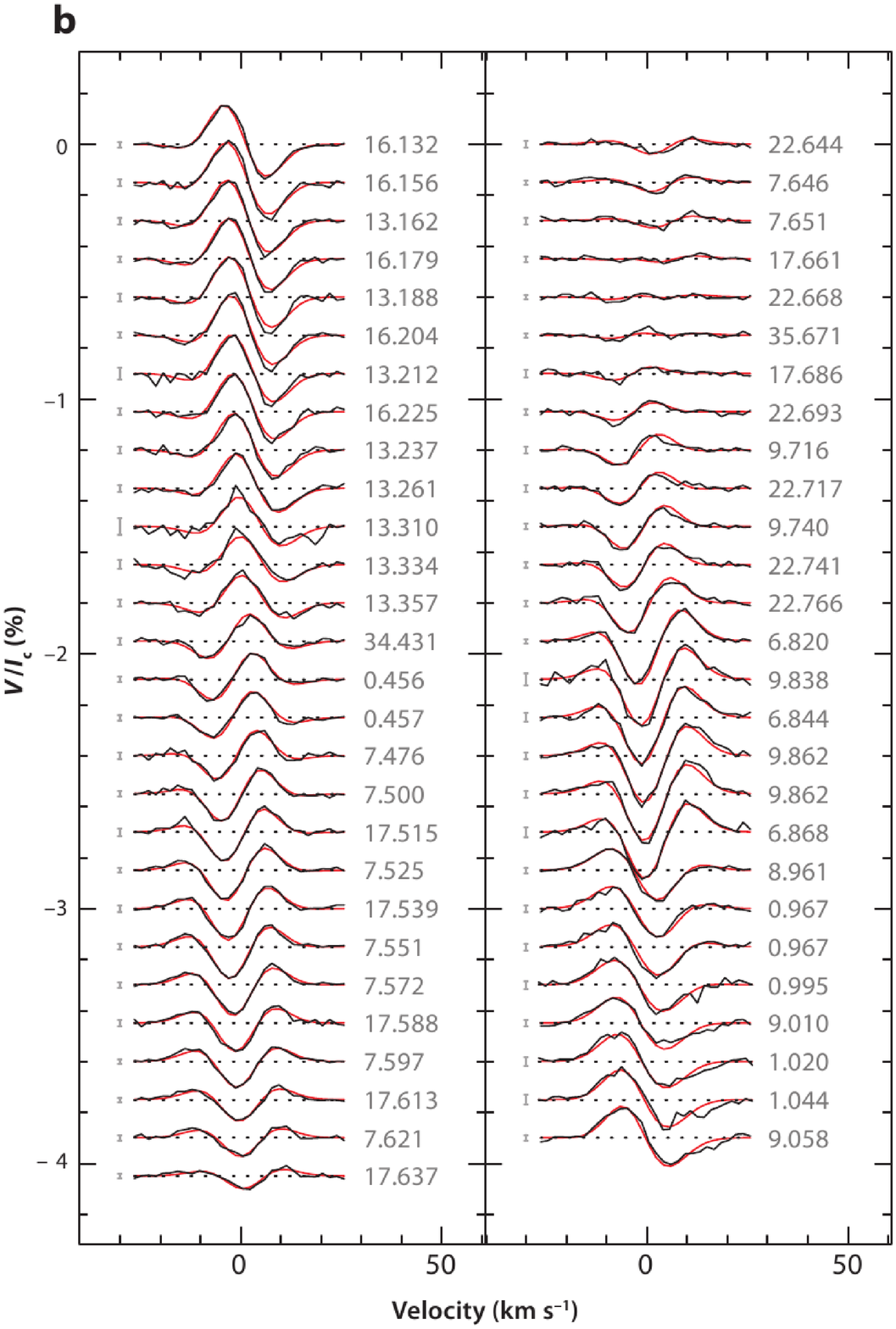}}
\caption{Large-scale magnetic topology of the young early B-star
  $\tau$ Sco ({\bf right}) derived with ZDI from a time series of
  circular polarization (Stokes V) Zeeman signatures covering the whole
  rotation cycle ({\bf left}). The Figure is taken
  from~\citet{2009ARA&A..47..333D}, their Figure~5. }
\label{fig:don_land_fig5}
\end{figure}

\begin{figure}[t]
\centerline{
\includegraphics[height=5.5cm]{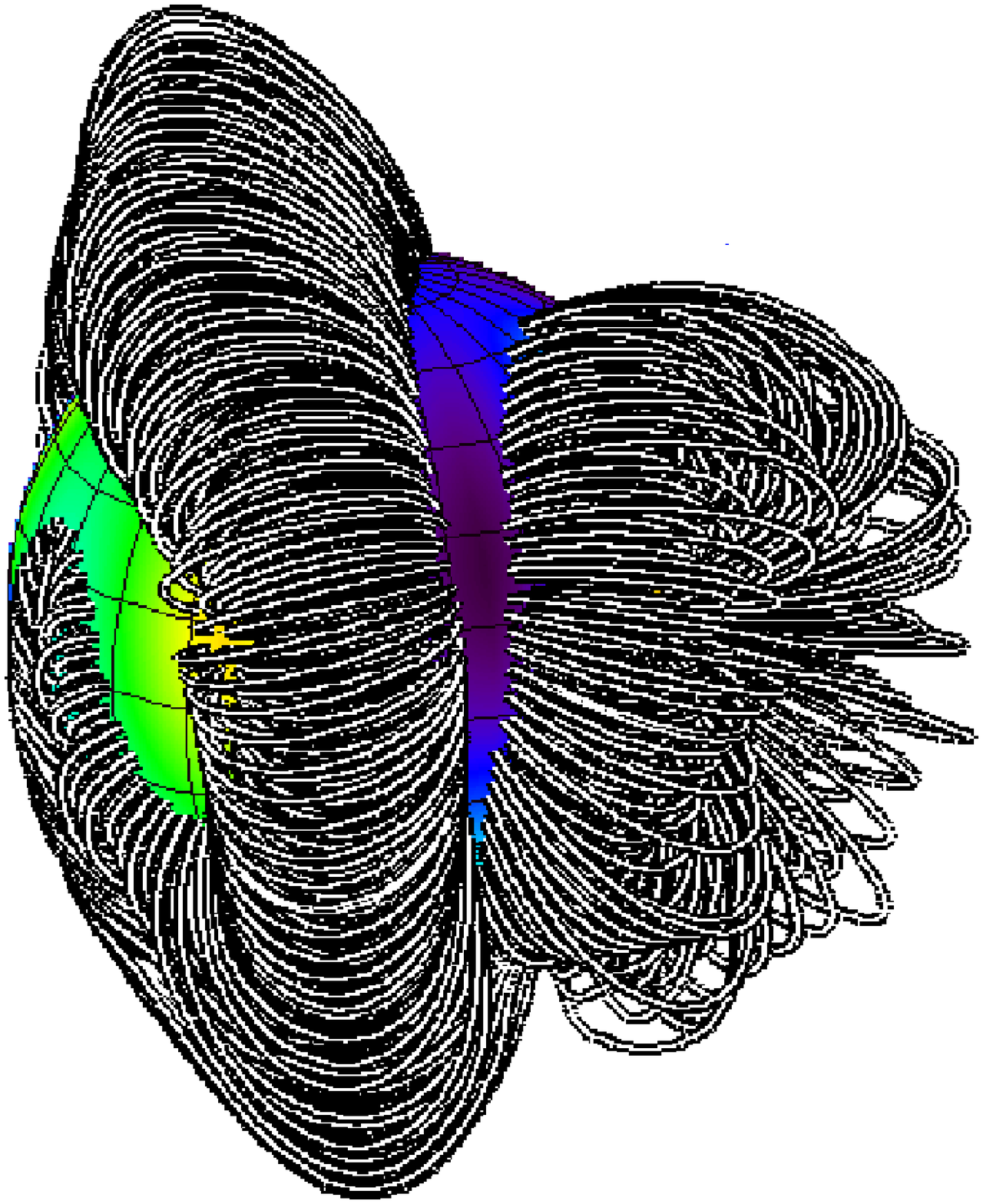}
\hspace{1cm}
\includegraphics[height=5.5cm]{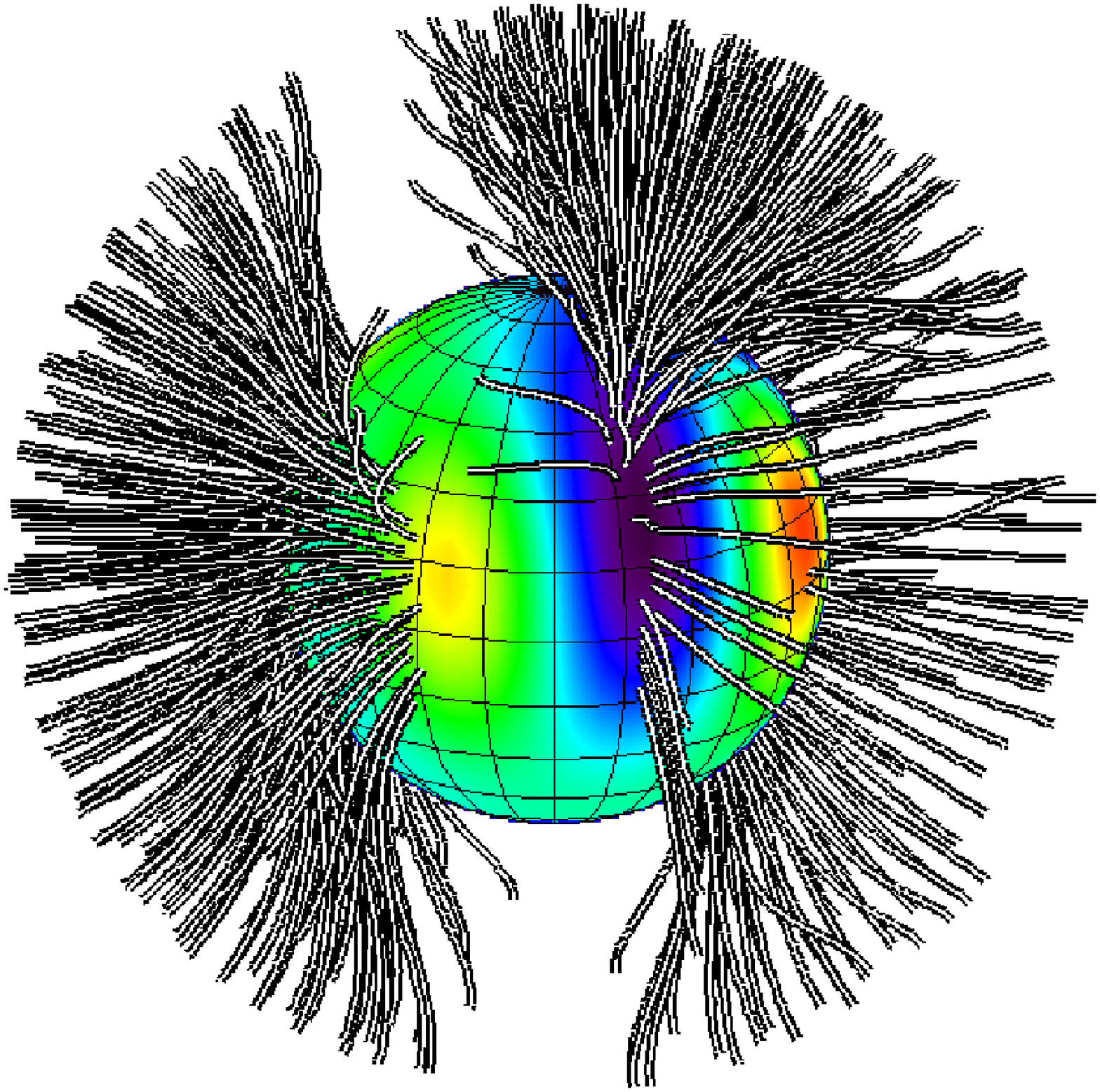}
}
\caption{Closed ({\bf left}) and open ({\bf right}) magnetic-field
  lines of the extended magnetic field configuration of $\tau$ Sco,
  extrapolated from photospheric maps using the technique
  by~\citet{1999MNRAS.305L..35J}. The star is shown at phase 0.83. The
  Figure is taken from~\citet{2006MNRAS.370..629D}, their Figures~11
  and~12. }
\label{fig:don_et_al_fig12}
\end{figure}

To obtain information not only on the line of sight magnetic field
component but on its 3D structure it would be necessary to measure
also Stokes Q and U, in addition to Stokes V. However, these
components carry an even weaker signature than Stokes V. Therefore,
they are hardly explored so far in the context of massive stars. An
alternative approach is followed instead, by exploring the time
dependence of the Zeeman signatures. To illustrate the basic idea,
assume a star with a bipolar magnetic field, the dipole not being
aligned with the rotation axis of the star. A distant observer will
see the Zeeman signature to vary with time as the dipole axis
processes about the rotation axis of the star. Note that the
interpretation of such a signal requires the inclination $i$ of the
rotation axis of the star to be known. The time dependent Zeeman
signature carries the imprint of the magnetic dipole field.
Information on the dipole (or higher moments) can be retrieved again
from this signal by decomposing the observational data in terms of
spherical harmonics. This technique today is known under the name
Zeeman Doppler imaging (ZDI). An example of ZDI is shown in
Figure~\ref{fig:don_land_fig5}, derived field lines using the
field-extrapolation technique by~\citet{1999MNRAS.305L..35J} are shown
in Figure~\ref{fig:don_et_al_fig12}.

Note that the term ZDI originally stood for the detection of a Zeeman
signature thanks to the Doppler effect~\citep{1989A&A...225..456S,
2009A&A...504.1003S}. In fast rotating stars, the later can help to
disentangle the contributions from opposite magnetic polarities and
their respective opposite polarization which otherwise may cancel. The
combination of Zeeman and Doppler effects is one of the reasons why
the detection of the Zeeman effect in fast rotating active solar type
stars became possible.
\subsection{A and late B stars}
\label{sec:obs_ab}
A and B stars account for around 0.6 \% and 0.1 \%, respectively, of
all main sequence stars in the HR diagram. Their masses range between
1.4 - 2.1 M$_{\odot}$ (A stars) and 2.1 - 16 M$_{\odot}$ (B stars). A
small fraction of these stars distinguishes themselves from normal A-
and B-type stars by chemically peculiar atmospheric compositions, for
example high abundances of Si, Cr, Sr, or
Eu~\citep{1974ARA&A..12..257P, 2007A&A...470..685L}. These stars are
commonly referred to as CP stars. A subset of these stars are the
Ap/Bp stars. In the following, {\it we will collectively refer to these
stars as Ap stars}.

Typically, Ap stars display periodic variability on time scales that
range from about half a day up to several decades and that are
inversely correlated with $v\cdot \mathrm{sin} i$, where $i$ denotes
the inclination of the rotational axis of the star and $v$ the
rotation velocity. The variability is commonly attributed to a
magnetic dipole field, whose axis is not aligned with the rotation
axis of the star. In fact, the strongest magnetic fields observed to
date in non-degenerated stars belong Ap stars: HD 215441 (34
kG)~\citep{1960ApJ...132..521B}, HD 75049 (30
kG)~\citep{2008MNRAS.389..441F, 2010MNRAS.402.1883E}, HD 137509 (29
kG)~\citep{1995A&A...293..746M, 2006A&A...454..321K}, HD 154708 (24.5
kG)~\citep{2005A&A...440L..37H}.

Assuming the field topology to be bipolar generally gives reasonable
agreement with the observational data~\citep{1967ApJ...150..547P,
1992A&ARv...4...35L, 2005LNP...664..183M, 2007A&A...475.1053A}.
Evidence for additional higher-order multi-polar contributions is,
however, found in most cases when searched
for~\citep{1969ApJ...156..967P, 1988ApJ...326..967L,
2000A&A...359..213L, 2007A&A...475.1053A}. Nevertheless, magnetic
fields of Ap stars are often specified in terms of the surface
strength of the dipole field.  The relative frequency of magnetic Ap
stars compared to the number of A- and B-type stars of similar mass is
below 10\% and may be decreasing with mass from about 10\% at 3
M$_{\odot}$ to around 0\% at about 1.6 M$_{\odot}$
\citep{2008CoSka..38..443P, 2009ARA&A..47..333D}.

Whether all Ap stars are magnetic remains controversial in the sense
that no direct field measurements are available for the majority of
the known Ap stars. A recently published catalog
by~\citet{2009A&A...498..961R} lists 3652 Ap stars. By contrast, an
equally recent catalog by~\citet{2009MNRAS.394.1338B} of direct line
of sight magnetic field measurements contains 1223 stars of
spectral type O to M, 610 of which are chemically peculiar and only
410 belong to the classical Ap class as defined
by~\citet{1974ARA&A..12..257P}. To clarify the
question,~\citet{2006A&A...450..777B} used the low-resolution FORS1
instrument to monitor 97 Ap stars for magnetic fields. A clear
detection was obtained for only 41 stars. Meanwhile, it seems
reasonable to assume that the many non-detections in this study were
due to the low-resolution.

\citet{2007A&A...475.1053A} employed the high-resolution
spectropolarimeter ESPaDOnS to monitor 28 firmly established Ap stars.
They found magnetic fields well above the detection limit in all of
them. The result is all the more remarkable as the 28 stars were
deliberately selected to have a weak magnetic field, if one at all.
Nevertheless, for all but two stars the inferred surface dipole field
is larger than 300 G. The authors conclude that not only all Ap stars
have a detectable magnetic field but also that this field is always
larger than some magnetic threshold value of about 300 G. For a
possible explanation for the existence of such a threshold value they
refer to~\citet{1999A&A...349..189S}. The basic argument is that a
too weak magnetic field will be wound up due to differential rotation,
a pinch-type instability will set in, and the magnetic field will be
destroyed. In more formal terms, the order of magnitude estimate for the
critical field $B_{\mathrm{c}}$ given by the authors is
\begin{equation}
\frac{B_{\mathrm{c}}}{B_{\mathrm{eq}}} \simeq 
                   2 \left ( \frac{P_{rot}}{5_{\mathrm{day}}} \right )^{-1} 
                     \left ( \frac{r}{3 R_{{\odot}}} \right )
                     \left ( \frac{T}{10^{4} \mathrm{K}} \right )^{-1/2}
\end{equation}
with $B_{\mathrm{eq}}$ the equipartition field ($B_{\mathrm{eq}}^{2} =
8 \pi P$, $P$ the gas pressure) at the surface ($\tau_{5000} = 2/3$), and
$P_{rot}$, $r$, and $T$ the stellar rotation period, radius, and
temperature, respectively.  For a typical Ap star, $B_{\mathrm{eq}} =
170$ G and $B_{\mathrm{c}}
\approx 300$ G.

The authors stress that such a mechanism would not only explain the
apparent lower bound in the strength of magnetic fields in Ap
stars. If a large enough field strength is required in order that the
magnetic field does not decay within a short time, this may explain
why only a small fraction of A- and B-type stars have a magnetic
field. Moreover, that the critical field strength increases with
initial stellar mass (thus increasing temperature $T$ and stellar
radius $r$) would naturally explain the even greater scarcity of
magnetic field detections in more massive stars.

Additional high-resolution observations have further refined the
picture of magnetic Ap stars. \citet{2008CoSka..38..443P} identified
57 Ap stars within 100 pc from the Sun, by means of the Hypparcos
Catalog and additional sources. They find for the distribution of
surface dipole field strengths a plateau at 2.5 $\pm$ 0.5 kG, dropping
off to higher and lower field strengths. The mass distribution of
these stars peaks around 2.1 to 2.3 M$_{\odot}$ with a strong decrease
towards lower and higher masses. The lowest mass stars in the sample
are 1.6 $\pm 0.1$ M$_{\odot}$. The findings for the magnetic fields
are consistent with the weak field results obtained
by~\citet{2007A&A...475.1053A}. The picture is further supported
by~\citet{2010AstL...36..370K}, who carried out detailed statistical
analysis of the stars catalogued by~\citet{2009MNRAS.394.1338B}.

To detect a potential dependence of the magnetic field strength on the
main sequence age of the star, a number of observations of Ap stars in
open clusters where performed by~\citet{2006A&A...450..777B}
and~\citet{2007A&A...470..685L, 2008A&A...481..465L}. The
observational data suggests that the magnetic field of an Ap star
indeed decays with time, but that the time scale for the decay depends
strongly on the mass of the star. The time over which an initially
strong (about 1 kG) magnetic field essentially disappears ranges from
about 250 Myr for stars of 2 - 3 M$_{\odot}$ over 40 Myr for stars of
3 - 4 M$_{\odot}$ to 15 Myr for stars of 4 - 5 M$_{\odot}$.

Further peculiarities of magnetic Ap stars may or may not be related
to their magnetic field. Probably related is their comparatively slow
rotation, with periods on the order of a few days instead of a few
hours to one day. \citet{2000A&A...359..213L} find that most slow
rotators (periods longer than 25 days) have small inclination angles
of less than 20 degrees between the rotation axis and the axis of the
magnetic field, whereas stars with short rotation periods have larger
inclination angles. Maybe also related to the presence of magnetic
fields are chemical abundance variations in the observed spectra.
Their period is mostly identical to the rotation period, implying that
the variations stem from chemical inhomogeneities on the surface. Not
clear is the reason for the lack of short period binaries among
magnetic Ap stars. Based on a sample of 95 Ap
stars,~\citet{1998CoSka..27..179N} find 27\% to be spectroscopic
binaries, as compared to (47 $\pm$ 5) \% for normal A- and B-type
stars, particularly apparent is the lack of tight orbits as compared
to normal stars. \citet{2009A&A...498..961R} report 5 definite
detections of binary Ap stars with orbital periods below 3 days.
\citet{2010ARep...54..156T} ascribe this lack of close binaries to
mergers. Effective temperatures of Ap stars are reported to lie in a
range between 7000 K and 23000 K~\citep{2004ASPC..318..297N}. The
authors argue that for temperatures outside this range all chemical
peculiarities will be erased by either strong radiatively driven winds
(higher temperatures) or convection (lower temperatures).

The only types of Ap / Bp stars for which no large scale magnetic
fields have been detected yet are the helium weak Bp stars of type
PGaS, and the Bp HgMn stars~\citep{2004ASPC..318..297N}. More recent
observations of the HgMn star in AR Aur, the only eclipsing binary
known to contain a HgMn star, place an upper limit of 100 G and 400 G
on any longitudinal or dipole field,
respectively~\citep{2010MNRAS.tmp.1094F}. For the same
star,~\citet{2009IAUS..259..401S} report the detection of patchy
chemical inhomogeneities at the surface. As an explanation they
suggest a weak magnetic field, caused by differential rotation in the
wake of a depth dependent tidal torque.  For another HgMn star,
$\alpha$ And,~\citet{2006A&A...451..293W} find an upper limit on the
order of 100 G as well, depending on the assumed field geometry.
Studying a larger sample of 12 Am and 3 HgMn stars with NARVAL,
\citet{2010arXiv1008.3086A} confirm these upper limits. Together,
these observations support the existence of a dichotomy between
spectroscopically defined Ap stars and any other A-type stars, with
only the former harboring substantial magnetic fields.

Turning finally back to normal A- and B-type stars it is interesting
to note that~\citet{2006A&A...450..777B} examined 138 normal A- and
B-type stars but found no evidence for any magnetic field. It has to
be stressed, however, that this does not necessarily imply that normal
A- and B-stars do not have a magnetic field. The field may just be too
weak to be measured or its topology may be too complicated, leading to
cancellation effects. In fact,~\citet{2009A&A...500L..41L} recently
reported the detection of a weak magnetic field for the young (a few
hundred Myr) normal A-type star Vega. \citet{2010arXiv1006.5868P}
meanwhile confirmed the detection using the NARVAL and ESPaDOnS
instruments. The field distinguishes itself, however, in several
aspects from the field of Ap stars. Averaged over the visible part of
the stellar surface the field strength does not exceed 1 G. The field
topology is comparatively complex, putting it at odds with the fossil
field hypothesis for which more simple, bipolar like geometries are
expected.
\subsection{Herbig Ae / Be Stars}
\label{sec:obs_haebe}
Herbig Ae / Be stars (HAeBe) are the progenitors of main sequence A /
B stars. A fraction of 5 - 10\% of the later are known to have a
magnetic field, the magnetic Ap stars. Their field typically is well
organized with a dominant dipole component of around 300 G or more at
the surface. As the envelope of Ap stars is not convective but
radiative, two hypothesis have been brought forward to explain the
presence of the observed magnetic fields. One idea is that the field
is generated below the stellar surface, in the stellar core or
envelope, from where it is transported to the surface of the star (see
Sect.~\ref{SubSec:Models_Dynamos}). The other suggestion is that the
magnetic field is of fossil origin, that it existed already during the
formation of the Ap star (see Sect.~\ref{SubSec:Models_Fossil}). This
second hypothesis lead to intense measurement campaigns in recent
years to search for magnetic fields in HAeBe stars, the
progenitors of Ap / Bp stars~\citep{2004A&A...428L...1H,
2007MNRAS.376.1145W, 2007A&A...463.1039H, 2008MNRAS.385..391A,
2008CoSka..38..235A, 2008CoSka..38..443P, 2009A&A...502..283H,
2009arXiv0901.0347W}.

A synthesis of the results from the above campaigns can be found
in~\citet{2009arXiv0901.0347W}. Based on observational data obtained
with the ESPaDOnS instrument on the Canada-France-Hawaii Telescope,
this survey is the largest and most sensitive published survey for
magnetic fields in HAeBe stars we are aware of. The survey
covers about 130 HAeBe stars with masses from about 2 to 13
M$_{\odot}$. Those stars in which magnetic fields are observed
correspond to about 7\% of the observed stars. The detected magnetic
fields all show organization on large scales. Those three stars which
so far have been studied in greater detail, namely V380 Ori, HD
72106A, and HD 200775, display an important dipole component with
characteristic polar strength on the order of 1 kG, have ages between
0.1 and 10 Myr, and rotation periods between half a day and a few days
(see Table 1 in~\citet{2007IAUS..243...43A} for details).

The 7\% are remarkably similar to the 5 - 10\% of magnetized A / B
stars. Concerning the field strength,~\citet{2009arXiv0901.0347W}
argue that the dipole intensity derived by~\citet{2008MNRAS.385..391A,
2008CoSka..38..235A} and~\citet{2008MNRAS.391..901F} for the detected
magnetic HAeBe stars is compatible with the typical field strengths of
Ap stars (few kG) if flux conservation as described
in~\citet{2008CoSka..38..235A} applies when the HAeBe stars evolve
towards the main sequence. 

One may wonder whether the fraction of 7\% is detection limited or
physical in origin. Using Monte Carlo simulations to study
synthetic populations of HAeBe stars and their observational
signatures in FORS1 data,~\citet{2007MNRAS.376.1145W} place the
following limits on the non-detections of magnetic fields in HAeBe
stars: within statistical uncertainty, the observational data is
consistent with a distribution of non-magnetic stars; the observations
are inconsistent with a {\it uniform} population of magnetic stars
with dipole intensities above 300 G or 500 G in case the dipole axis
is aligned or perpendicular to the rotational axis of the star,
respectively. 

While the above Monte Carlo study was carried out for the FORS1
instrument,~\citet{2009arXiv0901.0347W} point out that the results
essentially apply to ESPaDOnS as well, where they have a 90\% chance of
detecting even weak fields of only 300 G. The authors conclude that,
on the one hand, it is likely that some magnetic HAeBe stars still
went undetected in their survey, because of a too weak or too
unorganized magnetic field.  On the other hand, it is highly probable
that they captured the majority of magnetic HAeBe stars in their
sample. Consequently, they do not expect the fraction of 7\% magnetic
stars among all HAeBe stars to change substantially with future
surveys.
 
Somewhat a by product of the above surveys is the finding that only
two of the detected magnetic HAeBe stars, HD 72106A and NGC 6611-601,
display chemical peculiarities similar to those of Ap
stars~\citep{2009arXiv0901.0347W}. According to the authors, this
aspect of the data has, however, not yet been investigated in detail.

The above results seem compatible with the idea of the fossil origin
of the magnetic fields observed in Ap stars. Arguments in favor of
this point of view are, in particular, the definite detection of
magnetic fields in the HAeBe stars, the progenitors of Ap stars, and
the observation that magnetic fields of Ap stars on the main sequence
apparently weaken with the age of the star. Whether the surface
field observed in HAeBe stars are directly related to the surface
fields of Ap stars remains, however, debated. 
\citet{2009A&A...502..283H} observe 21 HAeBe stars and detect 
a magnetic field in 6 of them, for which they further find indications
of a decline of surface field strength with age. The age range spanned
by the sample is 2 - 14 Myr. On this basis, the authors speculate that
the surface fields of HAeBe may ultimately vanish and thus need not be
direct progenitors of surface fields in Ap stars. Instead, they
suggest that the surface fields in both Ap stars and HAeBe stars are
due to emergence of interior fields in the wake of slow stellar
rotation and that, consequently, the fields observed in Ap and HAeBe
stars need not be directly related.

Also enigmatic remains the origin of the field in the HAeBe stars.
At least three hypothesis have been brought forward. One idea is that
the magnetic fields of HAeBe stars are generated by a convective
dynamo during the formation process of the star. A second suggestion
is that the magnetic field results in the course of the merger of two
stars during pre-main-sequence or early main-sequence evolution. Yet
another idea is that the magnetic fields stem from the molecular cloud
in which the stars are born. All ideas will have to explain the fact
that probably only a small number of HAeBe stars have a (detectable)
magnetic field. A possible mechanism might again be the critical field
strength introduced earlier on, below which the magnetic field is
wound up and destroyed by instabilities.
\subsection{O and early B Stars}
\label{sec:obs_o}
Up to now, less than 20 O-type and early B-type stars have definite
detections of surface magnetic fields. Analysis thus is mostly restricted to
case studies, unlike for intermediate mass magnetic stars where
several hundred detections~\citep{2009MNRAS.394.1338B} enable
statistical data analysis. Much more magnetic OB-stars might be
expected, based on indirect indicators. For example, cyclic wind
variability in OB stars has long been related to the potential
presence of magnetic fields, in particular dipole
fields~\citep{1981ApJ...251..133M, 1997A&A...327..281K,
2001A&A...366..585R, 2003ASPC..305..333F}. Searching the IUE archive
for corresponding signatures,~\citet{2005ASPC..337..114H} identify 100
O-stars with statistically significant variability, 60 of which show
periodic variability. This large fraction of stars showing periodic
variability contrasts with the fraction of stars with clear magnetic
field detections. On the other hand, a systematic search
by~\citet{2009IAUS..259..449P} for a direct link between field
strength and X-ray emission, as suggested
by~\citet{2005ApJS..160..557S}, yielded no confirmation of this
hypothesis. Another indication for a large number of magnetic
OB stars comes from non-thermal radio emission of colliding wind
binaries (see Section~\ref{Sec:MagneticBinaries}).

A potential explanation for the comparatively few detections lies
certainly in the difficulty of direct field measurements in O-type
stars as they have rather few suitable spectral lines in spectral
ranges amenable by today's instruments. \citet{2008A&A...483..857S}
use cyclic wind variability to select 25 OB candidate stars, for which
they then carried out observations with the MUSICOS instrument. No
clear field detection was obtained for any of these stars.

Definite field detection in early B-type stars (estimated field in
parentheses) exist for $\beta$ Cep (360 G), $\zeta$ Cas (335 G),
$\tau$ Sco (500 G), $\xi^{1}$ CMa (300 G, line of sight), Par 1772
(1150 G), and $\nu$ Ori (620 G)~\citep{2000ASPC..214..324H,
  2003A&A...406.1019N, 2006MNRAS.370..629D, 2006MNRAS.369L..61H,
  2008MNRAS.387L..23P}.

For O-type stars, five reliable detections of a magnetic field are
known: $\theta^{1}$ Ori C~\citep{2002MNRAS.333...55D}, HD
191612~\citep{2006MNRAS.365L...6D}, $\zeta$ Ori
A~\citep{2008MNRAS.389...75B}, HD 57682~\citep{2009MNRAS.400L..94G},
and HD 108~\citep{2010MNRAS.tmp..972M}. Additional detections that are
generally regarded as tentative, as they are based on FORS1
observations only, exist for HD 36879, HD 148937, HD 152408, and HD
164794~\citep{2008A&A...490..793H}. See~\citet{2009MNRAS.398.1505S}
for a more detailed discussion.

$\theta^{1}$ Ori C is a very young star, upper age limits ranging
between 0.2 and 0.6 Myr~\citep{2002MNRAS.333...55D,
2006MNRAS.365L...6D}. Observed Stokes V signatures are consistent with
a dipole field of 1.1 $\pm$ 0.1 kG~\citep{2002MNRAS.333...55D}.  The
magnetic wind confinement parameter $\eta_{*}$ (see
Section~\ref{Sec:MagneticWind}) is about 20. The star has a rotational
period of 15.4 days, a mass of about 40 M$_{\odot}$, and its spectral
type varies between O4 and O6.  $\theta^{1}$ Ori C is a binary with a
period of 11 - 26 yr~\citep{2008RMxAA..44..331N}.

The Of?p star HD 191612 varies between spectral types O6 and
O8~\citep{2004ApJ...617L..61W}. The dipole strength inferred from
ESPaDOnS data is 1.5 kG, the magnetic confinement parameter is
$\eta_{*} \sim~10$~\citep{2006MNRAS.365L...6D}. The star has a
rotation period of 538 days.  The estimated age is between 3 and 4
Myr. Its mass is similar to that of $\theta^{1}$ Ori C, about 40
M$_{\odot}$. Like $\theta^{1}$ Ori C, HD 191612 is a binary with a
period of 1542 days~\citep{2008RMxAA..44..331N}.

$\zeta$ Ori A, an O9.7 super-giant, has by far the weakest magnetic
field detected so far in a hot massive star. Observations were done
by~\citet{2008MNRAS.389...75B} using NARVAL. Their data analysis
indicates a much more intricate field topology than a simple dipole
field.  Local surface magnetic fluxes are estimated at only a few tens
of Gauss, certainly not exceeding 100 G. Values for the magnetic
confinement parameter $\eta_{*}$ are between 0.03 and 0.07. A rotation
period of 7 days is found, as well as a mass of 40 M$_{\odot}$ and an
age of 5 - 6 Myr.

HD~57682 is an O9 sub-giant. Based on ESPaDOnS
data,~\citet{2009MNRAS.400L..94G} inferred a magnetic dipole field of
about 1680 G. The authors point out that a single inclined dipole
field is equally consistent with the observational data as individual
dipole configurations. The rotational period is estimated at about
31.5 days, the mass at about 17 M$_{\odot}$. The wind confinement
parameter $\eta_{*}$ is estimated to lie in a range $4 \times 10^{3}$
to $2 \times 10^{4}$.

For HD 108, the second Of?p star with a clear field
detection,~\citet{2010MNRAS.tmp..972M} infer a bipolar large-scale
field of at least 0.5 kG and most likely on the order of 1 - 2 kG,
based on longitudinal field measurements of 100 - 150 G obtained with
ESPaDOnS and NARVAL. For the magnetic confinement parameter $\eta_{*}
\ge 100$ is obtained. The rotation period is several decades, probably
between 50 - 60 years, the age is about 4 Myr. Estimates for the
stellar mass range of 35 - 43M$_{\odot}$~\citep{2008RMxAA..44..331N,
2010MNRAS.tmp..972M}.

Some properties of the above five O-type stars fit nicely with
theoretical expectations. For example, older stars are expected to
rotate more slowly as magnetic braking has had more time to act. The
young and fast rotating star $\theta^{1}$ Ori C, as well as the older
and more slowly rotating stars HD 191612 and HD 108 seem to confirm
this idea~\citep{2010MNRAS.tmp..972M}. Why $\zeta$ Ori A does not fit
into this scheme one can only speculate~\citep{2008MNRAS.389...75B}.
Support for theoretical models of rotational braking due to a
magnetized line-driven wind came recently
from~\citet{2008A&A...485..585M}. Analyzing 31 years of observational
data of the helium-strong star HD 37776, they found a lengthening of
the 1.5387 d period by 17.7$\pm$0.7
s. Similarly,~\citet{2010ApJ...714L.318T} find for the Bp star $\sigma$
Ori E a linear decrease of the rotational period of 77 ms per year,
based on 30 years of observational data. The corresponding spin-down
time of 1.34 Myr corresponds to theoretical expectations.

Yet another star, HD 148937, deserves mentioning here although field
detection is only tentative so far: $-276 \pm 88$ G, based on FORS1
data~\citep{2008A&A...490..793H}.  Spectropolarimetric observations
are not yet available. HD 148937 is interesting as it is the third of
only three known galactic Of?p stars, the other two stars having
definitive field detections (HD 191612 and HD 108). The Of?p class was
originally introduced by~\citet{1972AJ.....77..312W} and recently
slightly refined~\citep{2008RMxAA..44..331N, 2010ApJ...711L.143W}. A
particularity of HD 148937 is the bipolar circumstellar nebula around
it, NGC 6164-6165. As the nebula displays similar chemical anomalies
as HD 148937 (nitrogen 4 times overabundant as compared to solar) it
was suggested to have been formed by an eruption of the Of?p central
star~\citep{1987A&A...175..208L, 1988ApJ...327..859D}.  All three
galactic Of?p stars show nitrogen over-abundances as compared to solar
which, together with other spectral features,
lead~\citet{2003ApJ...588.1025W} to suggest that there may exist a
similarity between Of?p stars and WN9 objects.

HD~148937 has a rotational period of 7
days~\citep{2008RMxAA..44..331N, 2010arXiv1006.2054N}. Based on its
position in the HR diagram, its mass and age are estimated at
55~M$_{\odot}$ and 2 -- 4~Myr,
respectively~\citep{2008RMxAA..44..331N}.  With these properties -
magnetic field, old age, yet short rotation period - HD~148937 would
be much more similar to $\zeta$~Ori~A than to the other two galactic
Of?p stars HD 191612 and HD 108.
\subsection{Late Stages of Massive Star Evolution}
\label{sec:latestages}
Late-type super giants (spectral type F and later) represent the late
evolutionary stages of most massive stars. Characteristics of these
stars are extended radii, helium-burning core, a convective
hydrogen-burning envelope, and slow rotation. A recent systematic
survey of more than 30 late-type super-giants using ESPaDOns revealed
clear detection of magnetic fields in one third of all
stars~\citep{2010arXiv1006.5891G}. Clear detections were obtained for
the following stars (\citet{2010arXiv1006.5891G}, Table 1): $\alpha$
Lep, $\alpha$ Per, $\eta$ Aql, $\beta$ Dra, $\xi$ Pup, $\epsilon$ Gem,
$c$ Pup, 32 Cyg, $\lambda$ Vel. The most massive and also the hottest
star among this sample is $\alpha$ Lep ($~$ 15 M$_{\odot}$,
$T_{\mathrm{eff}} \approx 7200$ K). The lowest mass star is $\beta$ Dra ($~$
5 M$_{\odot}$), the coolest $c$ Pup ($T_{\mathrm{eff}} \approx 3700$ K).
The observations suggest topologically complex fields, longitudinal
magnetic fields are generally below 1 G. The authors speculate that
probably even a larger fraction of late-type super-giants or even all
of them may have a magnetic field, but that the signal to noise ratio
of their measurements is insufficient to detect these fields. They see
their point confirmed by the detection of a magnetic field in
Betelgeuse~\citep{2010A&A...516L...2A} where a signal to noise (S/N)
about twice as high as in their study was crucial.

The case of Betelgeuse and other late-type super-giants is of interest
as these stars occupy an extreme position in parameter space. Their
rotation is too slow that a solar like dynamo could be at work. An
alternative idea for the origin of the observed magnetic fields are
large convection cells~\citep{1975ApJ...195..137S}. The idea obtains
further support from 3D convection
simulations~\citep{2002AN....323..213F, 2004A&A...423.1101D}.

Very high mass stars with initial masses between 25~M$_{\odot}$ and
40~M$_{\odot}$ evolve into Wolf-Rayet (WR) stars towards the end of
their life~\citep{2010ASPC..425...13H}. To date, no direct magnetic
field detection for any WR exists. However, WR 142 recently gained
attention because of the detection of weak ($L_{\mathrm{X}} \approx
7\cdot 10^{30}$~ergs s$^{-1}$ or $L_{\mathrm{X}} / L_{\mathrm{bol}}
\le 10^{-8}$) but hard X-rays with
XMM-Newton~\citep{2009ApJ...693L..44O} and
Chandra~\citep{2010ApJ...715.1327S}. The inferred plasma temperature
is on the order of $10^{8}$ K~\citep{2009ApJ...693L..44O}, which is
difficult to explain in terms of internal shocks in the WR wind. The
authors suggest that the X-rays may be due to the presence of a
stellar magnetic field that deviates the stellar wind and creates
magnetically confined wind shocks (MCWS, see
Section~\ref{Sec:MagneticWind}). The authors estimate that a surface
field strength of some 10 kG would be required, a value they consider
compatible with the order 1 kG measured magnetic fields in the much
larger O-type stars. Alternative explanations have been suggested as
well, most notably inverse Compton scattering in colliding winds in a
(yet unresolved) binary~\citep{2010ApJ...715.1327S}. In fact, the
'colliding winds community' has been assembling indirect evidence for
the existence of magnetic fields in WR+OB and O+OB binaries for
several decades already, mostly in the form of non-thermal radio
emission (see Section~\ref{Sec:MagneticBinaries}).
\subsection{Summary}
\label{sec:summary_obs}
Measurements of Zeeman signatures clearly demonstrate the presence of
dipole fields of about 1 kG strength in four O-type stars. One more
O-type star, $\zeta$ Ori A, has a clear field detection as well, but
the field is much weaker (tens of Gauss) and less ordered. A clear
field detection also exists for the late-type supergiant
Betelgeuse. For WR stars only indirect evidence for a magnetic field
exists, in the form of hard X-ray emission and non-thermal radio
emission from WR+O binaries.

Much more data is available for intermediate mass Ap / Bp stars.
Evidence is growing that probably all Ap / Bp stars have a magnetic
field. The field of these stars often is dipole like with strength
between about 300~G and 30~kG. The lower limit of 300~G is probably
not detection limited but constitutes a physical lower limit
instead. Only one normal A-type star, Vega, is currently known to have
a magnetic field. The field is weak and unordered, resembling in this
sense the field of $\zeta$ Ori A.

Observations from open clusters suggest the field strength in Ap / Bp
stars to decrease with the age of the stars. Apart from field
detections in main sequence Ap / Bp stars, there exist a few field
detections in pre-main-sequence Herbig Ae / Be stars. Measured field
strenghts are compatible with those of Ap / Bp stars if stellar
evolution is taken into account.
\section{Theoretical models for magnetic fields in massive stars}
\label{Sec:Models}
\subsection{Introduction}
%
even strong -- magnetic fields on their surfaces demands for
explanation in terms of analytical and numerical models. Beyond the
interpretation of observational data, a better physical understanding
of magnetic fields in massive stars will allow a more consistent
inclusion of such fields in stellar evolution models\footnote{A very
  good presentation of the physics of rotating stars, including an
  elusive chapter on the role of magnetism, can be found in the book
  by~\citet{2009pfer.book.....M}. We also refer to the book by
  ~\citet{1999stma.book.....M}, which is perhaps the most complete
  compilation of stellar magnetism.}. For the potentially far reaching
consequences of magnetic fields for stellar evolution we refer to
Sect.~\ref{Sec:Implications}. Here we only stress that magnetic fields
relevant for stellar evolution need not necessarily emerge on the
stellar surface. It is sufficient that they dynamically link the
(convective) core and (radiative) envelope of the star. This poses
some difficulty as current observations of magnetic fields in massive
stars (see Section~\ref{Sec:Observations}) deal for the most part with
surface fields only.

The basic challenge associated with magnetic fields in early type
massive stars is that the envelope of such stars is for the most part
radiative. This excludes a solar
type dynamo as the origin of the magnetic field. Two groups of
alternative explanations are favored instead: some other kind of
dynamo, the precise nature of which is under debate, or fossil fields,
which have been preserved since the formation time of the star and
whose precise origin is under debate as well. Both groups of
mechanisms may be required to explain observed surface magnetism in
the upper main sequence: fossil fields for the kG dipole fields
typical of Ap-stars, dynamos for weak (1G) and unordered fields as in
the normal A-type star Vega~\citep{2009A&A...500L..41L,
2010arXiv1006.5868P}.

A common issue in both explanations, dynamo or fossil field, is the
stability or instability of the involved magnetic field. Some dynamo
theories rely on instabilities for closing the dynamo loop, while
fossil field theories rely on a sufficiently stable field
configuration for the field to last over the lifetime of the star. The
relevance of the question is reflected in a quite exhaustive
literature. We sketch here only the basic picture. Further details
will be given later on as needed.

A first point to note is that Ohmic dissipation alone is not efficient
in destroying a stellar magnetic field. The corresponding destruction
time scale is on the order of the life time of the star. Turning to
instabilities instead, it has long been known that the field topology
plays a decisive role. Only mixed poloidal-toroidal topologies may be
sufficiently stable. This hypothesis got recent support from 3D
simulations by~\citet{2004Natur.431..819B} and~\citet{2006A&A...450.1077B} which
show that initially unordered fields often relax to a long-lasting
mixed topology. Several decades earlier, \citet{1956ApJ...123..498P,
1958ApJ...128..361P} showed already that an exact solution of the
hydromagnetic equilibrium exists if the field is of mixed topology,
has a toroidal and a poloidal component. Probably one of the
earliest results in this direction is the
Ferraro-law~\citep{1937MNRAS..97..458F}:{\it "The magnetic field of a
star can only remain steady if it is symmetrical about the axis of
rotation and each line of force lies wholly in a surface which is
symmetrical about the axis and rotates with uniform angular
velocity"}.

By contrast, purely poloidal fields were shown to be unstable if
some~\citep{1973MNRAS.163...77M, 1973MNRAS.162..339W,
1974MNRAS.168..505M} or all~\citep{1977ApJ...215..302F} field lines
are closed outside of the star. A purely toroidal field is unstable to
non-axisymmetric perturbations of low azimuthal order, with a strong
dominance of the $m = 1$ mode~\citep{1973MNRAS.161..365T,
1973MNRAS.162..339W,
1981Ap&SS..75..521G}. Rotation~\citep{1985MNRAS.216..139P} or the
addition of thermal and Ohmic diffusion~\citep{1978RSPTA.289..459A}
help to stabilize such a situation, but cannot suppress the $m=1$
instability. Note that the above considerations refer to the full
field, inside and outside the star. A bipolar field outside of the
star can still exist, provided that some toroidal component exists
within the star.

In the rest of this Section, we review dynamo models proposed for the
different layers of massive stars (in
Sect.~\ref{SubSec:Models_Dynamos}). This includes the question whether
such fields can raise to the surface, for example by buoyancy or due
to meridional circulation. In Sect.~\ref{SubSec:Models_Fossil} we
review fossil field models, which propose that the field observed in
massive stars was enclosed during the process of their formation. A summary
concludes the section (Sect.~\ref{SubSec:Models_Discussion}).
\subsection{Dynamo generated fields and their appearance at the surface}
\label{SubSec:Models_Dynamos}
\subsubsection{Convective core}

An obvious candidate for a place in a massive star where magnetic
fields may be created is its convective core. By mean field dynamo
theory, \citet{2001ApJ...559.1094C} found that $\alpha 2$-, $\alpha 2
\Omega$- and $\alpha \Omega$ dynamos can operate in such cores. In
addition, the authors investigate whether meridional circulation could
bring the field to the surface. They conclude, however, that this is
not the case. The main argument is: {\it In all models with strong
core-to-envelope magnetic diffusivity contrast (presumably closest to
reality) whenever circulation is strong enough to carry a significant
magnetic flux, it is also strong enough to prevent dynamo action.}

In a follow-up paper, \citet{2003ApJ...586..480M} looked at
alternative transport mechanisms -- buoyant, centrifugal, Coriolis,
magnetic tension, and aerodynamic drag forces -- whether they may be
able to bring the field generated in the convective core to the
surface. They found that thin, axisymmetric, toroidal flux tubes can
be advected from the outer boundary of the convective core to the
atmosphere within a life-time of a 9 solar mass star. The trajectories
of the raising ring are more or less parallel to the rotation axis of
the star. They also found that the rise-time for smaller rings is
faster. \citet{2005MNRAS.356.1139M} argue, however, that correct
inclusion of yield gradients may suppress the buoyancy.

The idea of substantial field generation in the convective cores of
A-type stars is further supported by more recent 3D simulations
by~\citet{2004ApJ...601..512B} and~\citet{2005ApJ...629..461B}. Mega-Gauss
fields are found in the core, at its boundary layer to the radiative envelope.
\citet{2009ApJ...705.1000F} find, by means of 3D simulations of the 
innermost 30\% of a 2 M$_{\odot}$  A-type star, that the
presence of a fossil field in the radiative envelope which threads 
into the convective core can considerably augment the field
generated by a core dynamo. Apart from pointing out that stronger
fields generally have a better chance of reaching the surface, the
authors do not comment on whether or not the core dynamo generated
fields may emerge at the stellar surface.

Taken together, the above papers demonstrate that dynamo action in the
convective cores of upper main sequence stars is likely to exist, but
that it is unclear whether the generated fields can reach the stellar
surface.
\subsubsection{Radiative envelope}
Dynamos that can operate without a convection-related
$\alpha$-mechanism have been known for several years. An example is
the dynamo generated by differential rotation and the
magneto-rotational instability in
accretion-disks~\citep{1996ApJ...464..690H}. The recognition that a
similar dynamo mechanism, one without a convection-related
$\alpha$-mechanism, may also take place in the stably stratified
radiative regions of stars had far reaching consequences. The two
corresponding papers~\citep{1999A&A...349..189S, 2002A&A...381..923S}
are seminal for two reasons. Firstly, because they re-shape the theory
on stably stratified radiative regions and, secondly, because of their
impact on stellar evolution models, the topic of
Sect.~\ref{Sec:Implications}. Whether such a dynamo can indeed operate
in real stars of the upper main sequence has been
questioned~\citep{2007A&A...474..145Z} and is the topic of a still
ongoing debate.

One starting point of the above developments were results of
helioseismology, which found that the radiative solar core is
uniformly rotating. At the tachocline, this uniform rotation switches
rather abruptly, in only 0.05 R$_{\odot}$, to the differential
rotation of the envelope, which itself shows a 30 percent contrast in
its rotation between equator and poles. The observed uniform rotation
of the solar core was puzzling. A potential agent for the necessary
efficient transport of angular momentum could be magnetic fields, but
under the prevailing radiative conditions a conventional 'convective
dynamo' -- and thus magnetic fields -- would not exist.  The
connection of this problem with the radiative envelopes of upper main
sequence stars is obvious.

\begin{figure}[t]
\centerline{
\includegraphics[width=8cm]{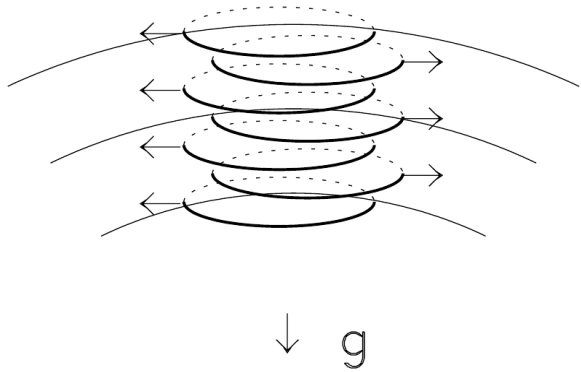}
}
\caption{Unstable displacements in an azimuthal field near the pole.
  Shown is the $m=1$ mode, which occurs under the widest range of
  conditions. The displacements are along horizontal surfaces
  (indicated by arcs). The Figure is taken
  from~\citet{1999A&A...349..189S}, his Figure~1.}
\label{fig:spruit}
\end{figure}

\citet{1999A&A...349..189S} re-investigated the stability of magnetic
fields in stably stratified, but differentially rotating stars. He
reviews the processes which contribute to the evolution of an
initially weak magnetic field in a differentially rotating star:
rotational smoothing and five instabilities, among them
magnetorotational instability, buoyancy instability, and pinch-type
instabilities. The Pitts-Taylor instability, a pinch-instability
driven by the free energy of the field itself, is found to set in
first. It generically occurs in a region near the pole, in the form of
an m = 1 displacement of the field lines along horizontal
surfaces (see Fig.~\ref{fig:spruit}). The analysis is complemented by an investigation of the
relevance of thermal and magnetic diffusion on these
instabilities. Based on heuristic arguments, concrete numbers are
provided that characterize the behavior of the instabilities.

In a follow up paper, \citet{2002A&A...381..923S} argues that on the
basis of the Pitts-Taylor instability a dynamo can operate in the
stably stratified radiative regions of stars. He states {\it 'the
generation of a magnetic field in a star requires only one essential
ingredient: a sufficiently powerful differential rotation. The
recreation of poloidal field components which is needed to close the
dynamo loop can be achieved by an instability in the toroidal
field.'}. The paper gives concrete formulas for the field
strength in radial and azimuthal direction produced by a dynamo of
this type. The crucial parameter is the differential rotation,
which of course will be affected by the presence of a field.

Numerical confirmation of the dominance of the Pitts-Taylor
instability and the associated growth rate predicted
by~\citet{2002A&A...381..923S} is reported
by~\citet{2006A&A...453..687B}. The $m = 1$ kink-mode is indeed found
to be the dominant instability in a toroidal field, where the field
strength is proportional to the distance from the axis, such as the
field formed by the winding up of a weak field by differential
rotation. The growth rate of the instability for initially weak fields
is found to agree with the analytic predictions.

\citet{2006A&A...449..451B} further report that he finds the 
dynamo mechanism to work: an initial small poloidal field is wound
up. The resulting toroidal field is subject to the Pitts-Taylor
instability, thus producing a poloidal field again, which is then
wound up again. This process continues up to a certain saturation
level, at which {\it 'the field is being wound up by differential
rotation at the same rate as it is decaying through its inherent
magneto-hydrodynamic instability.'} 

Doubts on the above results were raised
by~\citet{2007A&A...474..145Z}. While numerically investigating the
solar tachocline they found inconsistencies between their results and
results reported by~\citet{1999A&A...349..189S, 2002A&A...381..923S}
and~\citet{2006A&A...449..451B}. In their paper,
\citet{2007A&A...474..145Z} first provide a more rigorous analysis
than the more heuristic argumentation
by~\citet{1999A&A...349..189S}. They confirmed the scaling of Spruit
but give better bounds and factors (differences are of order
1). Comparing their analytical results with numerical results from the
ASH code \citet{2007A&A...474..145Z} find excellent agreement. In
strong contrast to~\citet{2006A&A...449..451B} they observe: {\it
  'Although the instability generated field reaches an energy
  comparable to that of the mean poloidal field, that field seems
  unaffected by the instability: it undergoes Ohmic decline, and is
  neither eroded nor regenerated by the instability.'}\,\, They
conclude that {\it 'In our simulations we observe no sign of dynamo
  action, of either mean field or fluctuation type, up to a magnetic
  Reynolds number of $10^5$.'}\,\, The authors also doubt that a large
scale dynamo-loop can be closed in the way suggested
by~\citet{2002A&A...381..923S} and~\citet{2006A&A...453..687B}, but
suggest an alternative instead.

The contradictory numerical results demand for some details, given the
importance of the question. The two simulation codes employed differ
in several respects, including magnetic boundary conditions and the
way differential rotation is enforced (see~\citet{2007A&A...474..145Z}
for details). Maybe the most important difference is the way the
equations are solved. The ASH code used by~\citet{2007A&A...474..145Z}
is of pseudo-spectral type and should allow to reach a magnetic
Reynolds number of $10^{5}$ in the simulations under debate, which the
authors consider enough to detect any dynamo action if
existent. \citet{2006A&A...453..687B} use a 6$^{\mbox{th}}$ order
finite difference code on a Cartesian domain. Whether or not dynamo
action sets in in the model by~\citet{2006A&A...453..687B} crucially
depends on the artificially enforced differential rotation. Adding a
uniform rotation does not suppress the dynamo but the field strength
shows an oscillatory behavior.  Further investigations are clearly
needed to disentangle the issue.

Apart from generating a magnetic field by dynamo in the radiative
envelope - if possible at all - there is again the question of whether
such a field may eventually become visible at the surface. Two
possibilities come to mind: the field is generated near the surface or
the field is generated further below and transported to the surface.
As to the first possibility, \citet{2007A&A...474..145Z} agree
with~\citet{2005A&A...440.1041M} who pointed out that there is too
little differential rotation in the layers immediately below the
surface to operate the Taylor-Spruit-dynamo (see
Sect.~\ref{Sec:Implications}). The second possibility is rejected as
well, as meridional circulation very likely does not reach the surface
layers and as the field generated would be too weak for the magnetic
buoyancy instability. The raising fields reported in numerical
simulations by~\citet{2005MNRAS.356.1139M} rely on field strengths on
the order of 1 MG. These field strengths, in turn stem from their
stellar structure model~\citep{2004MNRAS.348..702M}, which neglects
the back-coupling of the magnetic field on the rotation rate, thus
resulting in a very high differential rotation. In fact, as will be
discussed in Sect.~\ref{Sec:Implications}, even a weak field will
suppress differential rotation to a high degree and the consistent
rotation rate is nearly flat. Since the performance of the dynamo
depends on the root of $d \Omega/dr$, this makes a huge difference.
\subsubsection{Giant Convection Cells and Surface Dynamos}
Yet another form of dynamo action has been proposed to explain observed
magnetic fields in red super giants (see Sect.~\ref{sec:latestages}),
the late stages of stellar evolution of massive stars. 
\citet{2002AN....323..213F} performed detailed 3D radiation hydrodynamic
simulations of the late-type super-giant Betelgeuse. The simulations,
in comparison with observations, convincingly demonstrated the
existence of giant convection cells. In a subsequent paper,
\citet{2004A&A...423.1101D} presented magneto-hydrodynamical
simulations of non-linear dynamo action in Betelgeuse and found that
the giant convection cells can indeed result in the formation of
surface magnetic fields of up to 500 G. More recent simulation results
further support the existence of giant convection
cells~\citep{2010A&A...515A..12C, 2009A&A...506.1351C}.
\subsection{Fossil fields}
\label{SubSec:Models_Fossil}
An alternative to dynamo generated fields are field configurations
which -- once formed -- are stable over an essential part of the
life-time of the star. Such fields are generally called {\it fossil
fields}. They are particularly attractive to explain the existence of
magnetic fields in Ap-stars and the like: stars with a static, strong,
large-scale magnetic field. Although, as discussed in the last
section, there are competing ideas which ascribe such fields to dynamo
action in the convective core with subsequent raise to the surface
\citep{2001ApJ...559.1094C, 2003ApJ...586..480M}, a fossil origin of
these fields seems more attractive. One main reason in favor of a
fossil origin are that no correlation is observed between field
strength and stellar rotation, as would be expected for a dynamo
dragging its force from the differential rotation of the star. Also,
the field strengths are so large that dynamo action is
suppressed~\citep{1999A&A...349..189S}. The two main questions
associated with fossil fields are their origin and their stability.

We already briefly touched the question of the stability of a magnetic
field over essentially the life time of a star at the beginning of
Sect.~\ref{Sec:Models}. There we pointed out that a mixed
poloidal-toroidal topology is a necessary prerequisite. The results
quoted there have been further refined in recent years. For an
overview see for example~\citet{2005LNP...664..183M}.
\begin{figure}[t]
\centerline{
\includegraphics[width=8cm]{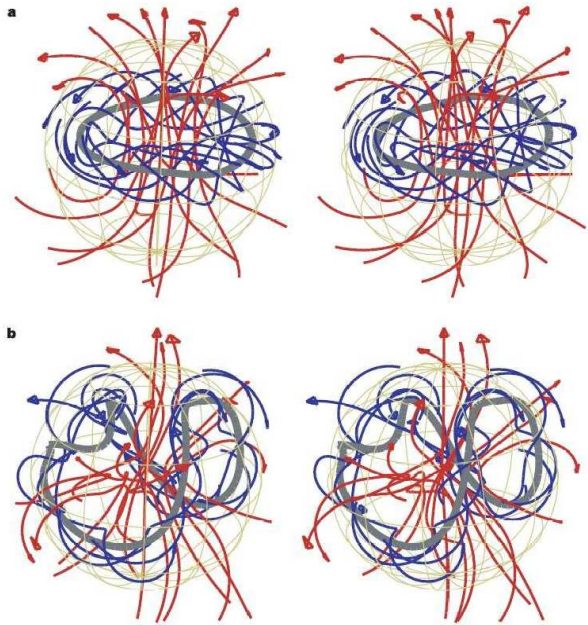}
}
\caption{Structure of stable magnetic fields, as found with
  three-dimensional numerical simulations. {\bf a:} Stereographic view
  of the long-lived magnetic field configuration as it evolves from a
  random initial condition. The stable core of the configuration is
  formed by a torus of twisted field lines inside the star (blue, with
  axis of torus shown in grey). Field lines that pass through the
  stellar surface (red) are stabilized by the torus. The configuration
  slowly evolves outwards by magnetic diffusion. When the torus
  reaches the surface it becomes unstable (stereographic view {\bf
    b}). From~\citet{2004Natur.431..819B}, their Figure~1.}
\label{fig:stable_field_config}
\end{figure}

More recently, \citet{2004Natur.431..819B} and
\cite{2006A&A...450.1077B} identified by means of numerical simulations 
a first concrete field topology that is stable over long time, illustrated in 
Fig.~\ref{fig:stable_field_config}. They
follow the evolution of a random field initially concentrated in a
sphere of radius $r_m < R^*$ within a star of radius $R^*$ and using a
polytrope with index $n=3$. For $r_m \lapprox 0.5R ^*$, the initial
configuration rapidly changes, within a few Alv\'en times the field
strength decays strongly before stabilizing into a configuration which
is always similar, independent of the concrete initialization: {\it
'the poloidal field component is very similar to that which would be
produced by an azimuthal current loop near the equator of the
star. The toroidal component then threads along this loop.'} The field
subsequently diffuses outwards, on a much longer time scale. While
roughly preserving the overall configuration, the poloidal component
gets stronger compared to the toroidal component. Once the imaginary
current associated with the toroidal field component reaches the
surface, the entire field rapidly decays.  For initial field
realizations of $r_m \gapprox 0.5$, the field directly enters this
final distorted state and decays rapidly.

The authors predict that the first appearance of the field on the
surface is after about 1/3 of the life-time of the star, that the
field strength subsequently should increase, and that contributions
from lower order modes should gradually add to the initially bipolar
surface field. Comparison of these predictions with observations shows
a mixed picture. \citet{2000ApJ...539..352H} find that magnetic Ap
stars are typically older than 30\% of their total main sequence life
span. By contrast, \citet{2008A&A...481..465L} find no such age
barrier. In addition, these authors find a decrease of magnetic field
strength with the main sequence age of the star. As possible
explanations for the discrepancy they suggest the non-consideration of
both, rotation and the changing stellar radius and structure with age,
in the simulations by~\citet{2004Natur.431..819B}.

The above picture essentially still holds today, although it has been
refined in details since~\citep{2010HiA....15..161B}. Using numerical
simulations, \citet{2009MNRAS.397..763B} found that for a stable field
topology the poloidal component must not be significantly stronger
than the toroidal component, while the toroidal may well be much
stronger than the poloidal component. \citet{2009A&A...499..557R}
investigated the effect of a non-barotropic stratification.
\citet{2010MNRAS.405.1845M} constructed a simple analytical model of
tori as those found in numerical models and show that radiative heat
transfer, Archimedes' principle, Lorentz force, and Ohmic decay all
play a significant role.  \citet{2010A&A...517A..58D} largely
generalized the Prendergast models~\citep{1956ApJ...123..498P,
  1958ApJ...128..361P} to barotropic magneto-hydrostatic equilibrium
states of realistic stellar interiors. The analytic solutions they
obtain strongly resemble the numerical results
by~\citet{2004Natur.431..819B} and~\cite{2006A&A...450.1077B}.
Although stability has not yet been proofed for the analytical model,
this combined effort between numerical and analytical work is
exemplary and has already proofed to be seminal. Note, however,
that~\citet{2010ApJ...724L..34D} tested the model numerically and
found no significant change of the configuration over 10 Alfv\'enic
times.

Turning now to the second corner stone of the fossil field hypothesis,
namely the physical origin of these fields, we find a much less
settled debate. At least three ideas have been put forward and are
being investigated today: field generation during the convective
pre-main sequence evolution, field generation in the wake of early
binary mergers~\citep{2009MNRAS.400L..71F}, and fields inherited from
the molecular cloud in which the stars are
born~\citep{2004MNRAS.355L..13T, 2006MNRAS.367.1323F}. Not
surprisingly, each idea has its pros and cons.

So far, only the most simple model, the flux-conservation model put
forward by \citet{2004MNRAS.355L..13T}
and~\citet{2006MNRAS.367.1323F}, is worked out to a stage which allows
confrontation with observations. The model relies on flux conservation
and assumes that the field of a star is built by the magnetic flux
trapped in the collapsing pre-stellar cloud. It then assumes that the
magnetic flux is given as the sum of two Gaussians in the logarithm
with dispersions $\sigma_i$ with appropriate weightings. The model
builds up on the remarkable fact that Ap and Bp-stars as well as
magnetic white dwarfs have magnetic fluxes of the same order.
Moreover, when fields of magnetic white dwarfs are appropriately
rescaled, the distribution of their strengths matches to a good degree
the distribution of the same quantity of neutron stars. They also
point out that the flux of $\theta$ Orion~C ($1.1 ×
10^{27}$~G~cm$^2$) and HD191612 ($7.5 × 1027$~G~cm$^2$) is remarkably
similar to the flux of the highest-field magnetar SGR 1806−20 ($5.7
× 10^{27}$~G~cm$^2$).

For massive main sequence stars, this model predicts a distribution
which peaks at 46 G with 5 percent of the stars having a field in
excess of 1 kG (originally 8 percent, but see the footnote
in~\citet{2008MNRAS.387L..23P}). The still very small sample of
massive stars in the Orion Nebular Cluster (ONC) allows a first test
of this prediction. As reported by~\citet{2008MNRAS.387L..23P}, the ONC
contains nine massive OB stars, ranging from B3 V ($\sim
8$~M$_{\odot}$) to O7 V ($\sim 40$~M$_{\odot}$), three of them showing
a very strong field. These authors find that, according to the
multinomial distribution of~\citet{2006MNRAS.367.1323F}, the
probability of obtaining the distribution of magnetic field strengths
observed in the ONC is about 1 per cent. As pointed out
by~\citet{2008MNRAS.387L..23P}, the sample is much too small to be
representative. Also, the ONC may just be particularly
magnetized, thus biasing the sample.  Nevertheless, this demonstrates
that observations now are at a level which starts to allow
discrimination between different models.
\subsection{Summary}
\label{SubSec:Models_Discussion}
Theory basically offers two hypothesis for the origin of magnetic
fields in intermediate mass and massive stars: fossil fields, which
existed already before the star reached the main sequence, and several
kinds of dynamo generated fields. Fossil fields are the preferred 
explanation for strong and ordered (dipole) fields, whereas
dynamo generated fields are believed to be less ordered and weaker.
In fact, dynamo generated fields may not even emerge on the surface.  

The details of both hypothesis are not well known so far. The origin
of the fossil fields -- from the parent molecular cloud, an early
merger, or pre-main-sequence convection -- are debated. Simulation
results disagree on whether or not a dynamo can operate in the
radiative envelope of a massive star. Similarly, it is unclear whether
fields from the convective core can reach the surface of the star.

A particular case are surface fields in late-type supergiants like
Betelgeuze, which numerical simulations indicate to be due to a
surface dynamo operating in the outer convection layers of the star.
\section{Magnetic fields in massive stars: implications for stellar evolution}
\label{Sec:Implications}
The role of magnetic fields in stellar evolution is an area of research
which largely remains to be explored and represents a top priority
challenge in astrophysics. This is illustrated by the fact that the
topic of magnetic field has been identified with rotation as one of
the areas of discovery potential in the report ``New Worlds, New
Horizons in Astronomy and Astrophysics'' written by the Committee for
a Decadal Survey of Astronomy and Astrophysics (2010).
 
Magnetic fields have probably a very crucial impact at the two extreme
points of the lifetime of a star, namely during their formation and
when they disappear in a supernova explosion.  But the magnetic field
has also likely an important impact, at least when strong enough,
during the hydrostatic phases of stellar evolution.

At the moment, stellar evolution computations taking into account the
effects of magnetic fields remain rare. The available models have
focused on exploring the consequences of the Tayler-Spruit dynamo
mentioned in Sect.~3.2.2 above. In the present Section we shall
briefly describe the main results of these computations.

Let us recall that in the Tayler-Spruit dynamo
\citep{1999A&A...349..189S, 2002A&A...381..923S}, a pristine poloidal
magnetic field is amplified through a dynamo effect. In a first step,
the poloidal field is sheared by differential rotation. This produces
a toroidal component which is sensitive to a pinch-type instability
(the Pitts \& Tayler instability). This generates an amplified
poloidal component. The process can then begin again and the dynamo
loop is closed.  The energy needed for this amplification is obtained
from the excess energy contained in differentially rotating layers.

The amplification will of course not be pursued indefinitely. The
regulating mechanism is due to two counteracting effects: on one hand
the source of the instability, which is differential rotation, will be
progressively eroded by the strong radial coupling exerted by the
increasing poloidal magnetic field, on the other hand meridional
currents, which become stronger when the angular velocity gradient
flattens, will rebuild the differential rotation. As a result of these
two counteracting effects an ``equilibrium profile'' of the angular
velocity in the star is reached.  This angular profile presents the
right degree of shear to sustain both the dynamo that continually
erodes the shear and the meridional currents which continually rebuild
it. Such a dynamo will produce internal magnetic fields of the the
order of 10$^4$ G in massive stars with an initial rotation of 300 km
s$^{-1}$ \citep{2005A&A...440.1041M}.

The theory and the equations describing these effects are presented in
\citep{1999A&A...349..189S, 2002A&A...381..923S} .  In this
theoretical frame, two modifications of the equations have been
brought by further works: 1) using the condition that the energy of
the magnetic field created by the Tayler-Spruit dynamo cannot be
larger than the energy excess present in the differential rotation,
\citet{2003A&A...411..543M} propose a criterion for the existence of
the magnetic field in stellar interior; 2) \citet{2004A&A...422..225M}
avoid the simplifying assumptions that either the mean molecular
weight- or the temperature-gradient dominates, but they treat the
general case and also account for the nonadiabatic effects, which
favor the growth of the magnetic field.

An interesting property of the Tayler-Spruit dynamo is that it can
explain the internal near solid body rotation of the Sun
\citep{2005A&A...440L...9E}. However some critics of the theory have
been presented: \citet{2007A&A...474..145Z}, using the 3-dimensional
ASH code conclude that, although the Pitts \& Tayler instability is
present in their simulations, the mean poloidal field remains
unaffected by it and thus that the dynamo loop is not closed.
\citet{2007ApJ...655.1157D} critic the fact that an important 
approximation used in the derivation of the equations is only valid
for small length scales while, for obtaining the equations presented
in \cite{2002A&A...381..923S}, it is applied for scales of the order
of the radius of the star.

Despite these difficulties, the effects of the Spruit-Tayler dynamo are
still explored in stellar models.  A justification for using this
theory is that, as discussed in \citet{2009IAUS..259..311M}, except
for the expression of the growth rate of the instability, the
equations presented in \citet{2002A&A...381..923S} are quite general
and can be derived by simply imposing that the rate of dissipation of
the excess energy of the shear is equal to the production rate of
magnetic energy. Said in other words, everything can be computed once
the growth rate is known. As long as this growth rate is short
enough, magnetic field will produce solid body rotation during the
Main-Sequence phase.

\begin{figure}
\includegraphics[width=2.3in,height=2.6in]{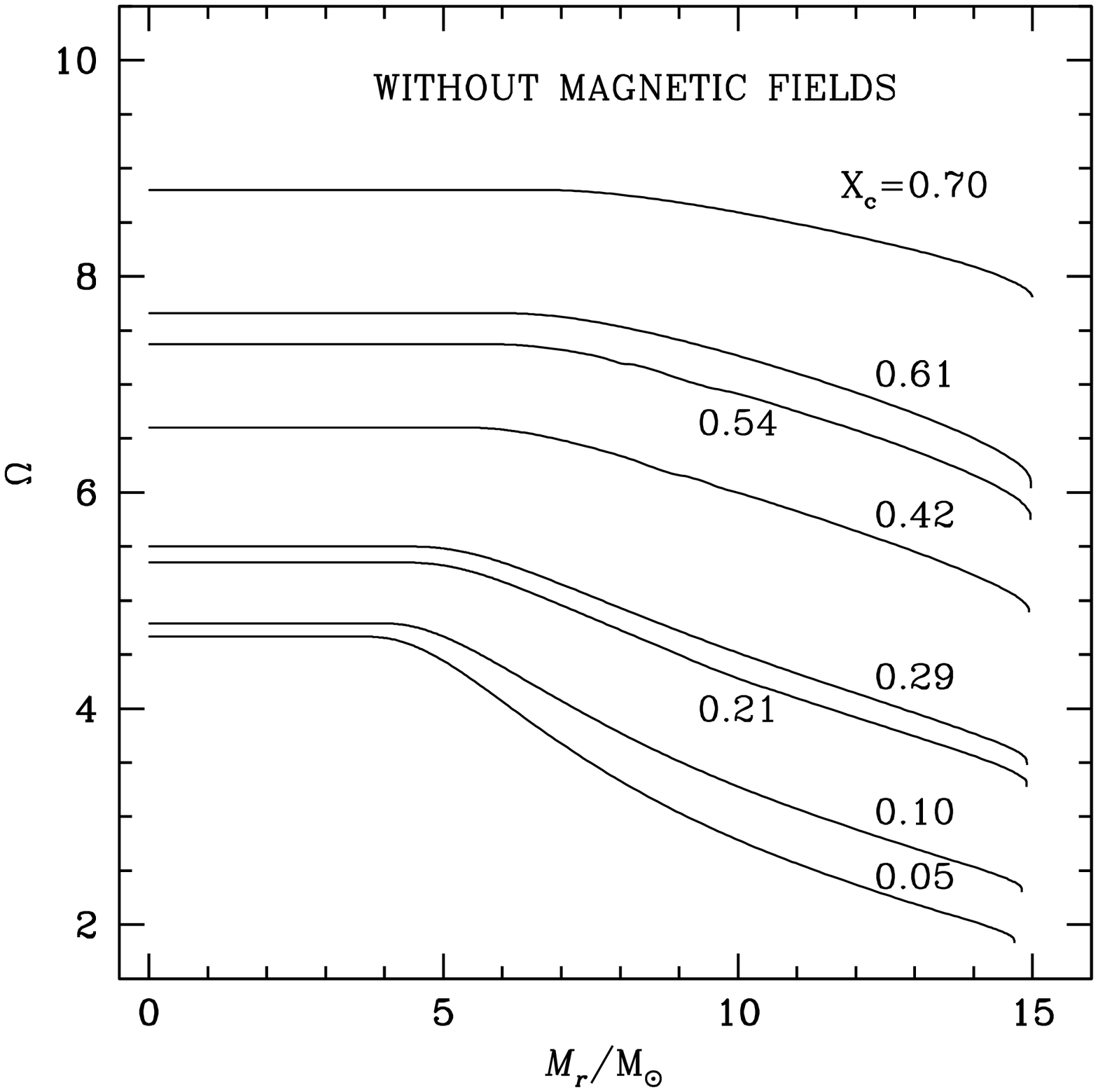}
\hfill
\includegraphics[width=2.3in,height=2.6in]{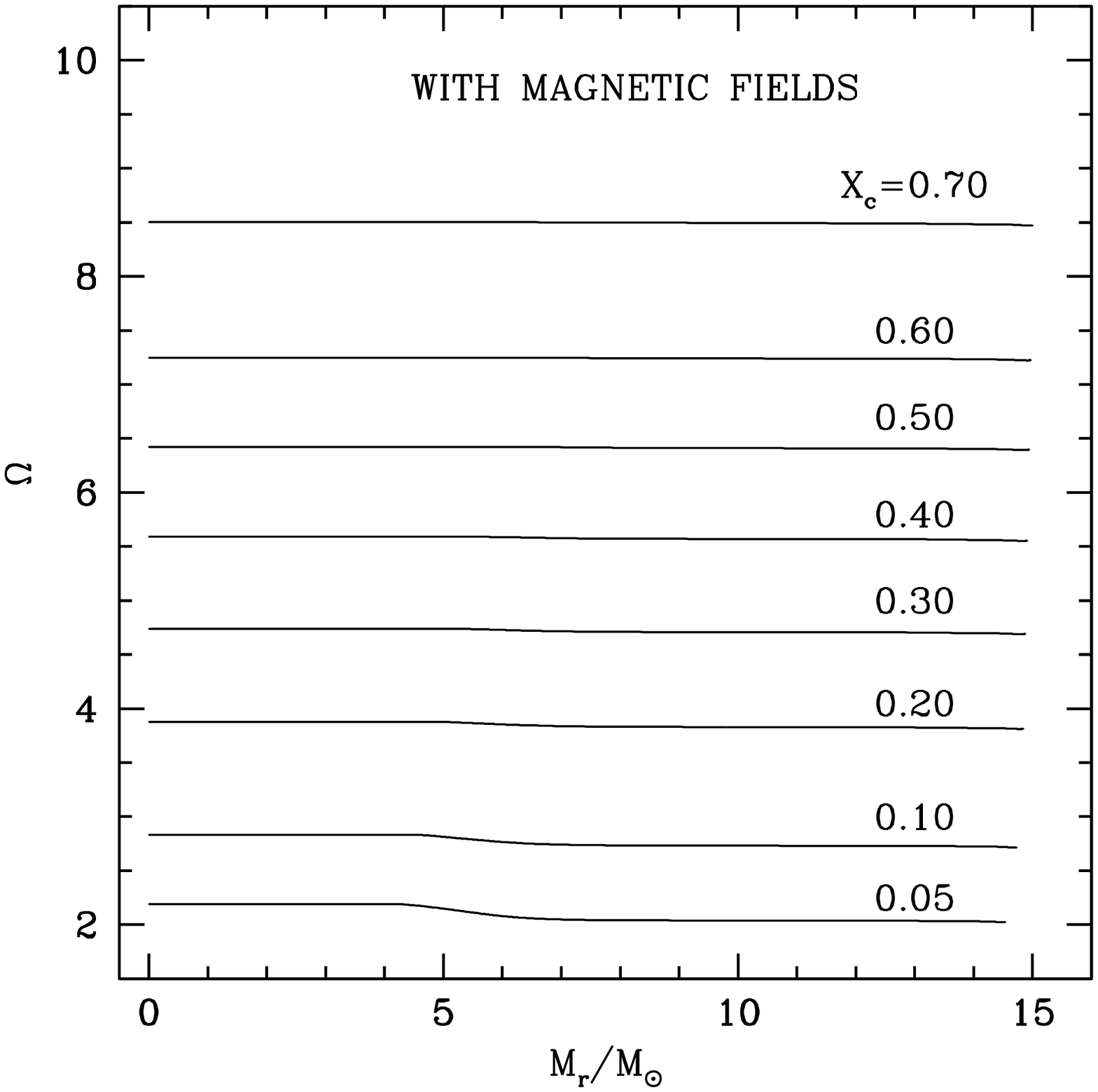}
\caption{{\it Left panel}: Internal distribution of the 
angular velocity $\Omega(r)$ as a function of the Lagrangian mass in
solar units in a 15 M$_\odot$ model, without magnetic fields, at
various stages of the model evolution indicated by the central
H-content $X_{\rm c}$ during the Main-Sequence phase. The initial
velocity is 300 km s$^{-1}$.  {\it Right panel}: Same a left panel but
with magnetic fields. Figures taken from
\citet{2005A&A...440.1041M}.  }
\label{rot}
\end{figure}

So, to first order, we can say that models accounting for the
Tayler-Spruit dynamo will differ from models with rotation only by the
fact they have a solid body rotation all along the MS phase while the
models with rotation only (and no magnetic fields) show a moderate
contrast (by a factor of a few) between the rotation rate of the core
and that of the surface (see Fig.~\ref{rot}).

An immediate consequence of the Tayler-Spruit dynamo is that mixing of
the chemical species by the shear is no longer efficient. However, as
shown by \citet{2005A&A...440.1041M}, meridional currents are more
rapid in the rigid rotating models than in those with differential
rotation. As a result, the deficit of mixing from the shear mechanism
is more than compensated by the increasing efficiency of the mixing
due to the meridional currents.  These authors thus obtain more
important surface enrichment (at equivalent age) in magnetic models
with respect to non-magnetic ones.

Another important consequence of the magnetic models is that they
predict for a given value of the initial rotation, higher surface
velocities during the MS phase.  This is because the magnetic field
imposes a strong coupling between the core and the envelope.  It
allows to extract angular momentum from the contracting still faster
rotating convective core and to bring it to the expanding and slower
rotating radiative envelope.

So from what has been said above, magnetic fields will produce, for
given initial conditions and a given age, faster rotating and more
surface enriched stars.  Also, the angular momentum of the core will
be decreased in magnetic models. \citet{2005ApJ...626..350H} find that
magnetic torques decrease the final rotation rate of the collapsing
iron core by about a factor of 30-50 when compared with the
nonmagnetic models.  From their 15 M$_\odot$ model, they predict
pulsar periods at birth near 15 ms which is in agreement with the
observed periods of the most rapidly rotating pulsars (see detailed
discussion in the above reference).  Thus we see that magnetic models
can reasonably well account for the observed rotation rate of young
pulsars, while models with rotation only would predict much too
fast rotating pulsars. Let us however mention that such arguments in
favor of magnetic models should be taken with some caution because
extraction of angular momentum from the core can also occur at the
time of the explosion and/or during the early life of the pulsars.

For very massive stars (above 40 M$_\odot$ at solar metallicity),
strong stellar winds during the Main-Sequence phase remove the outer
layers and thus angular momentum from the star. Since magnetic field,
by tightly coupling the core to the envelope increases the angular
momentum content of the outer layers, models with magnetic fields lose
more angular momentum by mass loss than their non-magnetic
counterparts (everything else being kept the same).  This feature of
magnetic models makes the production of a very fast spinning core more
difficult. As just seen above, this is a nice feature in the frame of
explaining the rotation rate of young pulsars, but it becomes a
difficulty when one wants to explain the progenitors of collapsars
proposed by \citet{1993ApJ...405..273W} to be at the origin of the
long soft Gamma Ray Bursts (GRB).  Let us briefly recall here that
collapsars are core collapsing stars producing a fast rotating black
hole. The fast rotation allows the formation of an accretion disk
around the black hole. Gravitational energy extracted from the
accretion disk is used (at least in part) to power polar jets. The
gamma ray burst comes from shocks in these jets characterized by very
high Lorentz factors.  In order for the gamma ray burst to be
observable, the star must have shed away its H-rich envelope.  Thus
models producing collapsars should present two properties which at
first sight do appear very difficult to reconcile: on the one hand, a
core with a high angular momentum content\footnote{Typically, the
  massive stars that give rise to gamma-ray bursts must have an amount
  of angular momentum in their inner regions, 1-2 orders of magnitude
  greater than the ones that make common pulsars
  \citep{2006ApJ...637..914W}.}, and at the same time no H-rich
envelope, which means that the model had to suffer strong mass loss
and hence strong angular momentum loss.

Non-magnetic models for high mass stars at low metallicity can easily
reach the needed conditions as discussed by
\citet{2005A&A...443..581H}.  However such models would predict a rate
of long soft gamma ray burst which is at the upper level of the
observed range. Magnetic models can reproduce the required condition
for having a collapsar but for a more restrained range of initial
velocities. Only stars initially very rapidly rotating have a chance
to give birth to a collapsar and then to a long soft gamma ray
burst. Magnetic fileds thus reduce the predicted rate of GRBs.

The initial velocity needed to produce a collapsar is such that the
star will follow a homogeneous evolution during the Main-Sequence
phase, i.e. the stellar composition of the star will be nearly equal
from the core to the surface at every time. Such an evolution has many
advantages for producing a collapsar: quasi homogeneity during the MS
phase allows the star to remove its H-rich envelope by nuclear
processing rather than by mass loss, moreover the tracks of such
homogeneous evolution in the Hertzsprung-Russel diagram remain in the
blue, reducing the mass loss rates during the pre-WR phases.
Evolution of these very fast rotating stars has been computed by
\citet{2005A&A...443..581H}.  These authors estimate that such stars
might comprise roughly 1\% of all stars above 10 M$_\odot$ and that
they can, under certain circumstances, retain enough angular momentum
to make GRBs. They underline the fact that the possibility to make
GRBs is very sensitive to mass loss and is favored in regions of low
metallicity.  A similar scenario has been explored by
\citet{2005A&A...443..643Y}.

The rapid rotation needed in such scenario may be obtained either as a
result of the initial very rapid rotation of the star or acquired by
mass transfer in a close binary system \citep{2005A&A...435..247P,
2008A&A...484..831D, 2007A&A...465L..29C}.

Let us end this section by saying a few words about the possibility
for massive stars to suffer magnetic braking. At the moment of writing
this review, there are no evolutionary computations accounting for
this effect in the range of massive stars\footnote{This effect is
taken into account in solar models on a standard basis.}.  We expect
that such computations will soon appear. Recently
\citet{2009MNRAS.392.1022U} have proposed a numerical recipe for
accounting for the angular momentum loss and associated rotational
spin-down for magnetic hot stars with a line-driven stellar wind and a
rotation-aligned dipole magnetic field.  The numerical scaling
relation that they obtain gives typical spin-down times of the order
of 1 Myr for several known magnetic massive stars.  This is quite in
line with the recent spin down time scale for $\sigma$ Ori E estimated to
be 1.34 Myr by \citet{2010ApJ...714L.318T}. These authors thus
conclude that the observations are consistent with $\sigma$ Ori E undergoing
rotational braking due to its magnetized line-driven wind.  Other
stars present wind and surface magnetic field characteristics which
are compatible with the existence of a magnetized line-driven
wind. This is the case for instance of the Of?p star HD108
\citep{2010MNRAS.tmp..972M} which has also a slow rotation rate (lower
than 50 km s$^{-1}$). How the surface chemical enrichments are
affected by such a magnetic braking remains to be seen and will
probably constitute an interesting line of research for the near
future.

\section{Magnetic massive stars in their environment}
\label{Sec:Environment}

Despite the comparatively strong wind of massive stars in all stages
of their evolution, the dynamics of the mass-loss in the vicinity of
the star will be affected by the magnetic field of the star if such a
field is present at all and if it is sufficiently strong. This likely
also affects photospheric emission and probably also leads to X-ray
emission (Sect.~\ref{Sec:MagneticWind}). If observed and if properly
understood, such effects may in turn be used to confine properties of
the stellar magnetic fields.

Going to larger scales, due to the faster decrease of the magnetic
energy density as compared to the kinetic energy density, the kinetic
energy of the stellar wind will dominate
again. \citet{2009EAS....39..223O} estimates that for the star with
the largest ratio of magnetic to kinetic energy density detected so
far, $\sigma$ Ori E, this transition from magnetically to kinetically
dominated flow takes place within less than 100 stellar radii,
assuming a dipole field. The transition radius is approximately the
radius where the first open field lines appear. Note that although the
magnetic field no longer dominates the flow, the flow remains
magnetized as the magnetic field is advected with the flow.

Observing consequences of the magnetic field at these larger scales,
like particle acceleration at shocks and non-thermal emission, could
again give a clue on the magnetism of the star. The general situation
is potentially similar to strong supernova-remnant shocks (see the
contributions in this volume: 'Magnetic Fields in Supernova Remnants
and Pulsar Wind Nebulae' and 'Magnetic Fields in Cosmic Particle
Acceleration Sources'). Although we are not aware of many
definitive results using this approach, we nevertheless will discuss
some points in this respect. The discussion is split into two parts,
which may roughly be described as single stars interacting with the
interstellar medium (Sect.~\ref{Sec:MagneticNebulae}) and multiple
stars interacting among themselves, in binaries, or open clusters
(Sect.~\ref{Sec:MagneticBinaries}).
\subsection{Magnetic wind structure and associated emission}
\label{Sec:MagneticWind}
An excellent review on stellar magnetospheres, their theoretical and
observational aspects especially also in early-type stars, has recently
been given by~\citet{2009EAS....39..223O}. Much of the following is
adopted from this source.

Typical wind speeds and mass losses in O-stars are on the order of
1000 km/s and $10^{-6}$ M$_{\odot}$/yr. The winds are driven by the
radiation field of the stars, by photon scattering in metal lines.
Analytical considerations and numerical simulations show this
acceleration mechanism to be subject to an instability, leading to the
formation of internal shocks in the O-star
wind~\citep{1984ApJ...284..337O, 1988ApJ...335..914O,
  1997A&A...322..878F}. The emission from these shocks is the
generally accepted explanation for the observed soft ($10^7$~K),
thermal, broad-lined X-ray emission ($L_X/L_{Bol} ~ 10^{-7}$) from
O-type stars.

Magnetic fields affecting the stellar winds of early-type stars have
been suggested as a possible explanation for a number of additional
observational peculiarities of these stars. Among these are strong
X-ray emission in combination with narrow spectral lines of some
stars~\citep{1997ApJ...485L..29B, 2009A&A...495..217F}, occasional
flares of very hard X-rays~\citep{2005MNRAS.357..251T,
2009ApJ...702..759M}, or the rotation modulated Balmer line emission
of Be stars~\citep{2005MNRAS.357..251T, 2010MNRAS.405L..51O}. The
concrete theoretical and numerical models invoked to explain these
observational data have become more and more elaborate over the years.

As a simple measure to estimate the relative importance of a magnetic
field for the wind of an early type star,~\citet{2002ApJ...576..413U}
introduced the wind magnetic confinement parameter
\begin{equation}
\eta_{*} = \frac{B_{*}^{2} R_{*}^{2}}{\dot{M}v_{\infty}}.
\end{equation}
Here, $B_{*}$ denotes the surface magnetic field of the star, $R_{*}$
the stellar radius, $\dot{M}$ the mass loss, and $v_{\infty}$ the
terminal wind speed. Basically, $\eta_{*}$ measures the ratio of the
magnetic to the kinetic energy densities~\citep{2002ApJ...576..413U,
  2009EAS....39..223O}. Values of $\eta_{*} << 1$ indicate that the
stellar wind dominates the magnetic field, and as the magnetic field
is carried with the wind a roughly radial field configuration results.
By contrast, if $\eta_{*} >> 1$ the magnetic field dominates, an
extreme example here being $\sigma$ Ori E with $\eta_{*}
\sim~10^{7}$. Closed magnetic field lines occur, at least in the
vicinity of the star, guiding the stellar wind. With increasing
distance from the star the stellar wind will generally dominate again,
as the magnetic field energy decreases faster than the kinetic energy
of the wind.

\begin{figure}[t]
\centerline{
\includegraphics[width=12.5cm]{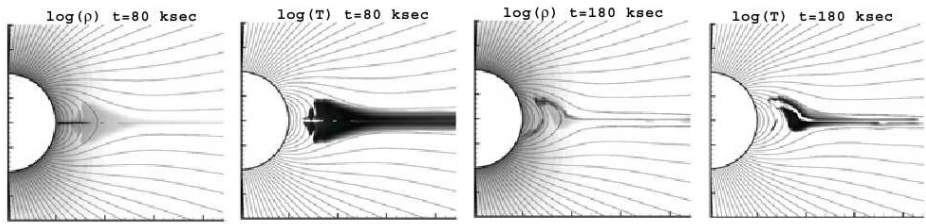}
}
\caption{MHD simulations of the MCWS for $\theta^{1}$ Ori C, compiled
  by~\citet{2009EAS....39..223O} on the basis of ideas developed
  in~\citet{2008MNRAS.385...97U}, his Figure~5. Shown is the
  logarithmic density $\rho$ and the temperature $T$ in a meridional
  plane at 80 ksec ({\bf left}) and 180 ksec ({\bf right}) after
  initialization. At the earlier time, the magnetic field has
  channeled wind material into a compressed, hot disk at the magnetic
  equator. At the later time, the cooled equatorial material is
  falling back toward the star, along field lines. The darkest areas
  in temperature correspond to about 10$^{7}$K.}
\label{fig:owocki_09_fig5}
\end{figure}

The basic effect of a strong enough dipole field on the wind structure
is to deflect the stellar wind towards the equatorial plane separating
the two magnetic hemispheres. There, collision flows results in strong
shocks and associated X-ray emission. This magnetically confined wind
shock model (MCWS) was first brought forward
by~\citet{1997A&A...323..121B}. In its semi-analytical formulation,
the model provided a natural explanation of the periodic X-ray
emission of $\theta^{1}$ Ori C~\citep{1997ApJ...485L..29B}. Subsequent
numerical simulations demonstrated that the X-ray emitting collision
zone is not stable~\citep{2002ApJ...576..413U, 2008MNRAS.385...97U}.
Shocked matter, once cooled, falls down along magnetic field lines,
unless supported by rapid rotation of the star and associated
centrifugal forces, as illustrated in Figure~\ref{fig:owocki_09_fig5}.
The simulations also demonstrated, however, that despite the much more
complicated, unstable flow structure the X-ray emission predicted
by~\citet{1997ApJ...485L..29B} is approximately recovered if
time-averages are considered.

Considering the effect of a strong dipole field in rapidly rotating
stars,~\citet{2002ApJ...578..951C} suggested that the coupling of
matter and field could result in a spin-up of the stellar wind and the
formation of a quasi-Keplerian magnetically torqued disk (MTD) at the
magnetic equator. The idea was not supported by subsequent numerical
simulations~\citep{2003ASPC..305..350O}. For the field strengths
required to spin up material to Keplerian velocities, the numerical
models showed a tendency for centrifugal mass ejection instead of disk
formation. Alternatively, a situation where material rotates at about
Keplerian velocities can be obtained if much stronger fields
($\eta_{*} \rightarrow \infty$) are invoked, as assumed in the rigidly
rotating magnetosphere (RRM) model~\citep{2005MNRAS.357..251T}.
Although the disk rotation in the RRM model is rigid body rather than
Keplerian, approximate Keplerian velocities result as the material
accumulates at the right distance from the star in the form of a thin
disk or of clouds. Applying the RRM model to $\sigma$ Ori
E,~\citet{2005ApJ...630L..81T} could well reproduce the periodic
modulations observed in the light curve, H$\alpha$ emission-line
profile, and longitudinal field strength.

Numerical simulations confirm the general working mechanism of RRM but
also reveal that the accumulation of matter cannot continue for
infinite time. Eventually, the centrifugal force will overcome the
magnetic tension forces and plasma ejection will result. During such a
centrifugal break-out~\citep{2004MNRAS.350..189T, 2006ApJ...640L.191U},
magnetic field lines will first be drawn away from the star before
they snap and reconnect again. The energy release associated with this
reconnection may explain the occasional observed hard X-ray flares in
some early-type stars.
\subsection{Single stars and their wind blown bubbles}
\label{Sec:MagneticNebulae}
We restrict ourselves to three kinds of wind blown bubbles and the
potential effect stellar magnetic fields may have on their shape and
emission: the case of an O-star wind blowing against the interstellar
medium, a Wolf Rayet (WR) wind blowing into the remnant flow
structures from earlier evolutionary phases, and planetary nebulae as
a phase in the life of an intermediate-mass A-type or late B-type star.

As the wind of an O-type star runs highly supersonically against the
interstellar medium, a strong wind-termination shock forms. There,
particles will be accelerated and non-thermal emission is likely to
take place. In principal, the observation of this shock and in
particular of the associated non-thermal emission should give access
to the field strength and topology in the wind. Unfortunately, if the
O-star wind runs against the low-density ISM ($\sim$1~cm$^{-3}$) the
termination shock is found at distances from the star, where wind
densities are well below 1/100 cm$^{-3}$. This makes detection of
non-thermal emission extremely difficult and we are actually not aware
of any detection of such emission so far. One should add perhaps that
-- from an astronomical point of view -- the observation of such a
shock has hardly been of interest at all so far. However, with the
increasing interest in the magnetism of massive stars this may change
now.

For the roughly 300 galactic WR stars, \citet{2001NewAR..45..135V,
2006A&A...458..453V} estimates that only about one quarter have a ring
nebula. These nebulae are commonly attributed to the fast WR wind
ploughing its way through the slow wind shed by the star when being in
its super-giant phase. The WR wind compresses the material from the
super-giant phase into a high density shell. Not all WR stars evolve,
however, through a super-giant phase and about 40\% of them live in
binaries~\citep{2001NewAR..45..135V, 2006A&A...458..453V} with
different evolutionary features. If the above wind collision scenario
applies, theoretical considerations by~\citet{1994ApJ...421..225C}
suggest that the field might even be able to affect the shape of the
resulting ring nebula.

The work by~\citet{1994ApJ...421..225C} is, however, not particularly
designed for WR ring nebulae but looks instead at the general
situation where a fast, powerful wind catches up with a slow, massive
wind. A thin interaction zone is produced, bounded by two shocks. The
inner shock is the termination shock of the magnetic wind from the
central star.  As they point out, the field in the wind gets
increasingly toroidal as this component drops in the free wind only
linearly with the distance $r$, in contrast to the radial component of
the field, which drops quadratically in $r$.  Also in contrast to the
radial field, the toroidal field is compressed at the termination
shock.  Thus, even if the magnetic field is not dynamically important
in the free wind, it can become dynamically very important in the
shocked wind bubble. A bipolar or elliptical nebula is a natural
consequence of the fact that magnetic tension exists in the equatorial
direction and the lack of such effects in the polar direction. As
derived in this paper, the shape of the nebula depends on two
parameters, namely, on $\lambda$, the ratio between the expansion
velocity of the nebula and the velocity of the slow wind, and $\sigma
= \eta_* (v_{rot}/v_{\infty})^{2}$.  The paper provides the shaping of
the nebulae for different combination of these parameters.

\citet{1999ApJ...517..767G} performed a numerical study of this effect 
for the case of planetary nebulae and also provided a series of
different shapes of nebulae depending on two parameters: rotation and
magnetic field strength. The authors point out that magnetic
hook-stresses can collimate the flow along the rotation axis and even
produce jet-like features and ansae which are indeed observed in
planetary nebulae of low-mass stars. 
\citet{2000ApJ...543L..53Z} point out that heat-conduction by
thermal electrons further contributes to the shape of such structures.
Heat conduction, in turn, is largely influenced by the presence of
magnetic fields, suppressing to a great deal conduction normal to the
field. 
%
%
%
%
\subsection{Wind Collision in Binaries and Open Clusters}
\label{Sec:MagneticBinaries}
To test the magnetization of the outflow from a massive star, and thus
the presence of a stellar magnetic field, wind collision regions in
binary star systems and open clusters offer much better conditions
than the wind blown bubbles of Sect.~\ref{Sec:MagneticNebulae}. While
both scenarios, wind collisions and wind blown bubbles, are
accompanied by strong shocks, densities are much higher in the wind
collision case and accompanying emissions thus are stronger. In fact,
\citet{1993ApJ...402..271E} predicted by analytical means that the
wind collision zone in early type binaries may be a strong source of
synchrotron generated, non-thermal radio emission. Key ingredients of
the model are the strong shocks confining the interaction zone, where
electrons can undergo Fermi acceleration and reach relativistic
speeds, and the availability of a magnetic field, which causes the
electrons to spin and emit synchrotron radiation.

A number of WR- and O-stars are indeed known to show non-thermal radio
emission due to synchrotron radiation, indicating the presence of a
magnetic field. Based on observational data,
\citet{1992ASPC...22..249V} noted that many WR-stars with non-thermal
emission were long-period binaries. The hypothesis that binarity is
even a pre-requisite for non-thermal emission of WR stars was put
forward by \citet{2000MNRAS.319.1005D}, based on observations of 9
non-thermal WR emitters, 7 of which known binaries. In a recent
review,~\citet{2007A&ARv..14..171D} lists 17 WR stars that show
non-thermal radio-emission. Of these, 13 live in binaries, for one
binarity is strongly suspected (as of November~2005), while for the
remaining three stars no companion has been detected to date. Binarity
remains, however, a possible explanation also for those non-thermal
emitters without a detected companion, as excluding binarity is very
difficult, especially if the companion is substantially less luminous.
The situation is similar for O-stars. \citet{2007A&ARv..14..171D}
lists 16 O-stars with non-thermal emission, 14 of them are confirmed
binaries or multiple systems.

That the inverse is not true, that binarity does not imply observable
non-thermal emission, was already pointed out
by~\citet{2000MNRAS.319.1005D}. Of the 11 WR+OB binary star systems
they observed, only those 7 systems with binary periods longer than
about one year showed non-thermal emissions, while the 4 short period
systems showed thermal emissions only. The authors argue that such
short periods result in densities that are so high as to be opaque to
radio emissions. Qualitatively the same conclusion is reached
by~\citet{1993ApJ...402..271E} on analytical grounds and
by~\citet{2006MNRAS.372..801P} and~\citet{2010ASPC..422..178B} using
numerical simulations.

\begin{figure}[t]
\centerline{
\includegraphics[width=10cm]{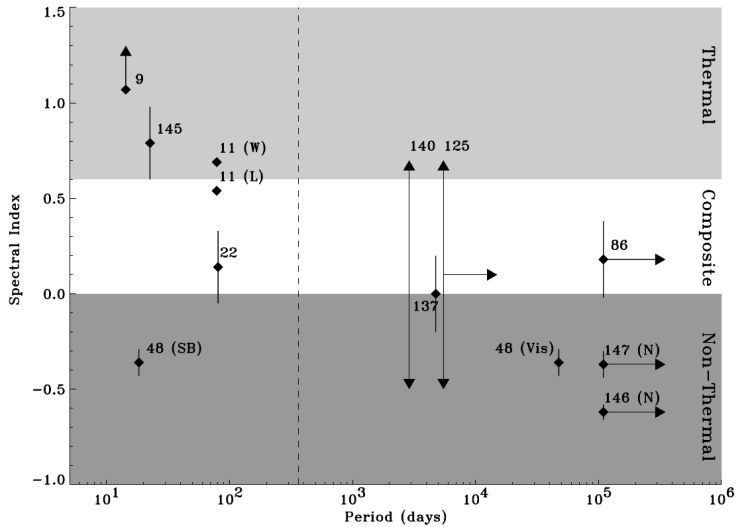}
}
\caption{Radio spectral index of WR+O binary systems against their
  binary period, taken from~\citet{2000MNRAS.319.1005D}, their
  Figure~1. While all long period binaries show some non-thermal
  emission, no such emission is found in short period binaries. WR 48
  is a particular case, for details see the original work.}
\label{fig:wil_dough_00_fig1}
\end{figure}
With binarity (or multiplicity in general) observationally established
to accompany most if not all non-thermal radio emission in WR- and
O-stars, the wind collision zone becomes the prime suspect for the
source of the observed emission, as suggested
by~\citet{1991MNRAS.252...49U} and \citet{1993ApJ...402..271E}.  That
the interaction zone of the two stellar winds indeed coincides with
the source of the non-thermal emission was demonstrated
by~\citet{1997MNRAS.289...10W} for the system WR147 (WN8 + B0.5V) by
means of spatially resolved radio observations. From the radio flux
they further derive a field strength between 1 and 9~mG in the WR
wind, corresponding to a field on the stellar surface in the range of
30-300 G.

Internal shocks embedded in the stellar wind have been discussed as an
alternative origin of the shocks needed for the Fermi acceleration of
the electrons. However, recent numerical simulations indicate that the
shocks are too weak when they reach the outer layers of the stellar
wind, from where radio emission would be observable at
all~\citep{2006A&A...452.1011V, 2010ASPC..422..157V,
2010arXiv1006.3540B}.  This makes the colliding wind interaction zone
not only one but possibly the source of non-thermal radio emission.

With regard to magnetism in massive stars, the above observational
results are interesting in at least two ways. First, the detection of
33 non-thermal emitters as listed by~\citet{2007A&ARv..14..171D}
indicates the presence of a stellar magnetic surface field in at least
33 and maybe up to 66 WR and O stars, thus outnumbering by far the
direct detection of fields in such stars (Sects.~\ref{sec:obs_o},
\ref{sec:latestages}). From the point of view that the observation 
of non-thermal radio emission requires a long enough binary period,
the number of 33 stars even constitutes a lower limit. 

\begin{figure}[t]
\centerline{
\includegraphics[width=6cm]{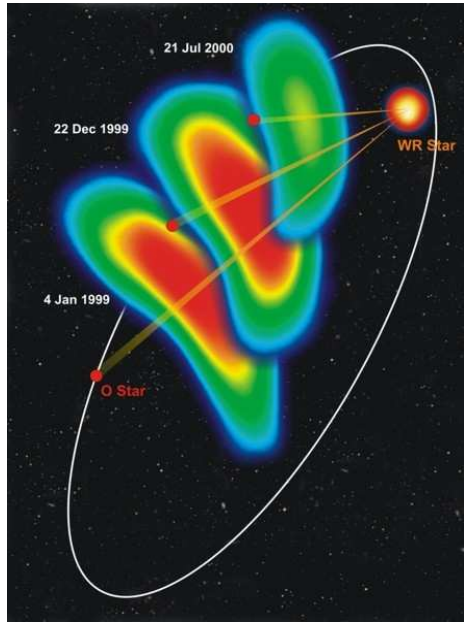}
}
\caption{A montage of 8.4-GHz VLBA observations of WR 140 at three
  orbital phases showing the rotation of the wind collision region as
  the orbit progresses.  The deduced orbit is
  superimposed. From~\citet{2006evn..confE..49D}, their Figure~4.}
\label{fig:wr140}
\end{figure}

Second, and more speculatively, one might argue
that~\citet{2000MNRAS.319.1005D} found indications for the presence of
a magnetic field (non-thermal radio emission) in all those WR stars
which fulfilled the two necessary conditions for such a field to leave
an observable signature: binarity to have a wind collision zone with
strong shocks, and a long enough binary period to avoid complete
absorption of the non-thermal emission within the system (see
Figure~\ref{fig:wil_dough_00_fig1}). Simply speaking, the limiting
factor may have been detectability and not the presence of a field as
such. One may then further speculate whether all WR+OB and O+OB star
binaries have a magnetic field. Or, if the existence of a stellar
magnetic field does not depend on the star living in a binary system
or not, whether all WR- and O-stars have a magnetic field. While it is
unlikely that these speculations hold true in this rigid form, they
indicate that it would be most interesting to search for non-thermal
emission in additional WR+O star binaries. This would allow to see to
what degree the above findings, based on 11 WR+O binaries only, carry
over to a larger sample of the around 300 known galactic WR-stars,
about 40\% of which are binaries~\citep{2001NewAR..45..135V,
  2006A&A...458..453V}.

Equally desirable is a better theoretical understanding of the wind
collision zone. The physics of the this zone is very complex, in
particular if the conditions allow for radiative cooling, and has been
studied extensively by analytical and numerical
means~\citep{1991MNRAS.252...49U, 1992ApJ...386..265S,
1993A&A...278..209N, 1994MNRAS.269..226S, 1995ApJ...454L.145O,
1995IAUS..163..420W, 1996A&A...315..265W, 1997ApJ...475..786G,
1998A&A...330L..21W, 2000Ap&SS.274..189F, 2000ASPC..204..267F,  
2000ApJ...543L..53Z, 2002ASPC..260..605F, 2003IAUS..212..139W,
2006A&A...459....1F, 2009MNRAS.396.1743P, 2010MNRAS.403.1633P,
2010MNRAS.403.1657P}. Positively speaking, shocks in colliding wind
binaries are an ideal laboratory to test the physics of particle
acceleration and other kinetic processes related to collisionless
shock waves. With comparatively high densities and large magnetic
fields as well as a very intense UV-photon field from the involved
stars, the parameter space of colliding winds in massive binaries
complements the region covered by shocks in supernova remnants (see
the contributions in this volume: 'Magnetic Fields in Supernova
Remnants and Pulsar Wind Nebulae' and 'Magnetic Fields in Cosmic
Particle Acceleration Sources').

The currently most detailed numerical studies of colliding winds in WR
binaries and associated non-thermal emission in radio, X-rays ,and
$\gamma$-rays are the series of papers by~\citet{2003A&A...409..217D},
\citet{2006A&A...446.1001P}, and~\citet{2006MNRAS.372..801P}.
Apart from hydrodynamical modeling of the collision zone, the authors
developed numerical tools to fit synthetic spectra in radio, X-ray,
and $\gamma$-rays to observed ones and applied these mainly to the two
systems WR147 and WR140. A composite of radio observations of the
latter is shown in Fig.~\ref{fig:wr140}. The model assumes diffuse shock acceleration
of particles at the confining shocks of the wind collision
zone. Corresponding particle aspects are described in terms of bulk
properties, like the total energy and the energy distribution of the
particles. Details on the models may be taken from a recent review of
one of the authors~\citep{2010ASPC..422..145P} or from corresponding,
more recent work on O+O star binaries~\citep{2009MNRAS.396.1743P,
2010MNRAS.403.1633P, 2010MNRAS.403.1657P}.

Noteworthy in the context of the present review, the results
by~\citet{2006MNRAS.372..801P} underline again that any non-thermal
emission in short-period binaries is unlikely to be observed. It is
absorbed already in the stellar winds, by inverse Compton cooling,
free-free absorption, and the Razin effect. Due to the strong
radiation field, inverse Compton cooling is found to be the by far
most effective cooling mechanism. Consequently, the authors expect
non-thermal X-ray and $\gamma$-ray emission to outpower the
synchrotron radio emission by orders of magnitude. The numerical
results even predict the emission of TeV photons due to $\pi^0$-decay,
the $\pi^0$ being produced by the collision of shock-accelerated
non-thermal hadrons (see also~\citet{1993ApJ...402..271E}).

Related to the 'classical' colliding wind scenario are processes
taking place in dense open clusters, where massive stars often live.
Such clusters blow strong galactic wind bubbles. Presumably, such
bubbles play a significant role in the chemical enrichment of
galaxies. By their energy impact, they furnish the turbulence of the
interstellar medium and they can induce secondary star formation.
Large bubbles cover a significant part of the galaxy and reach far out
of the galactic plane. This huge space is magnetized by the massive
star winds, which drive the bubble. As massive stars live very close
to each other (roughly 1000 AU), such a bubble is created by some
combined effect of single star nebulae and colliding winds. A network
of strong, magnetic shocks is established. Very likely, galactic super
bubbles are also a strong source of cosmic rays and another field to
study particle acceleration. We are not aware of a comprehensive study
of the role of magnetic fields in such bubbles.
\citet{2006MNRAS.372..801P} and~\citet{2009Natur.460..701B} discuss
some points, however.
\subsection{Summary}
In the immediate vicinity of a magnetic massive star, a magnetic
dipole field has a strong effect on the stellar wind, if the ratio of
the magnetic to the kinetic energy density, the wind magnetic
confinement parameter $\eta_*$, is much larger than one. A shock
compressed disk forms at the magnetic equator. Apart from periodic
variabilities of different kind, a prominent observational signature
expected from such a disk are hard X-rays.

For larger scales, further away from the star, much less is known on
the effect of a magnetic field on the stellar environment. The only
notable exception is the case of colliding winds in massive binary
star systems. Non-thermal radio emission from the wind collision zone
has been predicted and observed. The emission is considered to be
indicative of a magnetic field, which must be the advected, in the
wind frozen, magnetic field of the star (unless a dynamo is operating
in the wind-collision zone). While the numerical models so far do not
explicitly account for the stellar magnetic field, its strength and
geometry, the available numerical and observational results suggest
that many more massive stars (perhaps even a significant fraction)
have a magnetic field than is known from direct field detections
through Zeeman signatures. Performing further radio observations of
massive binaries thus seems worthwhile.
\section{Conclusion}
\label{Sec:Conclusions}
The research on magnetic fields in massive and intermediate mass stars
has seen tremendous progress in recent years. Reviewing corresponding
developments and results is an intimidating task, all the more so as
the author's backgrounds in modeling of astrophysical flows or stellar
evolution touch only partly on the subject. Impressing also the range
of perspectives taken on the issue: direct surface field measurements
by means of the Zeeman effect, from pre-main sequence to post-main
sequence stars; indirect indications of surface magnetic fields from
observation and numerical modeling of colliding winds and other
circumstellar phenomena; stellar evolution models exploring the
effects of internal magnetic fields; dynamo theories and
considerations on the stability of fossil fields to explain the origin
of magnetic fields. Concluding this review by trying to sketch a
general, emerging picture in this rapidly developing field is
notoriously difficult and prone to misjudgments.

Starting on firm grounds, current Zeeman observations leave no doubt
that at least some massive main-sequence stars poses a magnetic
field. Indirect evidence, especially the observed non-thermal radio
emission from a number of WR+OB and O+OB star binaries, hints at an
even much larger number of magnetic massive stars. More speculatively,
the only five direct field observations in main sequence O-stars may
suggest a dichotomy: relatively strong ($\sim$~1~kG) dipole fields on
the one hand, as for $\theta^{1}$ Ori C, and rather weak (few tens of
Gauss) and unordered fields on the other hand, as for $\zeta$ Ori A.

Observational data from intermediate mass main sequence stars seem to
support and enlarge this general picture. The 100\% detection rate in
a sample of 28 Ap / Bp stars suggests that maybe all Ap / Bp stars
have a relatively strong (up to 30 kG) magnetic dipole field and that
there exists, in addition, a lower limit to the magnetic dipole field
of about 300 G. The recent -- and first -- Zeeman detection of a weak
($\sim$~1~G) and unordered magnetic field in the normal A-type star
Vega may suggest a similar dichotomy as for the massive
stars. Observations of Ap / Bp stars in open clusters further suggest
that the strengths of their surface fields decrease with the age of the star.

Theory offers essentially two hypothesis on the origin of the observed
magnetic fields, but few firm conclusions. Compared to observations,
much fewer publications are available.

The fossil field hypothesis assumes that the field was present already
before the star reached the main sequence. It requires a field
configuration that is stable enough for the field survive for long
enough without being replenished by a dynamo. Analytical results
showed long ago that such a configuration may exist in the form
of a mixed poloidal-toroidal field configuration.  But it was only
recently that numerical simulations provided the first concrete
realization of such a stable field configuration.  The origin of
fossil fields generally is still speculative. Suggestions range from
the field being a relict of the parent molecular cloud, being
generated by convection in the forming star, or being due to the
merger of a pre-main-sequence binary. Whether the recently detected
magnetic fields in some pre-main sequence Herbig Ae / Be stars
constitute fossil fields that will eventually evolve into the field of
the main-sequence Ap / Bp stars is debated.

The other explanation for the origin of observed surface magnetic
fields is some sort of dynamo mechanism. There exists a lively debate
on whether the radiative envelope of massive and intermediate mass
stars can host a dynamo and how this would work. The only two
numerical simulations addressing the topic give contradictory
answers. Equally undecided is the debate on whether fields generated
in the convective core can reach the surface. For late stages of
massive star evolution, during the super giant phase, surface dynamos
may explain the observed magnetic field. One may speculate whether the
observed dichotomy between a 'strong and ordered field' and a 'weak
and unordered field', if such a dichotomy exists at all, is due to two
different origins, fossil or dynamo, of the field. 

Rather detailed numerical models exist for the consequences of
magnetic fields on the stellar evolution and environment. These models
are of particular interest as they often link the magnetism in massive
stars with a wider range of phenomena and, obviously, observation with
theory. Stellar evolution models link magnetism in massive stars with
the rotation rates of pulsars or the origin of gamma ray
bursts. Simulations of how stellar winds are affected by a stellar
magnetic field lead to the identification of indirect observational
evidence of the later.  Colliding wind models fulfill a similar task
and, in addition, allow to gain insight into the physics of particle
acceleration at collision-less magnetic shocks. The dependencies are often mutual and
illustrate that advances in the field of magnetism in massive stars go
hand in hand with advances in some other fields. In the future, one
may want to exploit these perspectives and mutual dependence's even
further.

To improve, for example, the statistics on what fraction of massive
stars have a magnetic field, it would be interesting to have an
observation based estimate of the fraction of WR+O star binaries
showing non-thermal emission. The interpretation of such data would
greatly benefit from improved colliding wind models that relate the
magnetic fields in the interaction zone and at the surface of the
star. Such a program would offer a complementary view on the certainly
anyway growing observational data base of direct magnetic surface
field detections in intermediate and high mass stars.

Given that stellar evolution is probably particularly sensitive to the
presence of a magnetic field when a star is formed or dies, another
interesting combination could be the observation of early stages of
stellar evolution or even molecular clouds with more elaborate models
of star formation in the presence of a magnetic field. Also most
promising in combination with stellar evolution models is the emerging
field of astroseismology of intermediate and high mass stars.  First
data and proof of concept studies have been published, which use
gravity modes that provide information on the stellar interior down to
the convective core~\citep{2008IAUS..250..237A, 2010Natur.464..259D,
2010arXiv1006.3139D, 2010arXiv1006.4013B}.

The above considerations make clear the necessity of combined
efforts of observation, analytical theory, and numerical simulations
to unravel the mysteries of magnetic fields in upper main sequence
stars.
\begin{acknowledgements}
We would like to thank Maxime Viallet for his comments on the
manus\-cript which helped to improve the text. RW and DF acknowledge
support from the French Stellar Astrophysics Program PNPS. We are
indebted to the people running the NASA's Astrophysics Data System
Abstract Service and arXiv of Cornell University Library. Both
data-bases have greatly facilitated our work.
\end{acknowledgements}
%
%
%
\bibliographystyle{aps-nameyear}      
\bibliography{WalderMassiveStars}     

\begin{thebibliography}{211}
\ifx \bisbn   \undefined \def \bisbn  #1{ISBN #1}\fi
\ifx \binits  \undefined \def \binits#1{#1} \fi
\ifx \bauthor  \undefined \def \bauthor#1{#1} \fi
\ifx \bjtitle  \undefined \def \bjtitle#1{\textrm{#1}}\fi
\ifx \batitle  \undefined \def \batitle#1{#1} \fi
\ifx \bctitle  \undefined \def \bctitle#1{#1} \fi
\ifx \bvolume  \undefined \def \bvolume#1{\textbf{#1}}\fi
\ifx \byear  \undefined \def \byear#1{#1} \fi
\ifx \bissue  \undefined \def \bissue#1{#1} \fi
\ifx \bfpage  \undefined \def \bfpage#1{#1} \fi
\ifx \blpage  \undefined \def \blpage #1{#1} \fi
\ifx \burl  \undefined \def \burl#1{#1} \fi
\ifx \doiurl  \undefined \def \doiurl#1{#1} \fi
\ifx \betal  \undefined \def \betal{et al.} \fi
\ifx \binstitute  \undefined \def \binstitute#1{#1} \fi
\ifx \beditor  \undefined \def \beditor#1{#1} \fi
\ifx \bpublisher  \undefined \def \bpublisher#1{#1} \fi
\ifx \bbtitle  \undefined \def \bbtitle#1{\textit{#1}} \fi
\ifx \bedition  \undefined \def \bedition#1{#1} \fi
\ifx \bseriesno  \undefined \def \bseriesno#1{#1} \fi
\ifx \blocation  \undefined \def \blocation#1{#1} \fi
\ifx \bsertitle  \undefined \def \bsertitle#1{#1} \fi
\ifx \bsnm \undefined \def \bsnm#1{#1} \fi
\ifx \bsuffix \undefined \def \bsuffix#1{#1} \fi
\ifx \bparticle \undefined \def \bparticle#1{#1} \fi
\ifx \barticle \undefined \def \barticle#1{#1} \fi
\ifx \botherref \undefined \def \botherref #1{#1} \fi
\ifx \url \undefined \def \url#1{#1} \fi
\ifx \bchapter \undefined \def \bchapter#1{#1} \fi
\ifx \bbook \undefined \def \bbook#1{#1} \fi
\ifx \bcomment \undefined \def \bcomment#1{#1} \fi
\ifx \oauthor \undefined \def \oauthor#1{#1} \fi
\ifx \citeauthoryear \undefined \def \citeauthoryear#1{#1} \fi
\ifx \texttildelow  \undefined \def \texttildelow{\symbol{126}} \fi
\def \endbibitem {}

\bibitem[\protect\citeauthoryear{{Acheson}}{1978}]{1978RSPTA.289..459A}
\begin{barticle}
\bauthor{\binits{D.J.} \bsnm{{Acheson}}},
\batitle{{On the instability of toroidal magnetic fields and differential
  rotation in stars}}.
\bjtitle{Royal Society of London Philosophical Transactions Series A}
\bvolume{289},
\bfpage{459}--\blpage{500}
(\byear{1978})
\end{barticle}
\endbibitem

\bibitem[\protect\citeauthoryear{{Aerts}}{2008}]{2008IAUS..250..237A}
\begin{botherref}
\oauthor{\binits{C.} \bsnm{{Aerts}}},
{Core Overshoot and Nonrigid Internal Rotation of Massive Stars: Current Status
  from Asteroseismology},
in \textit{IAU Symposium},
ed. by {F.~Bresolin, P.~A.~Crowther, \& J.~Puls}.
IAU Symposium,
vol. 250,
2008,
pp. 237--244
\end{botherref}
\endbibitem

\bibitem[\protect\citeauthoryear{{Alecian} et~al.}{2007}]{2007IAUS..243...43A}
\begin{botherref}
\oauthor{\binits{E.} \bsnm{{Alecian}}}, \oauthor{\binits{G.A.} \bsnm{{Wade}}},
  \oauthor{\binits{C.} \bsnm{{Catala}}}, \oauthor{\binits{C.} \bsnm{{Folsom}}},
  \oauthor{\binits{J.} \bsnm{{Grunhut}}}, \oauthor{\binits{J.}
  \bsnm{{Donati}}}, \oauthor{\binits{P.} \bsnm{{Petit}}}, \oauthor{\binits{S.}
  \bsnm{{Bagnulo}}}, \oauthor{\binits{T.} \bsnm{{Boehm}}}, \oauthor{\binits{J.}
  \bsnm{{Bouret}}}, \oauthor{\binits{J.D.} \bsnm{{Landstreet}}},
{Magnetism, rotation and accretion in Herbig Ae-Be stars},
in \textit{IAU Symposium},
ed. by {J.~Bouvier \& I.~Appenzeller}.
IAU Symposium,
vol. 243,
2007,
pp. 43--50
\end{botherref}
\endbibitem

\bibitem[\protect\citeauthoryear{{Alecian} et~al.}{2008a}]{2008MNRAS.385..391A}
\begin{barticle}
\bauthor{\binits{E.} \bsnm{{Alecian}}}, \bauthor{\binits{C.} \bsnm{{Catala}}},
  \bauthor{\binits{G.A.} \bsnm{{Wade}}}, \bauthor{\binits{J.} \bsnm{{Donati}}},
  \bauthor{\binits{P.} \bsnm{{Petit}}}, \bauthor{\binits{J.D.}
  \bsnm{{Landstreet}}}, \bauthor{\binits{T.} \bsnm{{B{\"o}hm}}},
  \bauthor{\binits{J.} \bsnm{{Bouret}}}, \bauthor{\binits{S.}
  \bsnm{{Bagnulo}}}, \bauthor{\binits{C.} \bsnm{{Folsom}}},
  \bauthor{\binits{J.} \bsnm{{Grunhut}}}, \bauthor{\binits{J.}
  \bsnm{{Silvester}}},
\batitle{{Characterization of the magnetic field of the Herbig Be star
  HD200775}}.
\bjtitle{\mnras}
\bvolume{385},
\bfpage{391}--\blpage{403}
(\byear{2008}a)
\end{barticle}
\endbibitem

\bibitem[\protect\citeauthoryear{{Alecian} et~al.}{2008b}]{2008CoSka..38..235A}
\begin{barticle}
\bauthor{\binits{E.} \bsnm{{Alecian}}}, \bauthor{\binits{G.A.} \bsnm{{Wade}}},
  \bauthor{\binits{C.} \bsnm{{Catala}}}, \bauthor{\binits{C.} \bsnm{{Folsom}}},
  \bauthor{\binits{J.} \bsnm{{Grunhut}}}, \bauthor{\binits{J.}
  \bsnm{{Donati}}}, \bauthor{\binits{P.} \bsnm{{Petit}}}, \bauthor{\binits{S.}
  \bsnm{{Bagnulo}}}, \bauthor{\binits{S.C.} \bsnm{{Marsden}}},
  \bauthor{\binits{J.C.} \bsnm{{Ramirez Velez}}}, \bauthor{\binits{J.D.}
  \bsnm{{Landstreet}}}, \bauthor{\binits{T.} \bsnm{{Boehm}}},
  \bauthor{\binits{J.} \bsnm{{Bouret}}}, \bauthor{\binits{J.}
  \bsnm{{Silvester}}},
\batitle{{Magnetism in pre-MS intermediate-mass stars and the fossil field
  hypothesis}}.
\bjtitle{Contributions of the Astronomical Observatory Skalnate Pleso}
\bvolume{38},
\bfpage{235}--\blpage{244}
(\byear{2008}b)
\end{barticle}
\endbibitem

\bibitem[\protect\citeauthoryear{{Auri{\`e}re}}{2003}]{2003EAS.....9..105A}
\begin{botherref}
\oauthor{\binits{M.} \bsnm{{Auri{\`e}re}}},
{Stellar Polarimetry with NARVAL},
in \textit{EAS Publications Series},
ed. by {J.~Arnaud \& N.~Meunier}.
EAS Publications Series,
vol. 9,
2003,
p. 105
\end{botherref}
\endbibitem

\bibitem[\protect\citeauthoryear{{Auri{\`e}re}
  et~al.}{2007}]{2007A&A...475.1053A}
\begin{barticle}
\bauthor{\binits{M.} \bsnm{{Auri{\`e}re}}}, \bauthor{\binits{G.A.}
  \bsnm{{Wade}}}, \bauthor{\binits{J.} \bsnm{{Silvester}}},
  \bauthor{\binits{F.} \bsnm{{Ligni{\`e}res}}}, \bauthor{\binits{S.}
  \bsnm{{Bagnulo}}}, \bauthor{\binits{K.} \bsnm{{Bale}}}, \bauthor{\binits{B.}
  \bsnm{{Dintrans}}}, \bauthor{\binits{J.F.} \bsnm{{Donati}}},
  \bauthor{\binits{C.P.} \bsnm{{Folsom}}}, \bauthor{\binits{M.}
  \bsnm{{Gruberbauer}}}, \bauthor{\binits{A.} \bsnm{{Hui Bon Hoa}}},
  \bauthor{\binits{S.} \bsnm{{Jeffers}}}, \bauthor{\binits{N.}
  \bsnm{{Johnson}}}, \bauthor{\binits{J.D.} \bsnm{{Landstreet}}},
  \bauthor{\binits{A.} \bsnm{{L{\`e}bre}}}, \bauthor{\binits{T.}
  \bsnm{{Lueftinger}}}, \bauthor{\binits{S.} \bsnm{{Marsden}}},
  \bauthor{\binits{D.} \bsnm{{Mouillet}}}, \bauthor{\binits{S.}
  \bsnm{{Naseri}}}, \bauthor{\binits{F.} \bsnm{{Paletou}}},
  \bauthor{\binits{P.} \bsnm{{Petit}}}, \bauthor{\binits{J.} \bsnm{{Power}}},
  \bauthor{\binits{F.} \bsnm{{Rincon}}}, \bauthor{\binits{S.}
  \bsnm{{Strasser}}}, \bauthor{\binits{N.} \bsnm{{Toqu{\'e}}}},
\batitle{{Weak magnetic fields in Ap/Bp stars. Evidence for a dipole field
  lower limit and a tentative interpretation of the magnetic dichotomy}}.
\bjtitle{\aap}
\bvolume{475},
\bfpage{1053}--\blpage{1065}
(\byear{2007})
\end{barticle}
\endbibitem

\bibitem[\protect\citeauthoryear{{Auriere} et~al.}{2010}]{2010arXiv1008.3086A}
\begin{botherref}
\oauthor{\binits{M.} \bsnm{{Auriere}}}, \oauthor{\binits{G.A.} \bsnm{{Wade}}},
  \oauthor{\binits{F.} \bsnm{{Lignieres}}}, \oauthor{\binits{A.}
  \bsnm{{Hui-Bon-Hoa}}}, \oauthor{\binits{J.D.} \bsnm{{Landstreet}}},
  \oauthor{\binits{I.} \bsnm{{Iliev}}}, \oauthor{\binits{J.} \bsnm{{Donati}}},
  \oauthor{\binits{P.} \bsnm{{Petit}}}, \oauthor{\binits{T.} \bsnm{{Roudier}}},
  \oauthor{\binits{S.} \bsnm{{Theado}}},
{No detection of large-scale magnetic fields at the surfaces of Am and HgMn
  stars}.
(
2010)
\end{botherref}
\endbibitem

\bibitem[\protect\citeauthoryear{{Auri{\`e}re}
  et~al.}{2010}]{2010A&A...516L...2A}
\begin{barticle}
\bauthor{\binits{M.} \bsnm{{Auri{\`e}re}}}, \bauthor{\binits{J.}
  \bsnm{{Donati}}}, \bauthor{\binits{R.} \bsnm{{Konstantinova-Antova}}},
  \bauthor{\binits{G.} \bsnm{{Perrin}}}, \bauthor{\binits{P.} \bsnm{{Petit}}},
  \bauthor{\binits{T.} \bsnm{{Roudier}}},
\batitle{{The magnetic field of Betelgeuse: a local dynamo from giant
  convection cells?}}
\bjtitle{\aap}
\bvolume{516},
\bfpage{2}
(\byear{2010})
\end{barticle}
\endbibitem

\bibitem[\protect\citeauthoryear{{Babcock}}{1947}]{1947ApJ...105..105B}
\begin{barticle}
\bauthor{\binits{H.W.} \bsnm{{Babcock}}},
\batitle{{Zeeman Effect in Stellar Spectra.}}
\bjtitle{\apj}
\bvolume{105},
\bfpage{105}
(\byear{1947})
\end{barticle}
\endbibitem

\bibitem[\protect\citeauthoryear{{Babcock}}{1960}]{1960ApJ...132..521B}
\begin{barticle}
\bauthor{\binits{H.W.} \bsnm{{Babcock}}},
\batitle{{The 34-KILOGAUSS Magnetic Field of HD 215441.}}
\bjtitle{\apj}
\bvolume{132},
\bfpage{521}
(\byear{1960})
\end{barticle}
\endbibitem

\bibitem[\protect\citeauthoryear{{Babel} and
  {Montmerle}}{1997a}]{1997ApJ...485L..29B}
\begin{barticle}
\bauthor{\binits{J.} \bsnm{{Babel}}}, \bauthor{\binits{T.} \bsnm{{Montmerle}}},
\batitle{{On the Periodic X-Ray Emission from the O7 V Star theta 1 Orionis
  C}}.
\bjtitle{\apjl}
\bvolume{485},
\bfpage{29}
(\byear{1997}a)
\end{barticle}
\endbibitem

\bibitem[\protect\citeauthoryear{{Babel} and
  {Montmerle}}{1997b}]{1997A&A...323..121B}
\begin{barticle}
\bauthor{\binits{J.} \bsnm{{Babel}}}, \bauthor{\binits{T.} \bsnm{{Montmerle}}},
\batitle{{X-ray emission from Ap-Bp stars: a magnetically confined wind-shock
  model for IQ Aur.}}
\bjtitle{\aap}
\bvolume{323},
\bfpage{121}--\blpage{138}
(\byear{1997}b)
\end{barticle}
\endbibitem

\bibitem[\protect\citeauthoryear{{Bagnulo} et~al.}{2006}]{2006A&A...450..777B}
\begin{barticle}
\bauthor{\binits{S.} \bsnm{{Bagnulo}}}, \bauthor{\binits{J.D.}
  \bsnm{{Landstreet}}}, \bauthor{\binits{E.} \bsnm{{Mason}}},
  \bauthor{\binits{V.} \bsnm{{Andretta}}}, \bauthor{\binits{J.}
  \bsnm{{Silaj}}}, \bauthor{\binits{G.A.} \bsnm{{Wade}}},
\batitle{{Searching for links between magnetic fields and stellar evolution. I.
  A survey of magnetic fields in open cluster A- and B-type stars with FORS1}}.
\bjtitle{\aap}
\bvolume{450},
\bfpage{777}--\blpage{791}
(\byear{2006})
\end{barticle}
\endbibitem

\bibitem[\protect\citeauthoryear{{Balona} et~al.}{2010}]{2010arXiv1006.4013B}
\begin{botherref}
\oauthor{\binits{L.} \bsnm{{Balona}}}, \oauthor{\binits{M.} \bsnm{{Cunha}}},
  \oauthor{\binits{D.} \bsnm{{Kurtz}}}, \oauthor{\binits{I.M.}
  \bsnm{{Brandao}}}, \oauthor{\binits{M.} \bsnm{{Gruberbauer}}},
  \oauthor{\binits{H.} \bsnm{{Saio}}}, \oauthor{\binits{R.} \bsnm{{Ostensen}}},
  \oauthor{\binits{V.} \bsnm{{Elkin}}}, \oauthor{\binits{W.} \bsnm{{Borucki}}},
  \oauthor{\binits{J.} \bsnm{{Christensen-Dalsgaard}}}, \oauthor{\binits{H.}
  \bsnm{{Kjeldsen}}}, \oauthor{\binits{D.} \bsnm{{Koch}}},
{Kepler observations of a roAp star: delta Scuti and gamma Doradus pulsations
  in Ap stars}.
(
2010)
\end{botherref}
\endbibitem

\bibitem[\protect\citeauthoryear{{Berdyugina}}{2009}]{2009IAUS..259..323B}
\begin{botherref}
\oauthor{\binits{S.V.} \bsnm{{Berdyugina}}},
{Stellar magnetic fields across the H-R diagram: observational evidence},
in \textit{IAU Symposium}.
IAU Symposium,
vol. 259,
2009,
pp. 323--332
\end{botherref}
\endbibitem

\bibitem[\protect\citeauthoryear{{Blomme}}{2010}]{2010ASPC..422..178B}
\begin{botherref}
\oauthor{\binits{R.} \bsnm{{Blomme}}},
{Non-Thermal Radio Emission from Colliding Wind Binaries},
in \textit{Astronomical Society of the Pacific Conference Series},
ed. by {J.~Mart{\'{\i}}, P.~L.~Luque-Escamilla, \& J.~A.~Combi}.
Astronomical Society of the Pacific Conference Series,
vol. 422,
2010,
p. 178
\end{botherref}
\endbibitem

\bibitem[\protect\citeauthoryear{{Blomme} et~al.}{2010}]{2010arXiv1006.3540B}
\begin{botherref}
\oauthor{\binits{R.} \bsnm{{Blomme}}}, \oauthor{\binits{M.} \bsnm{{De
  Becker}}}, \oauthor{\binits{D.} \bsnm{{Volpi}}}, \oauthor{\binits{G.}
  \bsnm{{Rauw}}},
{Non-thermal radio emission from O-type stars. IV. Cyg OB2 No. 8A}.
(
2010)
\end{botherref}
\endbibitem

\bibitem[\protect\citeauthoryear{{Bouret} et~al.}{2008}]{2008MNRAS.389...75B}
\begin{barticle}
\bauthor{\binits{J.} \bsnm{{Bouret}}}, \bauthor{\binits{J.} \bsnm{{Donati}}},
  \bauthor{\binits{F.} \bsnm{{Martins}}}, \bauthor{\binits{C.}
  \bsnm{{Escolano}}}, \bauthor{\binits{W.} \bsnm{{Marcolino}}},
  \bauthor{\binits{T.} \bsnm{{Lanz}}}, \bauthor{\binits{I.D.}
  \bsnm{{Howarth}}},
\batitle{{The weak magnetic field of the O9.7 supergiant {$\zeta$}OrionisA}}.
\bjtitle{\mnras}
\bvolume{389},
\bfpage{75}--\blpage{85}
(\byear{2008})
\end{barticle}
\endbibitem

\bibitem[\protect\citeauthoryear{{Braithwaite}}{2006a}]{2006A&A...449..451B}
\begin{barticle}
\bauthor{\binits{J.} \bsnm{{Braithwaite}}},
\batitle{{A differential rotation driven dynamo in a stably stratified star}}.
\bjtitle{\aap}
\bvolume{449},
\bfpage{451}--\blpage{460}
(\byear{2006}a)
\end{barticle}
\endbibitem

\bibitem[\protect\citeauthoryear{{Braithwaite}}{2006b}]{2006A&A...453..687B}
\begin{barticle}
\bauthor{\binits{J.} \bsnm{{Braithwaite}}},
\batitle{{The stability of toroidal fields in stars}}.
\bjtitle{\aap}
\bvolume{453},
\bfpage{687}--\blpage{698}
(\byear{2006}b)
\end{barticle}
\endbibitem

\bibitem[\protect\citeauthoryear{{Braithwaite}}{2009}]{2009MNRAS.397..763B}
\begin{barticle}
\bauthor{\binits{J.} \bsnm{{Braithwaite}}},
\batitle{{Axisymmetric magnetic fields in stars: relative strengths of poloidal
  and toroidal components}}.
\bjtitle{\mnras}
\bvolume{397},
\bfpage{763}--\blpage{774}
(\byear{2009})
\end{barticle}
\endbibitem

\bibitem[\protect\citeauthoryear{{Braithwaite} and
  {Nordlund}}{2006}]{2006A&A...450.1077B}
\begin{barticle}
\bauthor{\binits{J.} \bsnm{{Braithwaite}}}, \bauthor{\binits{{\AA}.}
  \bsnm{{Nordlund}}},
\batitle{{Stable magnetic fields in stellar interiors}}.
\bjtitle{\aap}
\bvolume{450},
\bfpage{1077}--\blpage{1095}
(\byear{2006})
\end{barticle}
\endbibitem

\bibitem[\protect\citeauthoryear{{Braithwaite} and
  {Spruit}}{2004}]{2004Natur.431..819B}
\begin{barticle}
\bauthor{\binits{J.} \bsnm{{Braithwaite}}}, \bauthor{\binits{H.C.}
  \bsnm{{Spruit}}},
\batitle{{A fossil origin for the magnetic field in A stars and white dwarfs}}.
\bjtitle{Nature}
\bvolume{431},
\bfpage{819}--\blpage{821}
(\byear{2004})
\end{barticle}
\endbibitem

\bibitem[\protect\citeauthoryear{{Braithwaite}
  et~al.}{2010}]{2010HiA....15..161B}
\begin{barticle}
\bauthor{\binits{J.} \bsnm{{Braithwaite}}}, \bauthor{\binits{T.}
  \bsnm{{Akg{\"u}n}}}, \bauthor{\binits{E.} \bsnm{{Alecian}}},
  \bauthor{\binits{A.F.} \bsnm{{Kholtygin}}}, \bauthor{\binits{J.D.}
  \bsnm{{Landstreet}}}, \bauthor{\binits{S.} \bsnm{{Mathis}}},
  \bauthor{\binits{G.} \bsnm{{Michaud}}}, \bauthor{\binits{J.}
  \bsnm{{Portnoy}}}, \bauthor{\binits{G.} \bsnm{{Alecian}}},
  \bauthor{\binits{V.D.} \bsnm{{Bychkov}}}, \bauthor{\binits{L.V.}
  \bsnm{{Bychkova}}}, \bauthor{\binits{N.} \bsnm{{Drake}}},
  \bauthor{\binits{S.N.} \bsnm{{Fabrika}}}, \bauthor{\binits{A.}
  \bsnm{{Reisenegger}}}, \bauthor{\binits{R.} \bsnm{{Steinitz}}},
  \bauthor{\binits{M.} \bsnm{{Vick}}},
\batitle{{CP and related phenomena in the context of Stellar Evolution}}.
\bjtitle{Highlights of Astronomy}
\bvolume{15},
\bfpage{161}--\blpage{171}
(\byear{2010}).
doi:\doiurl{10.1017/S1743921310008562}
\end{barticle}
\endbibitem

\bibitem[\protect\citeauthoryear{{Browning} et~al.}{2004}]{2004ApJ...601..512B}
\begin{barticle}
\bauthor{\binits{M.K.} \bsnm{{Browning}}}, \bauthor{\binits{A.S.}
  \bsnm{{Brun}}}, \bauthor{\binits{J.} \bsnm{{Toomre}}},
\batitle{{Simulations of Core Convection in Rotating A-Type Stars: Differential
  Rotation and Overshooting}}.
\bjtitle{\apj}
\bvolume{601},
\bfpage{512}--\blpage{529}
(\byear{2004})
\end{barticle}
\endbibitem

\bibitem[\protect\citeauthoryear{{Brun} et~al.}{2005}]{2005ApJ...629..461B}
\begin{barticle}
\bauthor{\binits{A.S.} \bsnm{{Brun}}}, \bauthor{\binits{M.K.}
  \bsnm{{Browning}}}, \bauthor{\binits{J.} \bsnm{{Toomre}}},
\batitle{{Simulations of Core Convection in Rotating A-Type Stars: Magnetic
  Dynamo Action}}.
\bjtitle{\apj}
\bvolume{629},
\bfpage{461}--\blpage{481}
(\byear{2005})
\end{barticle}
\endbibitem

\bibitem[\protect\citeauthoryear{{Butt}}{2009}]{2009Natur.460..701B}
\begin{barticle}
\bauthor{\binits{Y.} \bsnm{{Butt}}},
\batitle{{Beyond the myth of the supernova-remnant origin of cosmic rays}}.
\bjtitle{\nat}
\bvolume{460},
\bfpage{701}--\blpage{704}
(\byear{2009}).
doi:\doiurl{10.1038/nature08127}
\end{barticle}
\endbibitem

\bibitem[\protect\citeauthoryear{{Bychkov} et~al.}{2009}]{2009MNRAS.394.1338B}
\begin{barticle}
\bauthor{\binits{V.D.} \bsnm{{Bychkov}}}, \bauthor{\binits{L.V.}
  \bsnm{{Bychkova}}}, \bauthor{\binits{J.} \bsnm{{Madej}}},
\batitle{{Catalogue of averaged stellar effective magnetic fields - II.
  Re-discussion of chemically peculiar A and B stars}}.
\bjtitle{\mnras}
\bvolume{394},
\bfpage{1338}--\blpage{1350}
(\byear{2009})
\end{barticle}
\endbibitem

\bibitem[\protect\citeauthoryear{{Cantiello}
  et~al.}{2007}]{2007A&A...465L..29C}
\begin{barticle}
\bauthor{\binits{M.} \bsnm{{Cantiello}}}, \bauthor{\binits{S.} \bsnm{{Yoon}}},
  \bauthor{\binits{N.} \bsnm{{Langer}}}, \bauthor{\binits{M.} \bsnm{{Livio}}},
\batitle{{Binary star progenitors of long gamma-ray bursts}}.
\bjtitle{\aap}
\bvolume{465},
\bfpage{29}--\blpage{33}
(\byear{2007})
\end{barticle}
\endbibitem

\bibitem[\protect\citeauthoryear{{Cassinelli}
  et~al.}{2002}]{2002ApJ...578..951C}
\begin{barticle}
\bauthor{\binits{J.P.} \bsnm{{Cassinelli}}}, \bauthor{\binits{J.C.}
  \bsnm{{Brown}}}, \bauthor{\binits{M.} \bsnm{{Maheswaran}}},
  \bauthor{\binits{N.A.} \bsnm{{Miller}}}, \bauthor{\binits{D.C.}
  \bsnm{{Telfer}}},
\batitle{{A Magnetically Torqued Disk Model for Be Stars}}.
\bjtitle{\apj}
\bvolume{578},
\bfpage{951}--\blpage{966}
(\byear{2002})
\end{barticle}
\endbibitem

\bibitem[\protect\citeauthoryear{{Charbonneau} and
  {MacGregor}}{2001}]{2001ApJ...559.1094C}
\begin{barticle}
\bauthor{\binits{P.} \bsnm{{Charbonneau}}}, \bauthor{\binits{K.B.}
  \bsnm{{MacGregor}}},
\batitle{{Magnetic Fields in Massive Stars. I. Dynamo Models}}.
\bjtitle{\apj}
\bvolume{559},
\bfpage{1094}--\blpage{1107}
(\byear{2001})
\end{barticle}
\endbibitem

\bibitem[\protect\citeauthoryear{{Chevalier} and
  {Luo}}{1994}]{1994ApJ...421..225C}
\begin{barticle}
\bauthor{\binits{R.A.} \bsnm{{Chevalier}}}, \bauthor{\binits{D.} \bsnm{{Luo}}},
\batitle{{Magnetic shaping of planetary nebulae and other stellar wind
  bubbles}}.
\bjtitle{\apj}
\bvolume{421},
\bfpage{225}--\blpage{235}
(\byear{1994})
\end{barticle}
\endbibitem

\bibitem[\protect\citeauthoryear{{Chiavassa}
  et~al.}{2009}]{2009A&A...506.1351C}
\begin{barticle}
\bauthor{\binits{A.} \bsnm{{Chiavassa}}}, \bauthor{\binits{B.} \bsnm{{Plez}}},
  \bauthor{\binits{E.} \bsnm{{Josselin}}}, \bauthor{\binits{B.}
  \bsnm{{Freytag}}},
\batitle{{Radiative hydrodynamics simulations of red supergiant stars. I.
  interpretation of interferometric observations}}.
\bjtitle{\aap}
\bvolume{506},
\bfpage{1351}--\blpage{1365}
(\byear{2009})
\end{barticle}
\endbibitem

\bibitem[\protect\citeauthoryear{{Chiavassa}
  et~al.}{2010}]{2010A&A...515A..12C}
\begin{barticle}
\bauthor{\binits{A.} \bsnm{{Chiavassa}}}, \bauthor{\binits{X.}
  \bsnm{{Haubois}}}, \bauthor{\binits{J.S.} \bsnm{{Young}}},
  \bauthor{\binits{B.} \bsnm{{Plez}}}, \bauthor{\binits{E.} \bsnm{{Josselin}}},
  \bauthor{\binits{G.} \bsnm{{Perrin}}}, \bauthor{\binits{B.}
  \bsnm{{Freytag}}},
\batitle{{Radiative hydrodynamics simulations of red supergiant stars. II.
  Simulations of convection on Betelgeuse match interferometric observations}}.
\bjtitle{\aap}
\bvolume{515},
\bfpage{12}
(\byear{2010})
\end{barticle}
\endbibitem

\bibitem[\protect\citeauthoryear{{De Becker}}{2007}]{2007A&ARv..14..171D}
\begin{barticle}
\bauthor{\binits{M.} \bsnm{{De Becker}}},
\batitle{{Non-thermal emission processes in massive binaries}}.
\bjtitle{\aapr}
\bvolume{14},
\bfpage{171}--\blpage{216}
(\byear{2007})
\end{barticle}
\endbibitem

\bibitem[\protect\citeauthoryear{{Degroote}
  et~al.}{2010a}]{2010arXiv1006.3139D}
\begin{botherref}
\oauthor{\binits{P.} \bsnm{{Degroote}}}, \oauthor{\binits{M.}
  \bsnm{{Briquet}}}, \oauthor{\binits{M.} \bsnm{{Auvergne}}},
  \oauthor{\binits{S.} \bsnm{{Simon-Diaz}}}, \oauthor{\binits{C.}
  \bsnm{{Aerts}}}, \oauthor{\binits{A.} \bsnm{{Noels}}}, \oauthor{\binits{M.}
  \bsnm{{Rainer}}}, \oauthor{\binits{M.} \bsnm{{Hareter}}},
  \oauthor{\binits{E.} \bsnm{{Poretti}}}, \oauthor{\binits{L.} \bsnm{{Mahy}}},
  \oauthor{\binits{R.} \bsnm{{Oreiro}}}, \oauthor{\binits{M.}
  \bsnm{{Vuckovic}}}, \oauthor{\binits{K.} \bsnm{{Smolders}}},
  \oauthor{\binits{A.} \bsnm{{Baglin}}}, \oauthor{\binits{F.} \bsnm{{Baudin}}},
  \oauthor{\binits{C.} \bsnm{{Catala}}}, \oauthor{\binits{E.} \bsnm{{Michel}}},
  \oauthor{\binits{R.} \bsnm{{Samadi}}},
{Detection of frequency spacings in the young O-type binary HD 46149 from CoRoT
  photometry}.
(
2010a)
\end{botherref}
\endbibitem

\bibitem[\protect\citeauthoryear{{Degroote}
  et~al.}{2010b}]{2010Natur.464..259D}
\begin{barticle}
\bauthor{\binits{P.} \bsnm{{Degroote}}}, \bauthor{\binits{C.} \bsnm{{Aerts}}},
  \bauthor{\binits{A.} \bsnm{{Baglin}}}, \bauthor{\binits{A.} \bsnm{{Miglio}}},
  \bauthor{\binits{M.} \bsnm{{Briquet}}}, \bauthor{\binits{A.} \bsnm{{Noels}}},
  \bauthor{\binits{E.} \bsnm{{Niemczura}}}, \bauthor{\binits{J.}
  \bsnm{{Montalban}}}, \bauthor{\binits{S.} \bsnm{{Bloemen}}},
  \bauthor{\binits{R.} \bsnm{{Oreiro}}}, \bauthor{\binits{M.} \bsnm{{Vu{\v
  c}kovi{\'c}}}}, \bauthor{\binits{K.} \bsnm{{Smolders}}}, \bauthor{\binits{M.}
  \bsnm{{Auvergne}}}, \bauthor{\binits{F.} \bsnm{{Baudin}}},
  \bauthor{\binits{C.} \bsnm{{Catala}}}, \bauthor{\binits{E.} \bsnm{{Michel}}},
\batitle{{Deviations from a uniform period spacing of gravity modes in a
  massive star}}.
\bjtitle{Nature}
\bvolume{464},
\bfpage{259}--\blpage{261}
(\byear{2010}b)
\end{barticle}
\endbibitem

\bibitem[\protect\citeauthoryear{{Denissenkov} and
  {Pinsonneault}}{2007}]{2007ApJ...655.1157D}
\begin{barticle}
\bauthor{\binits{P.A.} \bsnm{{Denissenkov}}}, \bauthor{\binits{M.}
  \bsnm{{Pinsonneault}}},
\batitle{{A Revised Prescription for the Tayler-Spruit Dynamo: Magnetic Angular
  Momentum Transport in Stars}}.
\bjtitle{\apj}
\bvolume{655},
\bfpage{1157}--\blpage{1165}
(\byear{2007})
\end{barticle}
\endbibitem

\bibitem[\protect\citeauthoryear{{Detmers} et~al.}{2008}]{2008A&A...484..831D}
\begin{barticle}
\bauthor{\binits{R.G.} \bsnm{{Detmers}}}, \bauthor{\binits{N.}
  \bsnm{{Langer}}}, \bauthor{\binits{P.} \bsnm{{Podsiadlowski}}},
  \bauthor{\binits{R.G.} \bsnm{{Izzard}}},
\batitle{{Gamma-ray bursts from tidally spun-up Wolf-Rayet stars?}}
\bjtitle{\aap}
\bvolume{484},
\bfpage{831}--\blpage{839}
(\byear{2008})
\end{barticle}
\endbibitem

\bibitem[\protect\citeauthoryear{{Donati}}{2003}]{2003ASPC..307...41D}
\begin{botherref}
\oauthor{\binits{J.} \bsnm{{Donati}}},
{ESPaDOnS: An Echelle SpectroPolarimetric Device for the Observation of Stars
  at CFHT},
in \textit{Astronomical Society of the Pacific Conference Series},
ed. by {J.~Trujillo-Bueno \& J.~Sanchez Almeida}.
Astronomical Society of the Pacific Conference Series,
vol. 307,
2003,
p. 41
\end{botherref}
\endbibitem

\bibitem[\protect\citeauthoryear{{Donati} and
  {Landstreet}}{2009}]{2009ARA&A..47..333D}
\begin{barticle}
\bauthor{\binits{J.} \bsnm{{Donati}}}, \bauthor{\binits{J.D.}
  \bsnm{{Landstreet}}},
\batitle{{Magnetic Fields of Nondegenerate Stars}}.
\bjtitle{\araa}
\bvolume{47},
\bfpage{333}--\blpage{370}
(\byear{2009})
\end{barticle}
\endbibitem

\bibitem[\protect\citeauthoryear{{Donati} et~al.}{1997}]{1997MNRAS.291..658D}
\begin{barticle}
\bauthor{\binits{J.} \bsnm{{Donati}}}, \bauthor{\binits{M.} \bsnm{{Semel}}},
  \bauthor{\binits{B.D.} \bsnm{{Carter}}}, \bauthor{\binits{D.E.}
  \bsnm{{Rees}}}, \bauthor{\binits{A.} \bsnm{{Collier Cameron}}},
\batitle{{Spectropolarimetric observations of active stars}}.
\bjtitle{\mnras}
\bvolume{291},
\bfpage{658}
(\byear{1997})
\end{barticle}
\endbibitem

\bibitem[\protect\citeauthoryear{{Donati} et~al.}{1999}]{1999A&AS..134..149D}
\begin{barticle}
\bauthor{\binits{J.} \bsnm{{Donati}}}, \bauthor{\binits{C.} \bsnm{{Catala}}},
  \bauthor{\binits{G.A.} \bsnm{{Wade}}}, \bauthor{\binits{G.} \bsnm{{Gallou}}},
  \bauthor{\binits{G.} \bsnm{{Delaigue}}}, \bauthor{\binits{P.}
  \bsnm{{Rabou}}},
\batitle{{A dedicated polarimeter for the MuSiCoS {\'e}chelle spectrograph}}.
\bjtitle{\aaps}
\bvolume{134},
\bfpage{149}--\blpage{159}
(\byear{1999})
\end{barticle}
\endbibitem

\bibitem[\protect\citeauthoryear{{Donati} et~al.}{2002}]{2002MNRAS.333...55D}
\begin{barticle}
\bauthor{\binits{J.} \bsnm{{Donati}}}, \bauthor{\binits{J.} \bsnm{{Babel}}},
  \bauthor{\binits{T.J.} \bsnm{{Harries}}}, \bauthor{\binits{I.D.}
  \bsnm{{Howarth}}}, \bauthor{\binits{P.} \bsnm{{Petit}}}, \bauthor{\binits{M.}
  \bsnm{{Semel}}},
\batitle{{The magnetic field and wind confinement of ${\theta}^{1}$ Orionis
  C}}.
\bjtitle{\mnras}
\bvolume{333},
\bfpage{55}--\blpage{70}
(\byear{2002})
\end{barticle}
\endbibitem

\bibitem[\protect\citeauthoryear{{Donati} et~al.}{2006a}]{2006MNRAS.365L...6D}
\begin{barticle}
\bauthor{\binits{J.} \bsnm{{Donati}}}, \bauthor{\binits{I.D.}
  \bsnm{{Howarth}}}, \bauthor{\binits{J.} \bsnm{{Bouret}}},
  \bauthor{\binits{P.} \bsnm{{Petit}}}, \bauthor{\binits{C.} \bsnm{{Catala}}},
  \bauthor{\binits{J.} \bsnm{{Landstreet}}},
\batitle{{Discovery of a strong magnetic field on the O star HD 191612: new
  clues to the future of ${\theta}^{1}$ Orionis C$^{*}$}}.
\bjtitle{\mnras}
\bvolume{365},
\bfpage{6}--\blpage{10}
(\byear{2006}a)
\end{barticle}
\endbibitem

\bibitem[\protect\citeauthoryear{{Donati} et~al.}{2006b}]{2006MNRAS.370..629D}
\begin{barticle}
\bauthor{\binits{J.} \bsnm{{Donati}}}, \bauthor{\binits{I.D.}
  \bsnm{{Howarth}}}, \bauthor{\binits{M.M.} \bsnm{{Jardine}}},
  \bauthor{\binits{P.} \bsnm{{Petit}}}, \bauthor{\binits{C.} \bsnm{{Catala}}},
  \bauthor{\binits{J.D.} \bsnm{{Landstreet}}}, \bauthor{\binits{J.}
  \bsnm{{Bouret}}}, \bauthor{\binits{E.} \bsnm{{Alecian}}},
  \bauthor{\binits{J.R.} \bsnm{{Barnes}}}, \bauthor{\binits{T.}
  \bsnm{{Forveille}}}, \bauthor{\binits{F.} \bsnm{{Paletou}}},
  \bauthor{\binits{N.} \bsnm{{Manset}}},
\batitle{{The surprising magnetic topology of {$\tau$} Sco: fossil remnant or
  dynamo output?}}
\bjtitle{\mnras}
\bvolume{370},
\bfpage{629}--\blpage{644}
(\byear{2006}b)
\end{barticle}
\endbibitem

\bibitem[\protect\citeauthoryear{{Dorch}}{2004}]{2004A&A...423.1101D}
\begin{barticle}
\bauthor{\binits{S.B.F.} \bsnm{{Dorch}}},
\batitle{{Magnetic activity in late-type giant stars: Numerical MHD simulations
  of non-linear dynamo action in Betelgeuse}}.
\bjtitle{\aap}
\bvolume{423},
\bfpage{1101}--\blpage{1107}
(\byear{2004})
\end{barticle}
\endbibitem

\bibitem[\protect\citeauthoryear{{Dougherty} and
  {Pittard}}{2006}]{2006evn..confE..49D}
\begin{botherref}
\oauthor{\binits{S.M.} \bsnm{{Dougherty}}}, \oauthor{\binits{J.M.}
  \bsnm{{Pittard}}},
{Winds in collision: high-energy particles in massive binary systems},
in \textit{Proceedings of the 8th European VLBI Network Symposium},
2006
\end{botherref}
\endbibitem

\bibitem[\protect\citeauthoryear{{Dougherty} and
  {Williams}}{2000}]{2000MNRAS.319.1005D}
\begin{barticle}
\bauthor{\binits{S.M.} \bsnm{{Dougherty}}}, \bauthor{\binits{P.M.}
  \bsnm{{Williams}}},
\batitle{{Non-thermal emission in Wolf-Rayet stars: are massive companions
  required?}}
\bjtitle{\mnras}
\bvolume{319},
\bfpage{1005}--\blpage{1010}
(\byear{2000})
\end{barticle}
\endbibitem

\bibitem[\protect\citeauthoryear{{Dougherty}
  et~al.}{2003}]{2003A&A...409..217D}
\begin{barticle}
\bauthor{\binits{S.M.} \bsnm{{Dougherty}}}, \bauthor{\binits{J.M.}
  \bsnm{{Pittard}}}, \bauthor{\binits{L.} \bsnm{{Kasian}}},
  \bauthor{\binits{R.F.} \bsnm{{Coker}}}, \bauthor{\binits{P.M.}
  \bsnm{{Williams}}}, \bauthor{\binits{H.M.} \bsnm{{Lloyd}}},
\batitle{{Radio emission models of colliding-wind binary systems}}.
\bjtitle{\aap}
\bvolume{409},
\bfpage{217}--\blpage{233}
(\byear{2003})
\end{barticle}
\endbibitem

\bibitem[\protect\citeauthoryear{{Duez} and
  {Mathis}}{2010}]{2010A&A...517A..58D}
\begin{barticle}
\bauthor{\binits{V.} \bsnm{{Duez}}}, \bauthor{\binits{S.} \bsnm{{Mathis}}},
\batitle{{Relaxed equilibrium configurations to model fossil fields . I. A
  first family}}.
\bjtitle{\aap}
\bvolume{517},
\bfpage{58}
(\byear{2010}).
doi:\doiurl{10.1051/0004-6361/200913496}
\end{barticle}
\endbibitem

\bibitem[\protect\citeauthoryear{{Duez} et~al.}{2010}]{2010ApJ...724L..34D}
\begin{barticle}
\bauthor{\binits{V.} \bsnm{{Duez}}}, \bauthor{\binits{J.}
  \bsnm{{Braithwaite}}}, \bauthor{\binits{S.} \bsnm{{Mathis}}},
\batitle{{On the Stability of Non-force-free Magnetic Equilibria in Stars}}.
\bjtitle{\apjl}
\bvolume{724},
\bfpage{34}--\blpage{38}
(\byear{2010}).
doi:\doiurl{10.1088/2041-8205/724/1/L34}
\end{barticle}
\endbibitem

\bibitem[\protect\citeauthoryear{{Dufour} et~al.}{1988}]{1988ApJ...327..859D}
\begin{barticle}
\bauthor{\binits{R.J.} \bsnm{{Dufour}}}, \bauthor{\binits{R.A.R.}
  \bsnm{{Parker}}}, \bauthor{\binits{K.G.} \bsnm{{Henize}}},
\batitle{{The spectrophotometry and chemical composition of the oxygen-poor
  bipolar nebula NGC 6164-5}}.
\bjtitle{\apj}
\bvolume{327},
\bfpage{859}--\blpage{869}
(\byear{1988})
\end{barticle}
\endbibitem

\bibitem[\protect\citeauthoryear{{Eggenberger}
  et~al.}{2005}]{2005A&A...440L...9E}
\begin{barticle}
\bauthor{\binits{P.} \bsnm{{Eggenberger}}}, \bauthor{\binits{A.}
  \bsnm{{Maeder}}}, \bauthor{\binits{G.} \bsnm{{Meynet}}},
\batitle{{Stellar evolution with rotation and magnetic fields. IV. The solar
  rotation profile}}.
\bjtitle{\aap}
\bvolume{440},
\bfpage{9}--\blpage{12}
(\byear{2005})
\end{barticle}
\endbibitem

\bibitem[\protect\citeauthoryear{{Eichler} and
  {Usov}}{1993}]{1993ApJ...402..271E}
\begin{barticle}
\bauthor{\binits{D.} \bsnm{{Eichler}}}, \bauthor{\binits{V.} \bsnm{{Usov}}},
\batitle{{Particle acceleration and nonthermal radio emission in binaries of
  early-type stars}}.
\bjtitle{\apj}
\bvolume{402},
\bfpage{271}--\blpage{279}
(\byear{1993})
\end{barticle}
\endbibitem

\bibitem[\protect\citeauthoryear{{Elkin} et~al.}{2010}]{2010MNRAS.402.1883E}
\begin{barticle}
\bauthor{\binits{V.G.} \bsnm{{Elkin}}}, \bauthor{\binits{G.} \bsnm{{Mathys}}},
  \bauthor{\binits{D.W.} \bsnm{{Kurtz}}}, \bauthor{\binits{S.}
  \bsnm{{Hubrig}}}, \bauthor{\binits{L.M.} \bsnm{{Freyhammer}}},
\batitle{{A rival for Babcock's star: the extreme 30-kG variable magnetic field
  in the Ap star HD75049}}.
\bjtitle{\mnras}
\bvolume{402},
\bfpage{1883}--\blpage{1891}
(\byear{2010})
\end{barticle}
\endbibitem

\bibitem[\protect\citeauthoryear{{Favata} et~al.}{2009}]{2009A&A...495..217F}
\begin{barticle}
\bauthor{\binits{F.} \bsnm{{Favata}}}, \bauthor{\binits{C.} \bsnm{{Neiner}}},
  \bauthor{\binits{P.} \bsnm{{Testa}}}, \bauthor{\binits{G.} \bsnm{{Hussain}}},
  \bauthor{\binits{J.} \bsnm{{Sanz-Forcada}}},
\batitle{{Testing magnetically confined wind shock models for {$\beta$} Cephei
  using XMM-Newton and Chandra phase-resolved X-ray observations}}.
\bjtitle{\aap}
\bvolume{495},
\bfpage{217}--\blpage{229}
(\byear{2009})
\end{barticle}
\endbibitem

\bibitem[\protect\citeauthoryear{{Featherstone}
  et~al.}{2009}]{2009ApJ...705.1000F}
\begin{barticle}
\bauthor{\binits{N.A.} \bsnm{{Featherstone}}}, \bauthor{\binits{M.K.}
  \bsnm{{Browning}}}, \bauthor{\binits{A.S.} \bsnm{{Brun}}},
  \bauthor{\binits{J.} \bsnm{{Toomre}}},
\batitle{{Effects of Fossil Magnetic Fields on Convective Core Dynamos in
  A-type Stars}}.
\bjtitle{\apj}
\bvolume{705},
\bfpage{1000}--\blpage{1018}
(\byear{2009})
\end{barticle}
\endbibitem

\bibitem[\protect\citeauthoryear{{Feldmeier}
  et~al.}{1997}]{1997A&A...322..878F}
\begin{barticle}
\bauthor{\binits{A.} \bsnm{{Feldmeier}}}, \bauthor{\binits{J.} \bsnm{{Puls}}},
  \bauthor{\binits{A.W.A.} \bsnm{{Pauldrach}}},
\batitle{{A possible origin for X-rays from O stars.}}
\bjtitle{\aap}
\bvolume{322},
\bfpage{878}--\blpage{895}
(\byear{1997})
\end{barticle}
\endbibitem

\bibitem[\protect\citeauthoryear{{Ferrario} and
  {Wickramasinghe}}{2006}]{2006MNRAS.367.1323F}
\begin{barticle}
\bauthor{\binits{L.} \bsnm{{Ferrario}}}, \bauthor{\binits{D.}
  \bsnm{{Wickramasinghe}}},
\batitle{{Modelling of isolated radio pulsars and magnetars on the fossil field
  hypothesis}}.
\bjtitle{\mnras}
\bvolume{367},
\bfpage{1323}--\blpage{1328}
(\byear{2006})
\end{barticle}
\endbibitem

\bibitem[\protect\citeauthoryear{{Ferrario} et~al.}{2009}]{2009MNRAS.400L..71F}
\begin{barticle}
\bauthor{\binits{L.} \bsnm{{Ferrario}}}, \bauthor{\binits{J.E.}
  \bsnm{{Pringle}}}, \bauthor{\binits{C.A.} \bsnm{{Tout}}},
  \bauthor{\binits{D.T.} \bsnm{{Wickramasinghe}}},
\batitle{{The origin of magnetism on the upper main sequence}}.
\bjtitle{\mnras}
\bvolume{400},
\bfpage{71}--\blpage{74}
(\byear{2009})
\end{barticle}
\endbibitem

\bibitem[\protect\citeauthoryear{{Ferraro}}{1937}]{1937MNRAS..97..458F}
\begin{barticle}
\bauthor{\binits{V.C.A.} \bsnm{{Ferraro}}},
\batitle{{The non-uniform rotation of the Sun and its magnetic field}}.
\bjtitle{\mnras}
\bvolume{97},
\bfpage{458}
(\byear{1937})
\end{barticle}
\endbibitem

\bibitem[\protect\citeauthoryear{{Flowers} and
  {Ruderman}}{1977}]{1977ApJ...215..302F}
\begin{barticle}
\bauthor{\binits{E.} \bsnm{{Flowers}}}, \bauthor{\binits{M.A.}
  \bsnm{{Ruderman}}},
\batitle{{Evolution of pulsar magnetic fields}}.
\bjtitle{\apj}
\bvolume{215},
\bfpage{302}--\blpage{310}
(\byear{1977})
\end{barticle}
\endbibitem

\bibitem[\protect\citeauthoryear{{Folini} and
  {Walder}}{2000a}]{2000Ap&SS.274..189F}
\begin{barticle}
\bauthor{\binits{D.} \bsnm{{Folini}}}, \bauthor{\binits{R.} \bsnm{{Walder}}},
\batitle{{3D Hydrodynamical Simulations of Colliding Wind Binaries: Theory
  Confronts Observations}}.
\bjtitle{\apss}
\bvolume{274},
\bfpage{189}--\blpage{194}
(\byear{2000}a)
\end{barticle}
\endbibitem

\bibitem[\protect\citeauthoryear{{Folini} and
  {Walder}}{2000b}]{2000ASPC..204..267F}
\begin{botherref}
\oauthor{\binits{D.} \bsnm{{Folini}}}, \oauthor{\binits{R.} \bsnm{{Walder}}},
{Theory of Thermal and Ionization Effects in Colliding Winds of WR+O Binaries},
in \textit{Thermal and Ionization Aspects of Flows from Hot Stars},
ed. by {H.~Lamers \& A.~Sapar}.
Astronomical Society of the Pacific Conference Series,
vol. 204,
2000,
p. 267
\end{botherref}
\endbibitem

\bibitem[\protect\citeauthoryear{{Folini} and
  {Walder}}{2002}]{2002ASPC..260..605F}
\begin{botherref}
\oauthor{\binits{D.} \bsnm{{Folini}}}, \oauthor{\binits{R.} \bsnm{{Walder}}},
{Theoretical Predictions for the Cold Part of the Colliding-Wind
  Interaction-Zone},
in \textit{Interacting Winds from Massive Stars},
ed. by {A.~F.~J.~Moffat \& N.~St-Louis}.
Astronomical Society of the Pacific Conference Series,
vol. 260,
2002,
p. 605
\end{botherref}
\endbibitem

\bibitem[\protect\citeauthoryear{{Folini} and
  {Walder}}{2006}]{2006A&A...459....1F}
\begin{barticle}
\bauthor{\binits{D.} \bsnm{{Folini}}}, \bauthor{\binits{R.} \bsnm{{Walder}}},
\batitle{{Supersonic turbulence in shock-bound interaction zones. I. Symmetric
  settings}}.
\bjtitle{\aap}
\bvolume{459},
\bfpage{1}--\blpage{19}
(\byear{2006})
\end{barticle}
\endbibitem

\bibitem[\protect\citeauthoryear{{Folsom} et~al.}{2008}]{2008MNRAS.391..901F}
\begin{barticle}
\bauthor{\binits{C.P.} \bsnm{{Folsom}}}, \bauthor{\binits{G.A.} \bsnm{{Wade}}},
  \bauthor{\binits{O.} \bsnm{{Kochukhov}}}, \bauthor{\binits{E.}
  \bsnm{{Alecian}}}, \bauthor{\binits{C.} \bsnm{{Catala}}},
  \bauthor{\binits{S.} \bsnm{{Bagnulo}}}, \bauthor{\binits{T.}
  \bsnm{{B{\"o}hm}}}, \bauthor{\binits{J.} \bsnm{{Bouret}}},
  \bauthor{\binits{J.} \bsnm{{Donati}}}, \bauthor{\binits{J.}
  \bsnm{{Grunhut}}}, \bauthor{\binits{D.A.} \bsnm{{Hanes}}},
  \bauthor{\binits{J.D.} \bsnm{{Landstreet}}},
\batitle{{Magnetic fields and chemical peculiarities of the very young
  intermediate-mass binary system HD 72106}}.
\bjtitle{\mnras}
\bvolume{391},
\bfpage{901}--\blpage{914}
(\byear{2008})
\end{barticle}
\endbibitem

\bibitem[\protect\citeauthoryear{{Folsom} et~al.}{2010}]{2010MNRAS.tmp.1094F}
\begin{botherref}
\oauthor{\binits{C.P.} \bsnm{{Folsom}}}, \oauthor{\binits{O.}
  \bsnm{{Kochukhov}}}, \oauthor{\binits{G.A.} \bsnm{{Wade}}},
  \oauthor{\binits{J.} \bsnm{{Silvester}}}, \oauthor{\binits{S.}
  \bsnm{{Bagnulo}}},
{Magnetic field, chemical composition and line profile variability of the
  peculiar eclipsing binary star AR Aur}.
\mnras,
(
2010)
\end{botherref}
\endbibitem

\bibitem[\protect\citeauthoryear{{Freyhammer}
  et~al.}{2008}]{2008MNRAS.389..441F}
\begin{barticle}
\bauthor{\binits{L.M.} \bsnm{{Freyhammer}}}, \bauthor{\binits{V.G.}
  \bsnm{{Elkin}}}, \bauthor{\binits{D.W.} \bsnm{{Kurtz}}}, \bauthor{\binits{G.}
  \bsnm{{Mathys}}}, \bauthor{\binits{P.} \bsnm{{Martinez}}},
\batitle{{Discovery of 17 new sharp-lined Ap stars with magnetically resolved
  lines}}.
\bjtitle{\mnras}
\bvolume{389},
\bfpage{441}--\blpage{460}
(\byear{2008})
\end{barticle}
\endbibitem

\bibitem[\protect\citeauthoryear{{Freytag} et~al.}{2002}]{2002AN....323..213F}
\begin{barticle}
\bauthor{\binits{B.} \bsnm{{Freytag}}}, \bauthor{\binits{M.} \bsnm{{Steffen}}},
  \bauthor{\binits{B.} \bsnm{{Dorch}}},
\batitle{{Spots on the surface of Betelgeuse -- Results from new 3D stellar
  convection models}}.
\bjtitle{Astronomische Nachrichten}
\bvolume{323},
\bfpage{213}--\blpage{219}
(\byear{2002})
\end{barticle}
\endbibitem

\bibitem[\protect\citeauthoryear{{Fullerton}}{2003}]{2003ASPC..305..333F}
\begin{botherref}
\oauthor{\binits{A.W.} \bsnm{{Fullerton}}},
{Cyclical Wind Variability from O-Type Stars},
in \textit{Astronomical Society of the Pacific Conference Series},
ed. by {L.~A.~Balona, H.~F.~Henrichs, \& R.~Medupe}.
Astronomical Society of the Pacific Conference Series,
vol. 305,
2003,
p. 333
\end{botherref}
\endbibitem

\bibitem[\protect\citeauthoryear{{Garc{\'{\i}}a-Segura}
  et~al.}{1999}]{1999ApJ...517..767G}
\begin{barticle}
\bauthor{\binits{G.} \bsnm{{Garc{\'{\i}}a-Segura}}}, \bauthor{\binits{N.}
  \bsnm{{Langer}}}, \bauthor{\binits{M.} \bsnm{{R{\'o}{\.z}yczka}}},
  \bauthor{\binits{J.} \bsnm{{Franco}}},
\batitle{{Shaping Bipolar and Elliptical Planetary Nebulae: Effects of Stellar
  Rotation, Photoionization Heating, and Magnetic Fields}}.
\bjtitle{\apj}
\bvolume{517},
\bfpage{767}--\blpage{781}
(\byear{1999})
\end{barticle}
\endbibitem

\bibitem[\protect\citeauthoryear{{Gayley} et~al.}{1997}]{1997ApJ...475..786G}
\begin{barticle}
\bauthor{\binits{K.G.} \bsnm{{Gayley}}}, \bauthor{\binits{S.P.}
  \bsnm{{Owocki}}}, \bauthor{\binits{S.R.} \bsnm{{Cranmer}}},
\batitle{{Sudden Radiative Braking in Colliding Hot-Star Winds}}.
\bjtitle{\apj}
\bvolume{475},
\bfpage{786}
(\byear{1997})
\end{barticle}
\endbibitem

\bibitem[\protect\citeauthoryear{{Goossens} et~al.}{1981}]{1981Ap&SS..75..521G}
\begin{barticle}
\bauthor{\binits{M.} \bsnm{{Goossens}}}, \bauthor{\binits{D.} \bsnm{{Biront}}},
  \bauthor{\binits{R.J.} \bsnm{{Tayler}}},
\batitle{{Additional Results for Unstable Stratified Toroidal Magnetic Fields
  in Stars}}.
\bjtitle{\apss}
\bvolume{75},
\bfpage{521}--\blpage{526}
(\byear{1981})
\end{barticle}
\endbibitem

\bibitem[\protect\citeauthoryear{{Grunhut} et~al.}{2009}]{2009MNRAS.400L..94G}
\begin{barticle}
\bauthor{\binits{J.H.} \bsnm{{Grunhut}}}, \bauthor{\binits{G.A.}
  \bsnm{{Wade}}}, \bauthor{\binits{W.L.F.} \bsnm{{Marcolino}}},
  \bauthor{\binits{V.} \bsnm{{Petit}}}, \bauthor{\binits{H.F.}
  \bsnm{{Henrichs}}}, \bauthor{\binits{D.H.} \bsnm{{Cohen}}},
  \bauthor{\binits{E.} \bsnm{{Alecian}}}, \bauthor{\binits{D.}
  \bsnm{{Bohlender}}}, \bauthor{\binits{J.} \bsnm{{Bouret}}},
  \bauthor{\binits{O.} \bsnm{{Kochukhov}}}, \bauthor{\binits{C.}
  \bsnm{{Neiner}}}, \bauthor{\binits{N.} \bsnm{{St-Louis}}},
  \bauthor{\binits{R.H.D.} \bsnm{{Townsend}}},
\batitle{{Discovery of a magnetic field in the O9 sub-giant star HD 57682 by
  the MiMeS Collaboration}}.
\bjtitle{\mnras}
\bvolume{400},
\bfpage{94}--\blpage{98}
(\byear{2009})
\end{barticle}
\endbibitem

\bibitem[\protect\citeauthoryear{{Grunhut} et~al.}{2010}]{2010arXiv1006.5891G}
\begin{botherref}
\oauthor{\binits{J.H.} \bsnm{{Grunhut}}}, \oauthor{\binits{G.A.}
  \bsnm{{Wade}}}, \oauthor{\binits{D.A.} \bsnm{{Hanes}}}, \oauthor{\binits{E.}
  \bsnm{{Alecian}}},
{Systematic detection of magnetic fields in massive, late-type supergiants}.
(
2010)
\end{botherref}
\endbibitem

\bibitem[\protect\citeauthoryear{{Hawley} et~al.}{1996}]{1996ApJ...464..690H}
\begin{barticle}
\bauthor{\binits{J.F.} \bsnm{{Hawley}}}, \bauthor{\binits{C.F.}
  \bsnm{{Gammie}}}, \bauthor{\binits{S.A.} \bsnm{{Balbus}}},
\batitle{{Local Three-dimensional Simulations of an Accretion Disk
  Hydromagnetic Dynamo}}.
\bjtitle{\apj}
\bvolume{464},
\bfpage{690}
(\byear{1996})
\end{barticle}
\endbibitem

\bibitem[\protect\citeauthoryear{{Heger} et~al.}{2005}]{2005ApJ...626..350H}
\begin{barticle}
\bauthor{\binits{A.} \bsnm{{Heger}}}, \bauthor{\binits{S.E.} \bsnm{{Woosley}}},
  \bauthor{\binits{H.C.} \bsnm{{Spruit}}},
\batitle{{Presupernova Evolution of Differentially Rotating Massive Stars
  Including Magnetic Fields}}.
\bjtitle{\apj}
\bvolume{626},
\bfpage{350}--\blpage{363}
(\byear{2005})
\end{barticle}
\endbibitem

\bibitem[\protect\citeauthoryear{{Henrichs} et~al.}{2005}]{2005ASPC..337..114H}
\begin{botherref}
\oauthor{\binits{H.F.} \bsnm{{Henrichs}}}, \oauthor{\binits{R.S.}
  \bsnm{{Schnerr}}}, \oauthor{\binits{E.} \bsnm{{ten Kulve}}},
{Observed magnetism in massive stars},
in \textit{The Nature and Evolution of Disks Around Hot Stars},
ed. by {R.~Ignace \& K.~G.~Gayley}.
Astronomical Society of the Pacific Conference Series,
vol. 337,
2005,
p. 114
\end{botherref}
\endbibitem

\bibitem[\protect\citeauthoryear{{Henrichs} et~al.}{2000}]{2000ASPC..214..324H}
\begin{botherref}
\oauthor{\binits{H.F.} \bsnm{{Henrichs}}}, \oauthor{\binits{J.A.} \bsnm{{de
  Jong}}}, \oauthor{\binits{J.} \bsnm{{Donati}}}, \oauthor{\binits{C.}
  \bsnm{{Catala}}}, \oauthor{\binits{G.A.} \bsnm{{Wade}}},
  \oauthor{\binits{S.L.S.} \bsnm{{Shorlin}}}, \oauthor{\binits{P.M.}
  \bsnm{{Veen}}}, \oauthor{\binits{J.S.} \bsnm{{Nichols}}},
  \oauthor{\binits{L.} \bsnm{{Kaper}}},
{The Magnetic Field of {$\beta$} Cep and the Be Phenomenon},
in \textit{IAU Colloq. 175: The Be Phenomenon in Early-Type Stars},
ed. by {M.~A.~Smith, H.~F.~Henrichs, \& J.~Fabregat}.
Astronomical Society of the Pacific Conference Series,
vol. 214,
2000,
p. 324
\end{botherref}
\endbibitem

\bibitem[\protect\citeauthoryear{{Hirschi} et~al.}{2005}]{2005A&A...443..581H}
\begin{barticle}
\bauthor{\binits{R.} \bsnm{{Hirschi}}}, \bauthor{\binits{G.} \bsnm{{Meynet}}},
  \bauthor{\binits{A.} \bsnm{{Maeder}}},
\batitle{{Stellar evolution with rotation. XIII. Predicted GRB rates at various
  Z}}.
\bjtitle{\aap}
\bvolume{443},
\bfpage{581}--\blpage{591}
(\byear{2005})
\end{barticle}
\endbibitem

\bibitem[\protect\citeauthoryear{{Hirschi} et~al.}{2010}]{2010ASPC..425...13H}
\begin{botherref}
\oauthor{\binits{R.} \bsnm{{Hirschi}}}, \oauthor{\binits{G.} \bsnm{{Meynet}}},
  \oauthor{\binits{A.} \bsnm{{Maeder}}}, \oauthor{\binits{S.}
  \bsnm{{Ekstr{\"o}m}}}, \oauthor{\binits{C.} \bsnm{{Georgy}}},
{Stellar Evolution in the Upper HR Diagram},
in \textit{Astronomical Society of the Pacific Conference Series},
ed. by {C.~Leither, P.~Bennet, P.~Morris, \& J.~van Loon}.
Astronomical Society of the Pacific Conference Series,
vol. 425,
2010,
p. 13
\end{botherref}
\endbibitem

\bibitem[\protect\citeauthoryear{{Hubrig} et~al.}{2000}]{2000ApJ...539..352H}
\begin{barticle}
\bauthor{\binits{S.} \bsnm{{Hubrig}}}, \bauthor{\binits{P.} \bsnm{{North}}},
  \bauthor{\binits{G.} \bsnm{{Mathys}}},
\batitle{{Magnetic AP Stars in the Hertzsprung-Russell Diagram}}.
\bjtitle{\apj}
\bvolume{539},
\bfpage{352}--\blpage{363}
(\byear{2000})
\end{barticle}
\endbibitem

\bibitem[\protect\citeauthoryear{{Hubrig} et~al.}{2004}]{2004A&A...428L...1H}
\begin{barticle}
\bauthor{\binits{S.} \bsnm{{Hubrig}}}, \bauthor{\binits{M.}
  \bsnm{{Sch{\"o}ller}}}, \bauthor{\binits{R.V.} \bsnm{{Yudin}}},
\batitle{{Magnetic fields in Herbig Ae stars}}.
\bjtitle{\aap}
\bvolume{428},
\bfpage{1}--\blpage{4}
(\byear{2004})
\end{barticle}
\endbibitem

\bibitem[\protect\citeauthoryear{{Hubrig} et~al.}{2005}]{2005A&A...440L..37H}
\begin{barticle}
\bauthor{\binits{S.} \bsnm{{Hubrig}}}, \bauthor{\binits{N.} \bsnm{{Nesvacil}}},
  \bauthor{\binits{M.} \bsnm{{Sch{\"o}ller}}}, \bauthor{\binits{P.}
  \bsnm{{North}}}, \bauthor{\binits{G.} \bsnm{{Mathys}}},
  \bauthor{\binits{D.W.} \bsnm{{Kurtz}}}, \bauthor{\binits{B.} \bsnm{{Wolff}}},
  \bauthor{\binits{T.} \bsnm{{Szeifert}}}, \bauthor{\binits{M.S.}
  \bsnm{{Cunha}}}, \bauthor{\binits{V.G.} \bsnm{{Elkin}}},
\batitle{{Detection of an extraordinarily large magnetic field in the unique
  ultra-cool Ap star HD 154708}}.
\bjtitle{\aap}
\bvolume{440},
\bfpage{37}--\blpage{40}
(\byear{2005})
\end{barticle}
\endbibitem

\bibitem[\protect\citeauthoryear{{Hubrig} et~al.}{2006}]{2006MNRAS.369L..61H}
\begin{barticle}
\bauthor{\binits{S.} \bsnm{{Hubrig}}}, \bauthor{\binits{M.} \bsnm{{Briquet}}},
  \bauthor{\binits{M.} \bsnm{{Sch{\"o}ller}}}, \bauthor{\binits{P.} \bsnm{{De
  Cat}}}, \bauthor{\binits{G.} \bsnm{{Mathys}}}, \bauthor{\binits{C.}
  \bsnm{{Aerts}}},
\batitle{{Discovery of magnetic fields in the {$\beta$}Cephei star {$\xi^1$}
  CMa and in several slowly pulsating B stars}}.
\bjtitle{\mnras}
\bvolume{369},
\bfpage{61}--\blpage{65}
(\byear{2006}).
doi:\doiurl{10.1111/j.1745-3933.2006.00175.x}
\end{barticle}
\endbibitem

\bibitem[\protect\citeauthoryear{{Hubrig} et~al.}{2007}]{2007A&A...463.1039H}
\begin{barticle}
\bauthor{\binits{S.} \bsnm{{Hubrig}}}, \bauthor{\binits{M.A.}
  \bsnm{{Pogodin}}}, \bauthor{\binits{R.V.} \bsnm{{Yudin}}},
  \bauthor{\binits{M.} \bsnm{{Sch{\"o}ller}}}, \bauthor{\binits{R.S.}
  \bsnm{{Schnerr}}},
\batitle{{The magnetic field in the photospheric and circumstellar components
  of Herbig Ae stars}}.
\bjtitle{\aap}
\bvolume{463},
\bfpage{1039}--\blpage{1046}
(\byear{2007})
\end{barticle}
\endbibitem

\bibitem[\protect\citeauthoryear{{Hubrig} et~al.}{2008}]{2008A&A...490..793H}
\begin{barticle}
\bauthor{\binits{S.} \bsnm{{Hubrig}}}, \bauthor{\binits{M.}
  \bsnm{{Sch{\"o}ller}}}, \bauthor{\binits{R.S.} \bsnm{{Schnerr}}},
  \bauthor{\binits{J.F.} \bsnm{{Gonz{\'a}lez}}}, \bauthor{\binits{R.}
  \bsnm{{Ignace}}}, \bauthor{\binits{H.F.} \bsnm{{Henrichs}}},
\batitle{{Magnetic field measurements of O stars with FORS 1 at the VLT}}.
\bjtitle{\aap}
\bvolume{490},
\bfpage{793}--\blpage{800}
(\byear{2008})
\end{barticle}
\endbibitem

\bibitem[\protect\citeauthoryear{{Hubrig} et~al.}{2009}]{2009A&A...502..283H}
\begin{barticle}
\bauthor{\binits{S.} \bsnm{{Hubrig}}}, \bauthor{\binits{B.} \bsnm{{Stelzer}}},
  \bauthor{\binits{M.} \bsnm{{Sch{\"o}ller}}}, \bauthor{\binits{C.}
  \bsnm{{Grady}}}, \bauthor{\binits{O.} \bsnm{{Sch{\"u}tz}}},
  \bauthor{\binits{M.A.} \bsnm{{Pogodin}}}, \bauthor{\binits{M.}
  \bsnm{{Cur{\'e}}}}, \bauthor{\binits{K.} \bsnm{{Hamaguchi}}},
  \bauthor{\binits{R.V.} \bsnm{{Yudin}}},
\batitle{{Searching for a link between the magnetic nature and other observed
  properties of Herbig Ae/Be stars and stars with debris disks}}.
\bjtitle{\aap}
\bvolume{502},
\bfpage{283}--\blpage{301}
(\byear{2009})
\end{barticle}
\endbibitem

\bibitem[\protect\citeauthoryear{{Jardine} et~al.}{1999}]{1999MNRAS.305L..35J}
\begin{barticle}
\bauthor{\binits{M.} \bsnm{{Jardine}}}, \bauthor{\binits{J.R.}
  \bsnm{{Barnes}}}, \bauthor{\binits{J.} \bsnm{{Donati}}}, \bauthor{\binits{A.}
  \bsnm{{Collier Cameron}}},
\batitle{{The potential magnetic field of AB Doradus: comparison with
  Zeeman-Doppler images}}.
\bjtitle{mnras}
\bvolume{305},
\bfpage{35}--\blpage{39}
(\byear{1999}).
doi:\doiurl{10.1046/j.1365-8711.1999.02621.x}
\end{barticle}
\endbibitem

\bibitem[\protect\citeauthoryear{{Kaper} et~al.}{1997}]{1997A&A...327..281K}
\begin{barticle}
\bauthor{\binits{L.} \bsnm{{Kaper}}}, \bauthor{\binits{H.F.}
  \bsnm{{Henrichs}}}, \bauthor{\binits{A.W.} \bsnm{{Fullerton}}},
  \bauthor{\binits{H.} \bsnm{{Ando}}}, \bauthor{\binits{K.S.}
  \bsnm{{Bjorkman}}}, \bauthor{\binits{D.R.} \bsnm{{Gies}}},
  \bauthor{\binits{R.} \bsnm{{Hirata}}}, \bauthor{\binits{E.} \bsnm{{Kambe}}},
  \bauthor{\binits{D.} \bsnm{{McDavid}}}, \bauthor{\binits{J.S.}
  \bsnm{{Nichols}}},
\batitle{{Coordinated ultraviolet and H{$\alpha$} spectroscopy of bright O-type
  stars.}}
\bjtitle{\aap}
\bvolume{327},
\bfpage{281}--\blpage{298}
(\byear{1997})
\end{barticle}
\endbibitem

\bibitem[\protect\citeauthoryear{{Kholtygin}
  et~al.}{2010}]{2010AstL...36..370K}
\begin{barticle}
\bauthor{\binits{A.F.} \bsnm{{Kholtygin}}}, \bauthor{\binits{S.N.}
  \bsnm{{Fabrika}}}, \bauthor{\binits{N.A.} \bsnm{{Drake}}},
  \bauthor{\binits{V.D.} \bsnm{{Bychkov}}}, \bauthor{\binits{L.V.}
  \bsnm{{Bychkova}}}, \bauthor{\binits{G.A.} \bsnm{{Chountonov}}},
  \bauthor{\binits{T.E.} \bsnm{{Burlakova}}}, \bauthor{\binits{G.G.}
  \bsnm{{Valyavin}}},
\batitle{{Statistics of magnetic fields for OB stars}}.
\bjtitle{Astronomy Letters}
\bvolume{36},
\bfpage{370}--\blpage{379}
(\byear{2010}).
doi:\doiurl{10.1134/S1063773710050087}
\end{barticle}
\endbibitem

\bibitem[\protect\citeauthoryear{{Kochukhov}}{2006a}]{2006A&A...454..321K}
\begin{barticle}
\bauthor{\binits{O.} \bsnm{{Kochukhov}}},
\batitle{{Remarkable non-dipolar magnetic field of the Bp star HD 137509}}.
\bjtitle{\aap}
\bvolume{454},
\bfpage{321}--\blpage{325}
(\byear{2006}a)
\end{barticle}
\endbibitem

\bibitem[\protect\citeauthoryear{{Kochukhov}}{2006b}]{2006ASPC..358..345K}
\begin{botherref}
\oauthor{\binits{O.} \bsnm{{Kochukhov}}},
{Spectro-Polarimetry of Magnetic Hot Stars},
in \textit{Astronomical Society of the Pacific Conference Series},
ed. by {R.~Casini \& B.~W.~Lites}.
Astronomical Society of the Pacific Conference Series,
vol. 358,
2006,
p. 345
\end{botherref}
\endbibitem

\bibitem[\protect\citeauthoryear{{Kochukhov}
  et~al.}{2010}]{2010arXiv1008.5115K}
\begin{botherref}
\oauthor{\binits{O.} \bsnm{{Kochukhov}}}, \oauthor{\binits{V.}
  \bsnm{{Makaganiuk}}}, \oauthor{\binits{N.} \bsnm{{Piskunov}}},
{Least squares deconvolution of the stellar intensity and polarization
  spectra}.
(
2010)
\end{botherref}
\endbibitem

\bibitem[\protect\citeauthoryear{{Landstreet}}{1988}]{1988ApJ...326..967L}
\begin{barticle}
\bauthor{\binits{J.D.} \bsnm{{Landstreet}}},
\batitle{{The magnetic field and abundance distribution geometry of the
  peculiar A star 53 Camelopardalis}}.
\bjtitle{\apj}
\bvolume{326},
\bfpage{967}--\blpage{987}
(\byear{1988})
\end{barticle}
\endbibitem

\bibitem[\protect\citeauthoryear{{Landstreet}}{1992}]{1992A&ARv...4...35L}
\begin{barticle}
\bauthor{\binits{J.D.} \bsnm{{Landstreet}}},
\batitle{{Magnetic fields at the surfaces of stars}}.
\bjtitle{\aapr}
\bvolume{4},
\bfpage{35}--\blpage{77}
(\byear{1992})
\end{barticle}
\endbibitem

\bibitem[\protect\citeauthoryear{{Landstreet}}{2009}]{2009EAS....39....1L}
\begin{botherref}
\oauthor{\binits{J.D.} \bsnm{{Landstreet}}},
{Observing and Modelling Stellar Magnetic Fields 1. Basic Physics and Simple
  Models},
in \textit{EAS Publications Series},
ed. by {C.~Neiner \& J.-P.~Zahn}.
EAS Publications Series,
vol. 39,
2009,
pp. 1--20
\end{botherref}
\endbibitem

\bibitem[\protect\citeauthoryear{{Landstreet} and
  {Mathys}}{2000}]{2000A&A...359..213L}
\begin{barticle}
\bauthor{\binits{J.D.} \bsnm{{Landstreet}}}, \bauthor{\binits{G.}
  \bsnm{{Mathys}}},
\batitle{{Magnetic models of slowly rotating magnetic Ap stars: aligned
  magnetic and rotation axes}}.
\bjtitle{\aap}
\bvolume{359},
\bfpage{213}--\blpage{226}
(\byear{2000})
\end{barticle}
\endbibitem

\bibitem[\protect\citeauthoryear{{Landstreet}
  et~al.}{2007}]{2007A&A...470..685L}
\begin{barticle}
\bauthor{\binits{J.D.} \bsnm{{Landstreet}}}, \bauthor{\binits{S.}
  \bsnm{{Bagnulo}}}, \bauthor{\binits{V.} \bsnm{{Andretta}}},
  \bauthor{\binits{L.} \bsnm{{Fossati}}}, \bauthor{\binits{E.} \bsnm{{Mason}}},
  \bauthor{\binits{J.} \bsnm{{Silaj}}}, \bauthor{\binits{G.A.} \bsnm{{Wade}}},
\batitle{{Searching for links between magnetic fields and stellar evolution:
  II. The evolution of magnetic fields as revealed by observations of Ap stars
  in open clusters and associations}}.
\bjtitle{\aap}
\bvolume{470},
\bfpage{685}--\blpage{698}
(\byear{2007})
\end{barticle}
\endbibitem

\bibitem[\protect\citeauthoryear{{Landstreet}
  et~al.}{2008}]{2008A&A...481..465L}
\begin{barticle}
\bauthor{\binits{J.D.} \bsnm{{Landstreet}}}, \bauthor{\binits{J.}
  \bsnm{{Silaj}}}, \bauthor{\binits{V.} \bsnm{{Andretta}}},
  \bauthor{\binits{S.} \bsnm{{Bagnulo}}}, \bauthor{\binits{S.V.}
  \bsnm{{Berdyugina}}}, \bauthor{\binits{J.} \bsnm{{Donati}}},
  \bauthor{\binits{L.} \bsnm{{Fossati}}}, \bauthor{\binits{P.} \bsnm{{Petit}}},
  \bauthor{\binits{J.} \bsnm{{Silvester}}}, \bauthor{\binits{G.A.}
  \bsnm{{Wade}}},
\batitle{{Searching for links between magnetic fields and stellar evolution.
  III. Measurement of magnetic fields in open cluster Ap stars with ESPaDOnS}}.
\bjtitle{\aap}
\bvolume{481},
\bfpage{465}--\blpage{480}
(\byear{2008})
\end{barticle}
\endbibitem

\bibitem[\protect\citeauthoryear{{Leitherer} and
  {Chavarria-K.}}{1987}]{1987A&A...175..208L}
\begin{barticle}
\bauthor{\binits{C.} \bsnm{{Leitherer}}}, \bauthor{\binits{C.}
  \bsnm{{Chavarria-K.}}},
\batitle{{The O6.5f?p star HD 148937 and its interstellar environment}}.
\bjtitle{\aap}
\bvolume{175},
\bfpage{208}--\blpage{218}
(\byear{1987})
\end{barticle}
\endbibitem

\bibitem[\protect\citeauthoryear{{Ligni{\`e}res}
  et~al.}{2009}]{2009A&A...500L..41L}
\begin{barticle}
\bauthor{\binits{F.} \bsnm{{Ligni{\`e}res}}}, \bauthor{\binits{P.}
  \bsnm{{Petit}}}, \bauthor{\binits{T.} \bsnm{{B{\"o}hm}}},
  \bauthor{\binits{M.} \bsnm{{Auri{\`e}re}}},
\batitle{{First evidence of a magnetic field on <ASTROBJ>Vega</ASTROBJ>.
  Towards a new class of magnetic A-type stars}}.
\bjtitle{\aap}
\bvolume{500},
\bfpage{41}--\blpage{44}
(\byear{2009})
\end{barticle}
\endbibitem

\bibitem[\protect\citeauthoryear{{MacDonald} and
  {Mullan}}{2004}]{2004MNRAS.348..702M}
\begin{barticle}
\bauthor{\binits{J.} \bsnm{{MacDonald}}}, \bauthor{\binits{D.J.}
  \bsnm{{Mullan}}},
\batitle{{Magnetic fields in massive stars: dynamics and origin}}.
\bjtitle{\mnras}
\bvolume{348},
\bfpage{702}--\blpage{716}
(\byear{2004})
\end{barticle}
\endbibitem

\bibitem[\protect\citeauthoryear{{MacGregor} and
  {Cassinelli}}{2003}]{2003ApJ...586..480M}
\begin{barticle}
\bauthor{\binits{K.B.} \bsnm{{MacGregor}}}, \bauthor{\binits{J.P.}
  \bsnm{{Cassinelli}}},
\batitle{{Magnetic Fields in Massive Stars. II. The Buoyant Rise of Magnetic
  Flux Tubes through the Radiative Interior}}.
\bjtitle{\apj}
\bvolume{586},
\bfpage{480}--\blpage{494}
(\byear{2003})
\end{barticle}
\endbibitem

\bibitem[\protect\citeauthoryear{{Maeder}}{2009}]{2009pfer.book.....M}
\begin{botherref}
\oauthor{\binits{A.} \bsnm{{Maeder}}},
\textit{{Physics, Formation and Evolution of Rotating Stars}}.
\textit{Astronomy and Astrophysics Library, Volume .~ISBN
  978-3-540-76948-4.~Springer Berlin Heidelberg}
2009
\end{botherref}
\endbibitem

\bibitem[\protect\citeauthoryear{{Maeder} and
  {Meynet}}{2003}]{2003A&A...411..543M}
\begin{barticle}
\bauthor{\binits{A.} \bsnm{{Maeder}}}, \bauthor{\binits{G.} \bsnm{{Meynet}}},
\batitle{{Stellar evolution with rotation and magnetic fields. I. The relative
  importance of rotational and magnetic effects}}.
\bjtitle{\aap}
\bvolume{411},
\bfpage{543}--\blpage{552}
(\byear{2003})
\end{barticle}
\endbibitem

\bibitem[\protect\citeauthoryear{{Maeder} and
  {Meynet}}{2004}]{2004A&A...422..225M}
\begin{barticle}
\bauthor{\binits{A.} \bsnm{{Maeder}}}, \bauthor{\binits{G.} \bsnm{{Meynet}}},
\batitle{{Stellar evolution with rotation and magnetic fields. II. General
  equations for the transport by Tayler-Spruit dynamo}}.
\bjtitle{\aap}
\bvolume{422},
\bfpage{225}--\blpage{237}
(\byear{2004})
\end{barticle}
\endbibitem

\bibitem[\protect\citeauthoryear{{Maeder} and
  {Meynet}}{2005}]{2005A&A...440.1041M}
\begin{barticle}
\bauthor{\binits{A.} \bsnm{{Maeder}}}, \bauthor{\binits{G.} \bsnm{{Meynet}}},
\batitle{{Stellar evolution with rotation and magnetic fields. III. The
  interplay of circulation and dynamo}}.
\bjtitle{\aap}
\bvolume{440},
\bfpage{1041}--\blpage{1049}
(\byear{2005})
\end{barticle}
\endbibitem

\bibitem[\protect\citeauthoryear{{Maeder} et~al.}{2009}]{2009IAUS..259..311M}
\begin{botherref}
\oauthor{\binits{A.} \bsnm{{Maeder}}}, \oauthor{\binits{G.} \bsnm{{Meynet}}},
  \oauthor{\binits{C.} \bsnm{{Georgy}}}, \oauthor{\binits{S.}
  \bsnm{{Ekstr{\"o}m}}},
{The basic role of magnetic fields in stellar evolution},
in \textit{IAU Symposium}.
IAU Symposium,
vol. 259,
2009,
pp. 311--322
\end{botherref}
\endbibitem

\bibitem[\protect\citeauthoryear{{Markey} and
  {Tayler}}{1973}]{1973MNRAS.163...77M}
\begin{barticle}
\bauthor{\binits{P.} \bsnm{{Markey}}}, \bauthor{\binits{R.J.} \bsnm{{Tayler}}},
\batitle{{The adiabatic stability of stars containing magneticfields-II.
  Poloidal fields}}.
\bjtitle{\mnras}
\bvolume{163},
\bfpage{77}
(\byear{1973})
\end{barticle}
\endbibitem

\bibitem[\protect\citeauthoryear{{Markey} and
  {Tayler}}{1974}]{1974MNRAS.168..505M}
\begin{barticle}
\bauthor{\binits{P.} \bsnm{{Markey}}}, \bauthor{\binits{R.J.} \bsnm{{Tayler}}},
\batitle{{The adiabatic stability of stars containing magnetic fields-III.
  Additional results for poloidal fields}}.
\bjtitle{\mnras}
\bvolume{168},
\bfpage{505}--\blpage{514}
(\byear{1974})
\end{barticle}
\endbibitem

\bibitem[\protect\citeauthoryear{{Mart{\'{\i}}nez Gonz{\'a}lez}
  et~al.}{2008}]{2008A&A...486..637M}
\begin{barticle}
\bauthor{\binits{M.J.} \bsnm{{Mart{\'{\i}}nez Gonz{\'a}lez}}},
  \bauthor{\binits{A.} \bsnm{{Asensio Ramos}}}, \bauthor{\binits{T.A.}
  \bsnm{{Carroll}}}, \bauthor{\binits{M.} \bsnm{{Kopf}}},
  \bauthor{\binits{J.C.} \bsnm{{Ram{\'{\i}}rez V{\'e}lez}}},
  \bauthor{\binits{M.} \bsnm{{Semel}}},
\batitle{{PCA detection and denoising of Zeeman signatures in polarised stellar
  spectra}}.
\bjtitle{\aap}
\bvolume{486},
\bfpage{637}--\blpage{646}
(\byear{2008})
\end{barticle}
\endbibitem

\bibitem[\protect\citeauthoryear{{Martins} et~al.}{2010}]{2010MNRAS.tmp..972M}
\begin{botherref}
\oauthor{\binits{F.} \bsnm{{Martins}}}, \oauthor{\binits{J.} \bsnm{{Donati}}},
  \oauthor{\binits{W.} \bsnm{{Marcolino}}}, \oauthor{\binits{J.}
  \bsnm{{Bouret}}}, \oauthor{\binits{G.A.} \bsnm{{Wade}}}, \oauthor{\binits{C.}
  \bsnm{{Escolano}}}, \oauthor{\binits{I.} \bsnm{{Howarth}}},
{Detection of a magnetic field on HD108: clues to extreme magnetic braking and
  the Of?p phenomenon}.
\mnras,
(
2010)
\end{botherref}
\endbibitem

\bibitem[\protect\citeauthoryear{{Mathys}}{1989}]{1989FCPh...13..143M}
\begin{barticle}
\bauthor{\binits{G.} \bsnm{{Mathys}}},
\batitle{{The Observation of Magnetic Fields in Nondegenerate Stars}}.
\bjtitle{fcp}
\bvolume{13},
\bfpage{143}--\blpage{308}
(\byear{1989})
\end{barticle}
\endbibitem

\bibitem[\protect\citeauthoryear{{Mathys}}{1995}]{1995A&A...293..746M}
\begin{barticle}
\bauthor{\binits{G.} \bsnm{{Mathys}}},
\batitle{{Spectropolarimetry of magnetic stars. V. The mean quadratic magnetic
  field.}}
\bjtitle{\aap}
\bvolume{293},
\bfpage{746}--\blpage{763}
(\byear{1995})
\end{barticle}
\endbibitem

\bibitem[\protect\citeauthoryear{{Mathys} and
  {Hubrig}}{2006}]{2006A&A...453..699M}
\begin{barticle}
\bauthor{\binits{G.} \bsnm{{Mathys}}}, \bauthor{\binits{S.} \bsnm{{Hubrig}}},
\batitle{{The diagnosis of the mean quadratic magnetic field of Ap stars}}.
\bjtitle{\aap}
\bvolume{453},
\bfpage{699}--\blpage{715}
(\byear{2006})
\end{barticle}
\endbibitem

\bibitem[\protect\citeauthoryear{{Mestel}}{1999}]{1999stma.book.....M}
\begin{botherref}
\oauthor{\binits{L.} \bsnm{{Mestel}}},
\textit{{Stellar magnetism}}.
~Oxford : Clarendon, 1999.~(International series of monographs on physics ; 99)
1999
\end{botherref}
\endbibitem

\bibitem[\protect\citeauthoryear{{Mestel} and
  {Landstreet}}{2005}]{2005LNP...664..183M}
\begin{botherref}
\oauthor{\binits{L.} \bsnm{{Mestel}}}, \oauthor{\binits{J.D.}
  \bsnm{{Landstreet}}},
{Stellar Magnetic Fields},
in \textit{Cosmic Magnetic Fields},
ed. by {R.~Wielebinski \& R.~Beck}.
Lecture Notes in Physics, Berlin Springer Verlag,
vol. 664,
2005,
p. 183
\end{botherref}
\endbibitem

\bibitem[\protect\citeauthoryear{{Mestel} and
  {Moss}}{2010}]{2010MNRAS.405.1845M}
\begin{barticle}
\bauthor{\binits{L.} \bsnm{{Mestel}}}, \bauthor{\binits{D.} \bsnm{{Moss}}},
\batitle{{The evolution of stable magnetic fields in stars: an analytical
  approach}}.
\bjtitle{\mnras}
\bvolume{405},
\bfpage{1845}--\blpage{1853}
(\byear{2010})
\end{barticle}
\endbibitem

\bibitem[\protect\citeauthoryear{{Mikul{\'a}{\v s}ek}
  et~al.}{2008}]{2008A&A...485..585M}
\begin{barticle}
\bauthor{\binits{Z.} \bsnm{{Mikul{\'a}{\v s}ek}}}, \bauthor{\binits{J.}
  \bsnm{{Krti{\v c}ka}}}, \bauthor{\binits{G.W.} \bsnm{{Henry}}},
  \bauthor{\binits{J.} \bsnm{{Zverko}}}, \bauthor{\binits{J.} \bsnm{{{\v Z}i{\v
  z}{\aa}ovsk{\'y}}}}, \bauthor{\binits{D.} \bsnm{{Bohlender}}},
  \bauthor{\binits{I.I.} \bsnm{{Romanyuk}}}, \bauthor{\binits{J.}
  \bsnm{{Jan{\'{\i}}k}}}, \bauthor{\binits{H.} \bsnm{{Bo{\v z}i{\'c}}}},
  \bauthor{\binits{D.} \bsnm{{Kor{\v c}{\'a}kov{\'a}}}}, \bauthor{\binits{M.}
  \bsnm{{Zejda}}}, \bauthor{\binits{I.K.} \bsnm{{Iliev}}}, \bauthor{\binits{P.}
  \bsnm{{{\v S}koda}}}, \bauthor{\binits{M.} \bsnm{{{\v S}lechta}}},
  \bauthor{\binits{T.} \bsnm{{Gr{\'a}f}}}, \bauthor{\binits{M.}
  \bsnm{{Netolick{\'y}}}}, \bauthor{\binits{M.} \bsnm{{Ceniga}}},
\batitle{{The extremely rapid rotational braking of the magnetic helium-strong
  star HD 37776}}.
\bjtitle{\aap}
\bvolume{485},
\bfpage{585}--\blpage{597}
(\byear{2008})
\end{barticle}
\endbibitem

\bibitem[\protect\citeauthoryear{{Moffat} and
  {Michaud}}{1981}]{1981ApJ...251..133M}
\begin{barticle}
\bauthor{\binits{A.F.J.} \bsnm{{Moffat}}}, \bauthor{\binits{G.}
  \bsnm{{Michaud}}},
\batitle{{Zeta Puppis - an O-type oblique rotator}}.
\bjtitle{\apj}
\bvolume{251},
\bfpage{133}--\blpage{138}
(\byear{1981})
\end{barticle}
\endbibitem

\bibitem[\protect\citeauthoryear{{Mullan}}{2009}]{2009ApJ...702..759M}
\begin{barticle}
\bauthor{\binits{D.J.} \bsnm{{Mullan}}},
\batitle{{Flares on a Bp Star}}.
\bjtitle{\apj}
\bvolume{702},
\bfpage{759}--\blpage{766}
(\byear{2009})
\end{barticle}
\endbibitem

\bibitem[\protect\citeauthoryear{{Mullan} and
  {MacDonald}}{2005}]{2005MNRAS.356.1139M}
\begin{barticle}
\bauthor{\binits{D.J.} \bsnm{{Mullan}}}, \bauthor{\binits{J.}
  \bsnm{{MacDonald}}},
\batitle{{Dynamo-generated magnetic fields at the surface of a massive star}}.
\bjtitle{\mnras}
\bvolume{356},
\bfpage{1139}--\blpage{1148}
(\byear{2005})
\end{barticle}
\endbibitem

\bibitem[\protect\citeauthoryear{{Naz{\'e}} et~al.}{2008}]{2008RMxAA..44..331N}
\begin{barticle}
\bauthor{\binits{Y.} \bsnm{{Naz{\'e}}}}, \bauthor{\binits{N.R.}
  \bsnm{{Walborn}}}, \bauthor{\binits{F.} \bsnm{{Martins}}},
\batitle{{The mysterious Of?p class and the magnetic O-star $\theta^{1}$ Ori C:
  confronting observations}}.
\bjtitle{Revista Mexicana de Astronomia y Astrofisica}
\bvolume{44},
\bfpage{331}--\blpage{340}
(\byear{2008})
\end{barticle}
\endbibitem

\bibitem[\protect\citeauthoryear{{Naze} et~al.}{2010}]{2010arXiv1006.2054N}
\begin{botherref}
\oauthor{\binits{Y.} \bsnm{{Naze}}}, \oauthor{\binits{A.} \bsnm{{ud-Doula}}},
  \oauthor{\binits{M.} \bsnm{{Spano}}}, \oauthor{\binits{G.} \bsnm{{Rauw}}},
  \oauthor{\binits{M.} \bsnm{{De Becker}}}, \oauthor{\binits{N.R.}
  \bsnm{{Walborn}}},
{New findings on the prototypical Of?p stars}.
(
2010)
\end{botherref}
\endbibitem

\bibitem[\protect\citeauthoryear{{Neiner} et~al.}{2003}]{2003A&A...406.1019N}
\begin{barticle}
\bauthor{\binits{C.} \bsnm{{Neiner}}}, \bauthor{\binits{V.C.} \bsnm{{Geers}}},
  \bauthor{\binits{H.F.} \bsnm{{Henrichs}}}, \bauthor{\binits{M.}
  \bsnm{{Floquet}}}, \bauthor{\binits{Y.} \bsnm{{Fr{\'e}mat}}},
  \bauthor{\binits{A.} \bsnm{{Hubert}}}, \bauthor{\binits{O.} \bsnm{{Preuss}}},
  \bauthor{\binits{K.} \bsnm{{Wiersema}}},
\batitle{{Discovery of a magnetic field in the Slowly Pulsating B star
  <ASTROBJ>zeta Cassiopeiae</ASTROBJ>}}.
\bjtitle{\aap}
\bvolume{406},
\bfpage{1019}--\blpage{1031}
(\byear{2003})
\end{barticle}
\endbibitem

\bibitem[\protect\citeauthoryear{{North} and
  {Debernardi}}{2004}]{2004ASPC..318..297N}
\begin{botherref}
\oauthor{\binits{P.} \bsnm{{North}}}, \oauthor{\binits{Y.}
  \bsnm{{Debernardi}}},
{Chemically Peculiar stars in binaries},
in \textit{Spectroscopically and Spatially Resolving the Components of the
  Close Binary Stars},
ed. by {R.~W.~Hilditch, H.~Hensberge, \& K.~Pavlovski}.
Astronomical Society of the Pacific Conference Series,
vol. 318,
2004,
pp. 297--305
\end{botherref}
\endbibitem

\bibitem[\protect\citeauthoryear{{North} et~al.}{1998}]{1998CoSka..27..179N}
\begin{barticle}
\bauthor{\binits{P.} \bsnm{{North}}}, \bauthor{\binits{N.} \bsnm{{Ginestet}}},
  \bauthor{\binits{J.} \bsnm{{Carquillat}}}, \bauthor{\binits{F.}
  \bsnm{{Carrier}}}, \bauthor{\binits{S.} \bsnm{{Udry}}},
\batitle{{Binaries among AP and AM stars}}.
\bjtitle{Contributions of the Astronomical Observatory Skalnate Pleso}
\bvolume{27},
\bfpage{179}--\blpage{183}
(\byear{1998})
\end{barticle}
\endbibitem

\bibitem[\protect\citeauthoryear{{Nussbaumer} and
  {Walder}}{1993}]{1993A&A...278..209N}
\begin{barticle}
\bauthor{\binits{H.} \bsnm{{Nussbaumer}}}, \bauthor{\binits{R.}
  \bsnm{{Walder}}},
\batitle{{Modification of the nebular environment in symbiotic systems due to
  colliding winds}}.
\bjtitle{\aap}
\bvolume{278},
\bfpage{209}--\blpage{225}
(\byear{1993})
\end{barticle}
\endbibitem

\bibitem[\protect\citeauthoryear{{Oksala} et~al.}{2010}]{2010MNRAS.405L..51O}
\begin{barticle}
\bauthor{\binits{M.E.} \bsnm{{Oksala}}}, \bauthor{\binits{G.A.} \bsnm{{Wade}}},
  \bauthor{\binits{W.L.F.} \bsnm{{Marcolino}}}, \bauthor{\binits{J.}
  \bsnm{{Grunhut}}}, \bauthor{\binits{D.} \bsnm{{Bohlender}}},
  \bauthor{\binits{N.} \bsnm{{Manset}}}, \bauthor{\binits{R.H.D.}
  \bsnm{{Townsend}}},
\batitle{{Discovery of a strong magnetic field in the rapidly rotating B2Vn
  star HR 7355}}.
\bjtitle{\mnras}
\bvolume{405},
\bfpage{51}--\blpage{55}
(\byear{2010})
\end{barticle}
\endbibitem

\bibitem[\protect\citeauthoryear{{Oskinova} et~al.}{2009}]{2009ApJ...693L..44O}
\begin{barticle}
\bauthor{\binits{L.M.} \bsnm{{Oskinova}}}, \bauthor{\binits{W.}
  \bsnm{{Hamann}}}, \bauthor{\binits{A.} \bsnm{{Feldmeier}}},
  \bauthor{\binits{R.} \bsnm{{Ignace}}}, \bauthor{\binits{Y.} \bsnm{{Chu}}},
\batitle{{Discovery of X-Ray Emission from the Wolf-Rayet Star WR 142 of Oxygen
  Subtype}}.
\bjtitle{\apjl}
\bvolume{693},
\bfpage{44}--\blpage{48}
(\byear{2009})
\end{barticle}
\endbibitem

\bibitem[\protect\citeauthoryear{{Owocki}}{2009}]{2009EAS....39..223O}
\begin{botherref}
\oauthor{\binits{S.} \bsnm{{Owocki}}},
{Stellar Magnetospheres},
in \textit{EAS Publications Series},
ed. by {C.~Neiner \& J.-P.~Zahn}.
EAS Publications Series,
vol. 39,
2009,
pp. 223--254
\end{botherref}
\endbibitem

\bibitem[\protect\citeauthoryear{{Owocki} and
  {Ud-Doula}}{2003}]{2003ASPC..305..350O}
\begin{botherref}
\oauthor{\binits{S.} \bsnm{{Owocki}}}, \oauthor{\binits{A.} \bsnm{{Ud-Doula}}},
{Magnetic Spin-Up of Line-Driven Stellar Winds},
in \textit{Astronomical Society of the Pacific Conference Series},
ed. by {L.~A.~Balona, H.~F.~Henrichs, \& R.~Medupe}.
Astronomical Society of the Pacific Conference Series,
vol. 305,
2003,
p. 350
\end{botherref}
\endbibitem

\bibitem[\protect\citeauthoryear{{Owocki} and
  {Gayley}}{1995}]{1995ApJ...454L.145O}
\begin{barticle}
\bauthor{\binits{S.P.} \bsnm{{Owocki}}}, \bauthor{\binits{K.G.}
  \bsnm{{Gayley}}},
\batitle{{The Importance of Radiative Braking for the Wind Interaction in the
  Close WR+O Binary V444 Cygni}}.
\bjtitle{\apjl}
\bvolume{454},
\bfpage{145}
(\byear{1995})
\end{barticle}
\endbibitem

\bibitem[\protect\citeauthoryear{{Owocki} and
  {Rybicki}}{1984}]{1984ApJ...284..337O}
\begin{barticle}
\bauthor{\binits{S.P.} \bsnm{{Owocki}}}, \bauthor{\binits{G.B.}
  \bsnm{{Rybicki}}},
\batitle{{Instabilities in line-driven stellar winds. I - Dependence on
  perturbation wavelength}}.
\bjtitle{\apj}
\bvolume{284},
\bfpage{337}--\blpage{350}
(\byear{1984})
\end{barticle}
\endbibitem

\bibitem[\protect\citeauthoryear{{Owocki} et~al.}{1988}]{1988ApJ...335..914O}
\begin{barticle}
\bauthor{\binits{S.P.} \bsnm{{Owocki}}}, \bauthor{\binits{J.I.}
  \bsnm{{Castor}}}, \bauthor{\binits{G.B.} \bsnm{{Rybicki}}},
\batitle{{Time-dependent models of radiatively driven stellar winds. I -
  Nonlinear evolution of instabilities for a pure absorption model}}.
\bjtitle{\apj}
\bvolume{335},
\bfpage{914}--\blpage{930}
(\byear{1988})
\end{barticle}
\endbibitem

\bibitem[\protect\citeauthoryear{{Petit} et~al.}{2010}]{2010arXiv1006.5868P}
\begin{botherref}
\oauthor{\binits{P.} \bsnm{{Petit}}}, \oauthor{\binits{F.}
  \bsnm{{Ligni{\`e}res}}}, \oauthor{\binits{G.A.} \bsnm{{Wade}}},
  \oauthor{\binits{M.} \bsnm{{Auri{\`e}re}}}, \oauthor{\binits{T.}
  \bsnm{{B{\"o}hm}}}, \oauthor{\binits{S.} \bsnm{{Bagnulo}}},
  \oauthor{\binits{B.} \bsnm{{Dintrans}}}, \oauthor{\binits{A.}
  \bsnm{{Fumel}}}, \oauthor{\binits{J.} \bsnm{{Grunhut}}}, \oauthor{\binits{J.}
  \bsnm{{Lanoux}}}, \oauthor{\binits{A.} \bsnm{{Morgenthaler}}},
  \oauthor{\binits{V.} \bsnm{{Van Grootel}}},
{The rapid rotation and complex magnetic field geometry of Vega}.
(
2010)
\end{botherref}
\endbibitem

\bibitem[\protect\citeauthoryear{{Petit}}{2010}]{2010arXiv1010.2248P}
\begin{botherref}
\oauthor{\binits{V.} \bsnm{{Petit}}},
{Observations of magnetic fields in hot stars}.
(
2010)
\end{botherref}
\endbibitem

\bibitem[\protect\citeauthoryear{{Petit} et~al.}{2008}]{2008MNRAS.387L..23P}
\begin{barticle}
\bauthor{\binits{V.} \bsnm{{Petit}}}, \bauthor{\binits{G.A.} \bsnm{{Wade}}},
  \bauthor{\binits{L.} \bsnm{{Drissen}}}, \bauthor{\binits{T.}
  \bsnm{{Montmerle}}}, \bauthor{\binits{E.} \bsnm{{Alecian}}},
\batitle{{Discovery of two magnetic massive stars in the Orion Nebula Cluster:
  a clue to the origin of neutron star magnetic fields?}}
\bjtitle{\mnras}
\bvolume{387},
\bfpage{23}--\blpage{27}
(\byear{2008})
\end{barticle}
\endbibitem

\bibitem[\protect\citeauthoryear{{Petit} et~al.}{2009}]{2009IAUS..259..449P}
\begin{botherref}
\oauthor{\binits{V.} \bsnm{{Petit}}}, \oauthor{\binits{G.A.} \bsnm{{Wade}}},
  \oauthor{\binits{L.} \bsnm{{Drissen}}}, \oauthor{\binits{T.}
  \bsnm{{Montmerle}}}, \oauthor{\binits{E.} \bsnm{{Alecian}}},
{Magnetic fields, winds and X-rays of massive stars in the Orion Nebula
  Cluster},
in \textit{IAU Symposium}.
IAU Symposium,
vol. 259,
2009,
pp. 449--452
\end{botherref}
\endbibitem

\bibitem[\protect\citeauthoryear{{Petrovic} et~al.}{2005}]{2005A&A...435..247P}
\begin{barticle}
\bauthor{\binits{J.} \bsnm{{Petrovic}}}, \bauthor{\binits{N.} \bsnm{{Langer}}},
  \bauthor{\binits{S.} \bsnm{{Yoon}}}, \bauthor{\binits{A.} \bsnm{{Heger}}},
\batitle{{Which massive stars are gamma-ray burst progenitors?}}
\bjtitle{\aap}
\bvolume{435},
\bfpage{247}--\blpage{259}
(\byear{2005})
\end{barticle}
\endbibitem

\bibitem[\protect\citeauthoryear{{Pittard}}{2009}]{2009MNRAS.396.1743P}
\begin{barticle}
\bauthor{\binits{J.M.} \bsnm{{Pittard}}},
\batitle{{3D models of radiatively driven colliding winds in massive O+O star
  binaries - I. Hydrodynamics}}.
\bjtitle{\mnras}
\bvolume{396},
\bfpage{1743}--\blpage{1763}
(\byear{2009})
\end{barticle}
\endbibitem

\bibitem[\protect\citeauthoryear{{Pittard}}{2010a}]{2010MNRAS.403.1633P}
\begin{barticle}
\bauthor{\binits{J.M.} \bsnm{{Pittard}}},
\batitle{{3D models of radiatively driven colliding winds in massive O + O star
  binaries - II. Thermal radio to submillimetre emission}}.
\bjtitle{\mnras}
\bvolume{403},
\bfpage{1633}--\blpage{1656}
(\byear{2010}a)
\end{barticle}
\endbibitem

\bibitem[\protect\citeauthoryear{{Pittard}}{2010b}]{2010ASPC..422..145P}
\begin{botherref}
\oauthor{\binits{J.M.} \bsnm{{Pittard}}},
{Models of the Non-Thermal Emission from Early-Type Binaries},
in \textit{Astronomical Society of the Pacific Conference Series},
ed. by {J.~Mart{\'{\i}}, P.~L.~Luque-Escamilla, \& J.~A.~Combi}.
Astronomical Society of the Pacific Conference Series,
vol. 422,
2010,
p. 145
\end{botherref}
\endbibitem

\bibitem[\protect\citeauthoryear{{Pittard} and
  {Dougherty}}{2006}]{2006MNRAS.372..801P}
\begin{barticle}
\bauthor{\binits{J.M.} \bsnm{{Pittard}}}, \bauthor{\binits{S.M.}
  \bsnm{{Dougherty}}},
\batitle{{Radio, X-ray, and {$\gamma$}-ray emission models of the
  colliding-wind binary WR140}}.
\bjtitle{\mnras}
\bvolume{372},
\bfpage{801}--\blpage{826}
(\byear{2006})
\end{barticle}
\endbibitem

\bibitem[\protect\citeauthoryear{{Pittard} and
  {Parkin}}{2010}]{2010MNRAS.403.1657P}
\begin{barticle}
\bauthor{\binits{J.M.} \bsnm{{Pittard}}}, \bauthor{\binits{E.R.}
  \bsnm{{Parkin}}},
\batitle{{3D models of radiatively driven colliding winds in massive O + O star
  binaries - III. Thermal X-ray emission}}.
\bjtitle{\mnras}
\bvolume{403},
\bfpage{1657}--\blpage{1683}
(\byear{2010})
\end{barticle}
\endbibitem

\bibitem[\protect\citeauthoryear{{Pittard} et~al.}{2006}]{2006A&A...446.1001P}
\begin{barticle}
\bauthor{\binits{J.M.} \bsnm{{Pittard}}}, \bauthor{\binits{S.M.}
  \bsnm{{Dougherty}}}, \bauthor{\binits{R.F.} \bsnm{{Coker}}},
  \bauthor{\binits{E.} \bsnm{{O'Connor}}}, \bauthor{\binits{N.J.}
  \bsnm{{Bolingbroke}}},
\batitle{{Radio emission models of colliding-wind binary systems. Inclusion of
  IC cooling}}.
\bjtitle{\aap}
\bvolume{446},
\bfpage{1001}--\blpage{1019}
(\byear{2006})
\end{barticle}
\endbibitem

\bibitem[\protect\citeauthoryear{{Pitts} and
  {Tayler}}{1985}]{1985MNRAS.216..139P}
\begin{barticle}
\bauthor{\binits{E.} \bsnm{{Pitts}}}, \bauthor{\binits{R.J.} \bsnm{{Tayler}}},
\batitle{{The adiabatic stability of stars containing magnetic fields. IV - The
  influence of rotation}}.
\bjtitle{\mnras}
\bvolume{216},
\bfpage{139}--\blpage{154}
(\byear{1985})
\end{barticle}
\endbibitem

\bibitem[\protect\citeauthoryear{{Power} et~al.}{2008}]{2008CoSka..38..443P}
\begin{barticle}
\bauthor{\binits{J.} \bsnm{{Power}}}, \bauthor{\binits{G.A.} \bsnm{{Wade}}},
  \bauthor{\binits{M.} \bsnm{{Auri{\`e}re}}}, \bauthor{\binits{J.}
  \bsnm{{Silvester}}}, \bauthor{\binits{D.} \bsnm{{Hanes}}},
\batitle{{Propertiesofavolume-limitedsampleofAp-stars}}.
\bjtitle{Contributions of the Astronomical Observatory Skalnate Pleso}
\bvolume{38},
\bfpage{443}--\blpage{444}
(\byear{2008})
\end{barticle}
\endbibitem

\bibitem[\protect\citeauthoryear{{Prendergast}}{1956}]{1956ApJ...123..498P}
\begin{barticle}
\bauthor{\binits{K.H.} \bsnm{{Prendergast}}},
\batitle{{The Equilibrium of a Self-Gravitating Incompressible Fluid Sphere
  with a Magnetic Field. I.}}
\bjtitle{\apj}
\bvolume{123},
\bfpage{498}
(\byear{1956})
\end{barticle}
\endbibitem

\bibitem[\protect\citeauthoryear{{Prendergast}}{1958}]{1958ApJ...128..361P}
\begin{barticle}
\bauthor{\binits{K.H.} \bsnm{{Prendergast}}},
\batitle{{The Equilibrium of a Self-Gravitating Incompressible Fluid Sphere
  with a Magnetic Field. II.}}
\bjtitle{\apj}
\bvolume{128},
\bfpage{361}
(\byear{1958})
\end{barticle}
\endbibitem

\bibitem[\protect\citeauthoryear{{Preston}}{1967}]{1967ApJ...150..547P}
\begin{barticle}
\bauthor{\binits{G.W.} \bsnm{{Preston}}},
\batitle{{A Statistical Investigation of the Orientation of Magnetic Axes in
  the Periodic Magnetic Variables}}.
\bjtitle{\apj}
\bvolume{150},
\bfpage{547}
(\byear{1967})
\end{barticle}
\endbibitem

\bibitem[\protect\citeauthoryear{{Preston}}{1969}]{1969ApJ...156..967P}
\begin{barticle}
\bauthor{\binits{G.W.} \bsnm{{Preston}}},
\batitle{{The Magnetic Field of HD 215441}}.
\bjtitle{\apj}
\bvolume{156},
\bfpage{967}
(\byear{1969})
\end{barticle}
\endbibitem

\bibitem[\protect\citeauthoryear{{Preston}}{1974}]{1974ARA&A..12..257P}
\begin{barticle}
\bauthor{\binits{G.W.} \bsnm{{Preston}}},
\batitle{{The chemically peculiar stars of the upper main sequence}}.
\bjtitle{\araa}
\bvolume{12},
\bfpage{257}--\blpage{277}
(\byear{1974})
\end{barticle}
\endbibitem

\bibitem[\protect\citeauthoryear{{Rauw} et~al.}{2001}]{2001A&A...366..585R}
\begin{barticle}
\bauthor{\binits{G.} \bsnm{{Rauw}}}, \bauthor{\binits{N.D.} \bsnm{{Morrison}}},
  \bauthor{\binits{J.} \bsnm{{Vreux}}}, \bauthor{\binits{E.} \bsnm{{Gosset}}},
  \bauthor{\binits{C.L.} \bsnm{{Mulliss}}},
\batitle{{The spectral variability of HD 192639 and its implications for the
  star's wind structure}}.
\bjtitle{\aap}
\bvolume{366},
\bfpage{585}--\blpage{597}
(\byear{2001})
\end{barticle}
\endbibitem

\bibitem[\protect\citeauthoryear{{Reisenegger}}{2009}]{2009A&A...499..557R}
\begin{barticle}
\bauthor{\binits{A.} \bsnm{{Reisenegger}}},
\batitle{{Stable magnetic equilibria and their evolution in the upper main
  sequence, white dwarfs, and neutron stars}}.
\bjtitle{\aap}
\bvolume{499},
\bfpage{557}--\blpage{566}
(\byear{2009})
\end{barticle}
\endbibitem

\bibitem[\protect\citeauthoryear{{Renson} and
  {Manfroid}}{2009}]{2009A&A...498..961R}
\begin{barticle}
\bauthor{\binits{P.} \bsnm{{Renson}}}, \bauthor{\binits{J.} \bsnm{{Manfroid}}},
\batitle{{Catalogue of Ap, HgMn and Am stars}}.
\bjtitle{\aap}
\bvolume{498},
\bfpage{961}--\blpage{966}
(\byear{2009})
\end{barticle}
\endbibitem

\bibitem[\protect\citeauthoryear{{Savanov} et~al.}{2009}]{2009IAUS..259..401S}
\begin{botherref}
\oauthor{\binits{I.S.} \bsnm{{Savanov}}}, \oauthor{\binits{S.}
  \bsnm{{Hubrig}}}, \oauthor{\binits{J.F.} \bsnm{{Gonz{\'a}lez}}},
  \oauthor{\binits{M.} \bsnm{{Sch{\"o}ller}}},
{Searching for a link between the presence of chemical spots on the surface of
  HgMn stars and their weak magnetic fields},
in \textit{IAU Symposium}.
IAU Symposium,
vol. 259,
2009,
pp. 401--402
\end{botherref}
\endbibitem

\bibitem[\protect\citeauthoryear{{Schnerr} et~al.}{2008}]{2008A&A...483..857S}
\begin{barticle}
\bauthor{\binits{R.S.} \bsnm{{Schnerr}}}, \bauthor{\binits{H.F.}
  \bsnm{{Henrichs}}}, \bauthor{\binits{C.} \bsnm{{Neiner}}},
  \bauthor{\binits{E.} \bsnm{{Verdugo}}}, \bauthor{\binits{J.} \bsnm{{de
  Jong}}}, \bauthor{\binits{V.C.} \bsnm{{Geers}}}, \bauthor{\binits{K.}
  \bsnm{{Wiersema}}}, \bauthor{\binits{B.} \bsnm{{van Dalen}}},
  \bauthor{\binits{A.} \bsnm{{Tijani}}}, \bauthor{\binits{B.}
  \bsnm{{Plaggenborg}}}, \bauthor{\binits{K.L.J.} \bsnm{{Rygl}}},
\batitle{{Magnetic field measurements and wind-line variability of OB-type
  stars}}.
\bjtitle{\aap}
\bvolume{483},
\bfpage{857}--\blpage{867}
(\byear{2008})
\end{barticle}
\endbibitem

\bibitem[\protect\citeauthoryear{{Schwarzschild}}{1975}]{1975ApJ...195..137S}
\begin{barticle}
\bauthor{\binits{M.} \bsnm{{Schwarzschild}}},
\batitle{{On the scale of photospheric convection in red giants and
  supergiants}}.
\bjtitle{\apj}
\bvolume{195},
\bfpage{137}--\blpage{144}
(\byear{1975})
\end{barticle}
\endbibitem

\bibitem[\protect\citeauthoryear{{Semel}}{1989}]{1989A&A...225..456S}
\begin{barticle}
\bauthor{\binits{M.} \bsnm{{Semel}}},
\batitle{{Zeeman-Doppler imaging of active stars. I - Basic principles}}.
\bjtitle{\aap}
\bvolume{225},
\bfpage{456}--\blpage{466}
(\byear{1989})
\end{barticle}
\endbibitem

\bibitem[\protect\citeauthoryear{{Semel} and {Li}}{1996}]{1996SoPh..164..417S}
\begin{barticle}
\bauthor{\binits{M.} \bsnm{{Semel}}}, \bauthor{\binits{J.} \bsnm{{Li}}},
\batitle{{Zeeman-Doppler Imaging of Solar-Type Stars: Multi Line Technique}}.
\bjtitle{solphys}
\bvolume{164},
\bfpage{417}--\blpage{428}
(\byear{1996})
\end{barticle}
\endbibitem

\bibitem[\protect\citeauthoryear{{Semel} et~al.}{2006}]{2006ASPC..358..355S}
\begin{botherref}
\oauthor{\binits{M.} \bsnm{{Semel}}}, \oauthor{\binits{D.E.} \bsnm{{Rees}}},
  \oauthor{\binits{J.C.} \bsnm{{Ram{\'{\i}}rez V{\'e}lez}}},
  \oauthor{\binits{M.J.} \bsnm{{Stift}}}, \oauthor{\binits{F.} \bsnm{{Leone}}},
{Multi-Line Spectro-Polarimetry of Stellar Magnetic Fields Using Principal
  Components Analysis},
in \textit{Astronomical Society of the Pacific Conference Series},
ed. by {R.~Casini \& B.~W.~Lites}.
Astronomical Society of the Pacific Conference Series,
vol. 358,
2006,
p. 355
\end{botherref}
\endbibitem

\bibitem[\protect\citeauthoryear{{Semel} et~al.}{2009}]{2009A&A...504.1003S}
\begin{barticle}
\bauthor{\binits{M.} \bsnm{{Semel}}}, \bauthor{\binits{J.C.}
  \bsnm{{Ram{\'{\i}}rez V{\'e}lez}}}, \bauthor{\binits{M.J.}
  \bsnm{{Mart{\'{\i}}nez Gonz{\'a}lez}}}, \bauthor{\binits{A.} \bsnm{{Asensio
  Ramos}}}, \bauthor{\binits{M.J.} \bsnm{{Stift}}}, \bauthor{\binits{A.}
  \bsnm{{L{\'o}pez Ariste}}}, \bauthor{\binits{F.} \bsnm{{Leone}}},
\batitle{{Multiline Zeeman signatures through line addition}}.
\bjtitle{\aap}
\bvolume{504},
\bfpage{1003}--\blpage{1009}
(\byear{2009})
\end{barticle}
\endbibitem

\bibitem[\protect\citeauthoryear{{Silvester}
  et~al.}{2009}]{2009MNRAS.398.1505S}
\begin{barticle}
\bauthor{\binits{J.} \bsnm{{Silvester}}}, \bauthor{\binits{C.}
  \bsnm{{Neiner}}}, \bauthor{\binits{H.F.} \bsnm{{Henrichs}}},
  \bauthor{\binits{G.A.} \bsnm{{Wade}}}, \bauthor{\binits{V.} \bsnm{{Petit}}},
  \bauthor{\binits{E.} \bsnm{{Alecian}}}, \bauthor{\binits{A.} \bsnm{{Huat}}},
  \bauthor{\binits{C.} \bsnm{{Martayan}}}, \bauthor{\binits{J.}
  \bsnm{{Power}}}, \bauthor{\binits{O.} \bsnm{{Thizy}}},
\batitle{{On the incidence of magnetic fields in slowly pulsating B, {$\beta$}
  Cephei and B-type emission-line stars}}.
\bjtitle{\mnras}
\bvolume{398},
\bfpage{1505}--\blpage{1511}
(\byear{2009})
\end{barticle}
\endbibitem

\bibitem[\protect\citeauthoryear{{Sokal} et~al.}{2010}]{2010ApJ...715.1327S}
\begin{barticle}
\bauthor{\binits{K.R.} \bsnm{{Sokal}}}, \bauthor{\binits{S.L.}
  \bsnm{{Skinner}}}, \bauthor{\binits{S.A.} \bsnm{{Zhekov}}},
  \bauthor{\binits{M.} \bsnm{{G{\"u}del}}}, \bauthor{\binits{W.}
  \bsnm{{Schmutz}}},
\batitle{{Chandra Detects the Rare Oxygen-type Wolf-Rayet Star WR 142 and OB
  Stars in Berkeley 87}}.
\bjtitle{\apj}
\bvolume{715},
\bfpage{1327}--\blpage{1337}
(\byear{2010})
\end{barticle}
\endbibitem

\bibitem[\protect\citeauthoryear{{Spruit}}{1999}]{1999A&A...349..189S}
\begin{barticle}
\bauthor{\binits{H.C.} \bsnm{{Spruit}}},
\batitle{{Differential rotation and magnetic fields in stellar interiors}}.
\bjtitle{\aap}
\bvolume{349},
\bfpage{189}--\blpage{202}
(\byear{1999})
\end{barticle}
\endbibitem

\bibitem[\protect\citeauthoryear{{Spruit}}{2002}]{2002A&A...381..923S}
\begin{barticle}
\bauthor{\binits{H.C.} \bsnm{{Spruit}}},
\batitle{{Dynamo action by differential rotation in a stably stratified stellar
  interior}}.
\bjtitle{\aap}
\bvolume{381},
\bfpage{923}--\blpage{932}
(\byear{2002})
\end{barticle}
\endbibitem

\bibitem[\protect\citeauthoryear{{Stelzer} et~al.}{2005}]{2005ApJS..160..557S}
\begin{barticle}
\bauthor{\binits{B.} \bsnm{{Stelzer}}}, \bauthor{\binits{E.}
  \bsnm{{Flaccomio}}}, \bauthor{\binits{T.} \bsnm{{Montmerle}}},
  \bauthor{\binits{G.} \bsnm{{Micela}}}, \bauthor{\binits{S.}
  \bsnm{{Sciortino}}}, \bauthor{\binits{F.} \bsnm{{Favata}}},
  \bauthor{\binits{T.} \bsnm{{Preibisch}}}, \bauthor{\binits{E.D.}
  \bsnm{{Feigelson}}},
\batitle{{X-Ray Emission from Early-Type Stars in the Orion Nebula Cluster}}.
\bjtitle{\apjs}
\bvolume{160},
\bfpage{557}--\blpage{581}
(\byear{2005})
\end{barticle}
\endbibitem

\bibitem[\protect\citeauthoryear{{Stevens} and
  {Pollock}}{1994}]{1994MNRAS.269..226S}
\begin{botherref}
\oauthor{\binits{I.R.} \bsnm{{Stevens}}}, \oauthor{\binits{A.M.T.}
  \bsnm{{Pollock}}},
{Stagnation-Point Flow in Colliding Wind Binary Systems}.
\mnras
(
1994)
\end{botherref}
\endbibitem

\bibitem[\protect\citeauthoryear{{Stevens} et~al.}{1992}]{1992ApJ...386..265S}
\begin{barticle}
\bauthor{\binits{I.R.} \bsnm{{Stevens}}}, \bauthor{\binits{J.M.}
  \bsnm{{Blondin}}}, \bauthor{\binits{A.M.T.} \bsnm{{Pollock}}},
\batitle{{Colliding winds from early-type stars in binary systems}}.
\bjtitle{\apj}
\bvolume{386},
\bfpage{265}--\blpage{287}
(\byear{1992})
\end{barticle}
\endbibitem

\bibitem[\protect\citeauthoryear{{Tayler}}{1973}]{1973MNRAS.161..365T}
\begin{barticle}
\bauthor{\binits{R.J.} \bsnm{{Tayler}}},
\batitle{{The adiabatic stability of stars containing magnetic
  fields-I.Toroidal fields}}.
\bjtitle{\mnras}
\bvolume{161},
\bfpage{365}
(\byear{1973})
\end{barticle}
\endbibitem

\bibitem[\protect\citeauthoryear{{Tout} et~al.}{2004}]{2004MNRAS.355L..13T}
\begin{barticle}
\bauthor{\binits{C.A.} \bsnm{{Tout}}}, \bauthor{\binits{D.T.}
  \bsnm{{Wickramasinghe}}}, \bauthor{\binits{L.} \bsnm{{Ferrario}}},
\batitle{{Magnetic fields in white dwarfs and stellar evolution}}.
\bjtitle{\mnras}
\bvolume{355},
\bfpage{13}--\blpage{16}
(\byear{2004})
\end{barticle}
\endbibitem

\bibitem[\protect\citeauthoryear{{Townsend} and
  {Owocki}}{2005}]{2005MNRAS.357..251T}
\begin{barticle}
\bauthor{\binits{R.H.D.} \bsnm{{Townsend}}}, \bauthor{\binits{S.P.}
  \bsnm{{Owocki}}},
\batitle{{A rigidly rotating magnetosphere model for circumstellar emission
  from magnetic OB stars}}.
\bjtitle{\mnras}
\bvolume{357},
\bfpage{251}--\blpage{264}
(\byear{2005})
\end{barticle}
\endbibitem

\bibitem[\protect\citeauthoryear{{Townsend} et~al.}{2005}]{2005ApJ...630L..81T}
\begin{barticle}
\bauthor{\binits{R.H.D.} \bsnm{{Townsend}}}, \bauthor{\binits{S.P.}
  \bsnm{{Owocki}}}, \bauthor{\binits{D.} \bsnm{{Groote}}},
\batitle{{The Rigidly Rotating Magnetosphere of {$\sigma$} Orionis E}}.
\bjtitle{\apjl}
\bvolume{630},
\bfpage{81}--\blpage{84}
(\byear{2005})
\end{barticle}
\endbibitem

\bibitem[\protect\citeauthoryear{{Townsend} et~al.}{2004}]{2004MNRAS.350..189T}
\begin{barticle}
\bauthor{\binits{R.H.D.} \bsnm{{Townsend}}}, \bauthor{\binits{S.P.}
  \bsnm{{Owocki}}}, \bauthor{\binits{I.D.} \bsnm{{Howarth}}},
\batitle{{Be-star rotation: how close to critical?}}
\bjtitle{\mnras}
\bvolume{350},
\bfpage{189}--\blpage{195}
(\byear{2004})
\end{barticle}
\endbibitem

\bibitem[\protect\citeauthoryear{{Townsend} et~al.}{2010}]{2010ApJ...714L.318T}
\begin{barticle}
\bauthor{\binits{R.H.D.} \bsnm{{Townsend}}}, \bauthor{\binits{M.E.}
  \bsnm{{Oksala}}}, \bauthor{\binits{D.H.} \bsnm{{Cohen}}},
  \bauthor{\binits{S.P.} \bsnm{{Owocki}}}, \bauthor{\binits{A.}
  \bsnm{{ud-Doula}}},
\batitle{{Discovery of Rotational Braking in the Magnetic Helium-strong Star
  Sigma Orionis E}}.
\bjtitle{\apjl}
\bvolume{714},
\bfpage{318}--\blpage{322}
(\byear{2010})
\end{barticle}
\endbibitem

\bibitem[\protect\citeauthoryear{{Tutukov} and
  {Fedorova}}{2010}]{2010ARep...54..156T}
\begin{barticle}
\bauthor{\binits{A.V.} \bsnm{{Tutukov}}}, \bauthor{\binits{A.V.}
  \bsnm{{Fedorova}}},
\batitle{{Possible scenarios for the formation of Ap/Bp stars}}.
\bjtitle{Astronomy Reports}
\bvolume{54},
\bfpage{156}--\blpage{162}
(\byear{2010})
\end{barticle}
\endbibitem

\bibitem[\protect\citeauthoryear{{ud-Doula} and
  {Owocki}}{2002}]{2002ApJ...576..413U}
\begin{barticle}
\bauthor{\binits{A.} \bsnm{{ud-Doula}}}, \bauthor{\binits{S.P.}
  \bsnm{{Owocki}}},
\batitle{{Dynamical Simulations of Magnetically Channeled Line-driven Stellar
  Winds. I. Isothermal, Nonrotating, Radially Driven Flow}}.
\bjtitle{\apj}
\bvolume{576},
\bfpage{413}--\blpage{428}
(\byear{2002})
\end{barticle}
\endbibitem

\bibitem[\protect\citeauthoryear{{Ud-Doula} et~al.}{2008}]{2008MNRAS.385...97U}
\begin{barticle}
\bauthor{\binits{A.} \bsnm{{Ud-Doula}}}, \bauthor{\binits{S.P.}
  \bsnm{{Owocki}}}, \bauthor{\binits{R.H.D.} \bsnm{{Townsend}}},
\batitle{{Dynamical simulations of magnetically channelled line-driven stellar
  winds - II. The effects of field-aligned rotati on}}.
\bjtitle{\mnras}
\bvolume{385},
\bfpage{97}--\blpage{108}
(\byear{2008})
\end{barticle}
\endbibitem

\bibitem[\protect\citeauthoryear{{Ud-Doula} et~al.}{2009}]{2009MNRAS.392.1022U}
\begin{barticle}
\bauthor{\binits{A.} \bsnm{{Ud-Doula}}}, \bauthor{\binits{S.P.}
  \bsnm{{Owocki}}}, \bauthor{\binits{R.H.D.} \bsnm{{Townsend}}},
\batitle{{Dynamical simulations of magnetically channelled line-driven stellar
  winds - III. Angular momentum loss and rotation al spin-down}}.
\bjtitle{\mnras}
\bvolume{392},
\bfpage{1022}--\blpage{1033}
(\byear{2009})
\end{barticle}
\endbibitem

\bibitem[\protect\citeauthoryear{{ud-Doula} et~al.}{2006}]{2006ApJ...640L.191U}
\begin{barticle}
\bauthor{\binits{A.} \bsnm{{ud-Doula}}}, \bauthor{\binits{R.H.D.}
  \bsnm{{Townsend}}}, \bauthor{\binits{S.P.} \bsnm{{Owocki}}},
\batitle{{Centrifugal Breakout of Magnetically Confined Line-driven Stellar
  Winds}}.
\bjtitle{\apjl}
\bvolume{640},
\bfpage{191}--\blpage{194}
(\byear{2006})
\end{barticle}
\endbibitem

\bibitem[\protect\citeauthoryear{{Usov}}{1991}]{1991MNRAS.252...49U}
\begin{barticle}
\bauthor{\binits{V.V.} \bsnm{{Usov}}},
\batitle{{Stellar wind collision and dust formation in long-period, heavily
  interacting Wolf-Rayet binaries}}.
\bjtitle{\mnras}
\bvolume{252},
\bfpage{49}--\blpage{52}
(\byear{1991})
\end{barticle}
\endbibitem

\bibitem[\protect\citeauthoryear{{van der Hucht}}{2001}]{2001NewAR..45..135V}
\begin{barticle}
\bauthor{\binits{K.A.} \bsnm{{van der Hucht}}},
\batitle{{The VIIth catalogue of galactic Wolf-Rayet stars}}.
\bjtitle{\nar}
\bvolume{45},
\bfpage{135}--\blpage{232}
(\byear{2001})
\end{barticle}
\endbibitem

\bibitem[\protect\citeauthoryear{{van der Hucht}}{2006}]{2006A&A...458..453V}
\begin{barticle}
\bauthor{\binits{K.A.} \bsnm{{van der Hucht}}},
\batitle{{New Galactic Wolf-Rayet stars, and candidates. An annex to The VIIth
  Catalogue of Galactic Wolf-Rayet Stars}}.
\bjtitle{\aap}
\bvolume{458},
\bfpage{453}--\blpage{459}
(\byear{2006})
\end{barticle}
\endbibitem

\bibitem[\protect\citeauthoryear{{van der Hucht}
  et~al.}{1992}]{1992ASPC...22..249V}
\begin{botherref}
\oauthor{\binits{K.A.} \bsnm{{van der Hucht}}}, \oauthor{\binits{P.M.}
  \bsnm{{Williams}}}, \oauthor{\binits{T.A.T.} \bsnm{{Spoclstra}}},
  \oauthor{\binits{A.C.} \bsnm{{de Bruyn}}},
{Non-Thermal Radio Observations of Wolf-Rayet Stars: A Case for Long-Period
  Binaries (Invited Paper)},
in \textit{Nonisotropic and Variable Outflows from Stars},
ed. by {L.~Drissen, C.~Leitherer, \& A.~Nota}.
Astronomical Society of the Pacific Conference Series,
vol. 22,
1992,
p. 249
\end{botherref}
\endbibitem

\bibitem[\protect\citeauthoryear{{van Loo}}{2010}]{2010ASPC..422..157V}
\begin{botherref}
\oauthor{\binits{S.} \bsnm{{van Loo}}},
{Non-Thermal Radio Emission from {\^a}??Presumably{\^a}?? Single O Stars},
in \textit{Astronomical Society of the Pacific Conference Series},
ed. by {J.~Mart{\'{\i}}, P.~L.~Luque-Escamilla, \& J.~A.~Combi}.
Astronomical Society of the Pacific Conference Series,
vol. 422,
2010,
p. 157
\end{botherref}
\endbibitem

\bibitem[\protect\citeauthoryear{{van Loo} et~al.}{2006}]{2006A&A...452.1011V}
\begin{barticle}
\bauthor{\binits{S.} \bsnm{{van Loo}}}, \bauthor{\binits{M.C.}
  \bsnm{{Runacres}}}, \bauthor{\binits{R.} \bsnm{{Blomme}}},
\batitle{{Can single O stars produce non-thermal radio emission?}}
\bjtitle{\aap}
\bvolume{452},
\bfpage{1011}--\blpage{1019}
(\byear{2006})
\end{barticle}
\endbibitem

\bibitem[\protect\citeauthoryear{{Wade} and {the MiMeS
  Collaboration}}{2010}]{2010arXiv1012.2925W}
\begin{botherref}
\oauthor{\binits{G.A.} \bsnm{{Wade}}}, \oauthor{\bsnm{{the MiMeS
  Collaboration}}},
{The MiMeS Project: Current status and recent results}.
(
2010)
\end{botherref}
\endbibitem

\bibitem[\protect\citeauthoryear{{Wade} et~al.}{2006}]{2006A&A...451..293W}
\begin{barticle}
\bauthor{\binits{G.A.} \bsnm{{Wade}}}, \bauthor{\binits{M.}
  \bsnm{{Auri{\`e}re}}}, \bauthor{\binits{S.} \bsnm{{Bagnulo}}},
  \bauthor{\binits{J.} \bsnm{{Donati}}}, \bauthor{\binits{N.}
  \bsnm{{Johnson}}}, \bauthor{\binits{J.D.} \bsnm{{Landstreet}}},
  \bauthor{\binits{F.} \bsnm{{Ligni{\`e}res}}}, \bauthor{\binits{S.}
  \bsnm{{Marsden}}}, \bauthor{\binits{D.} \bsnm{{Monin}}}, \bauthor{\binits{D.}
  \bsnm{{Mouillet}}}, \bauthor{\binits{F.} \bsnm{{Paletou}}},
  \bauthor{\binits{P.} \bsnm{{Petit}}}, \bauthor{\binits{N.}
  \bsnm{{Toqu{\'e}}}}, \bauthor{\binits{E.} \bsnm{{Alecian}}},
  \bauthor{\binits{C.} \bsnm{{Folsom}}},
\batitle{{A search for magnetic fields in the variable HgMn star {$\alpha$}
  Andromedae}}.
\bjtitle{\aap}
\bvolume{451},
\bfpage{293}--\blpage{302}
(\byear{2006})
\end{barticle}
\endbibitem

\bibitem[\protect\citeauthoryear{{Wade} et~al.}{2007}]{2007MNRAS.376.1145W}
\begin{barticle}
\bauthor{\binits{G.A.} \bsnm{{Wade}}}, \bauthor{\binits{S.} \bsnm{{Bagnulo}}},
  \bauthor{\binits{D.} \bsnm{{Drouin}}}, \bauthor{\binits{J.D.}
  \bsnm{{Landstreet}}}, \bauthor{\binits{D.} \bsnm{{Monin}}},
\batitle{{A search for strong, ordered magnetic fields in Herbig Ae/Be stars}}.
\bjtitle{\mnras}
\bvolume{376},
\bfpage{1145}--\blpage{1161}
(\byear{2007})
\end{barticle}
\endbibitem

\bibitem[\protect\citeauthoryear{{Wade} et~al.}{2009a}]{2009arXiv0901.0347W}
\begin{botherref}
\oauthor{\binits{G.A.} \bsnm{{Wade}}}, \oauthor{\binits{E.} \bsnm{{Alecian}}},
  \oauthor{\binits{J.} \bsnm{{Grunhut}}}, \oauthor{\binits{C.}
  \bsnm{{Catala}}}, \oauthor{\binits{S.} \bsnm{{Bagnulo}}},
  \oauthor{\binits{C.P.} \bsnm{{Folsom}}}, \oauthor{\binits{J.D.}
  \bsnm{{Landstreet}}},
{Magnetism of Herbig Ae/Be stars}.
(
2009a)
\end{botherref}
\endbibitem

\bibitem[\protect\citeauthoryear{{Wade} et~al.}{2009b}]{2009IAUS..259..333W}
\begin{botherref}
\oauthor{\binits{G.A.} \bsnm{{Wade}}}, \oauthor{\binits{E.} \bsnm{{Alecian}}},
  \oauthor{\binits{D.A.} \bsnm{{Bohlender}}}, \oauthor{\binits{J.}
  \bsnm{{Bouret}}}, \oauthor{\binits{J.H.} \bsnm{{Grunhut}}},
  \oauthor{\binits{H.} \bsnm{{Henrichs}}}, \oauthor{\binits{C.}
  \bsnm{{Neiner}}}, \oauthor{\binits{V.} \bsnm{{Petit}}},
  \oauthor{\binits{N.S.} \bsnm{{Louis}}}, \oauthor{\binits{M.}
  \bsnm{{Auri{\`e}re}}}, \oauthor{\binits{O.} \bsnm{{Kochukhov}}},
  \oauthor{\binits{J.} \bsnm{{Silvester}}}, \oauthor{\binits{A.}
  \bsnm{{ud-Doula}}}, \oauthor{\bsnm{{ud-Doula}}},
{The MiMeS project: magnetism in massive stars},
in \textit{IAU Symposium}.
IAU Symposium,
vol. 259,
2009,
pp. 333--338
\end{botherref}
\endbibitem

\bibitem[\protect\citeauthoryear{{Walborn}}{1972}]{1972AJ.....77..312W}
\begin{barticle}
\bauthor{\binits{N.R.} \bsnm{{Walborn}}},
\batitle{{Spectral classification of OB stars in both hemispheres and the
  absolute-magnitude calibration.}}
\bjtitle{\aj}
\bvolume{77},
\bfpage{312}--\blpage{318}
(\byear{1972})
\end{barticle}
\endbibitem

\bibitem[\protect\citeauthoryear{{Walborn} et~al.}{2003}]{2003ApJ...588.1025W}
\begin{barticle}
\bauthor{\binits{N.R.} \bsnm{{Walborn}}}, \bauthor{\binits{I.D.}
  \bsnm{{Howarth}}}, \bauthor{\binits{A.} \bsnm{{Herrero}}},
  \bauthor{\binits{D.J.} \bsnm{{Lennon}}},
\batitle{{The Remarkable Alternating Spectra of the Of?p Star HD 191612}}.
\bjtitle{\apj}
\bvolume{588},
\bfpage{1025}--\blpage{1038}
(\byear{2003})
\end{barticle}
\endbibitem

\bibitem[\protect\citeauthoryear{{Walborn} et~al.}{2004}]{2004ApJ...617L..61W}
\begin{barticle}
\bauthor{\binits{N.R.} \bsnm{{Walborn}}}, \bauthor{\binits{I.D.}
  \bsnm{{Howarth}}}, \bauthor{\binits{G.} \bsnm{{Rauw}}},
  \bauthor{\binits{D.J.} \bsnm{{Lennon}}}, \bauthor{\binits{H.E.}
  \bsnm{{Bond}}}, \bauthor{\binits{I.} \bsnm{{Negueruela}}},
  \bauthor{\binits{Y.} \bsnm{{Naz{\'e}}}}, \bauthor{\binits{M.F.}
  \bsnm{{Corcoran}}}, \bauthor{\binits{A.} \bsnm{{Herrero}}},
  \bauthor{\binits{A.} \bsnm{{Pellerin}}},
\batitle{{A Period and a Prediction for the Of?p Spectrum Alternator HD
  191612}}.
\bjtitle{\apjl}
\bvolume{617},
\bfpage{61}--\blpage{64}
(\byear{2004})
\end{barticle}
\endbibitem

\bibitem[\protect\citeauthoryear{{Walborn} et~al.}{2010}]{2010ApJ...711L.143W}
\begin{barticle}
\bauthor{\binits{N.R.} \bsnm{{Walborn}}}, \bauthor{\binits{A.} \bsnm{{Sota}}},
  \bauthor{\binits{J.} \bsnm{{Ma{\'{\i}}z Apell{\'a}niz}}},
  \bauthor{\binits{E.J.} \bsnm{{Alfaro}}}, \bauthor{\binits{N.I.}
  \bsnm{{Morrell}}}, \bauthor{\binits{R.H.} \bsnm{{Barb{\'a}}}},
  \bauthor{\binits{J.I.} \bsnm{{Arias}}}, \bauthor{\binits{R.C.}
  \bsnm{{Gamen}}},
\batitle{{Early Results from the Galactic O-Star Spectroscopic Survey: C III
  Emission Lines in Of Spectra}}.
\bjtitle{\apjl}
\bvolume{711},
\bfpage{143}--\blpage{147}
(\byear{2010})
\end{barticle}
\endbibitem

\bibitem[\protect\citeauthoryear{{Walder}}{1995}]{1995IAUS..163..420W}
\begin{botherref}
\oauthor{\binits{R.} \bsnm{{Walder}}},
{Simulations of colliding winds in 3 dimensions},
in \textit{Wolf-Rayet Stars: Binaries; Colliding Winds; Evolution},
ed. by {K.~A.~van der Hucht \& P.~M.~Williams}.
IAU Symposium,
vol. 163,
1995,
p. 420
\end{botherref}
\endbibitem

\bibitem[\protect\citeauthoryear{{Walder} and
  {Folini}}{1996}]{1996A&A...315..265W}
\begin{barticle}
\bauthor{\binits{R.} \bsnm{{Walder}}}, \bauthor{\binits{D.} \bsnm{{Folini}}},
\batitle{{Radiative cooling instability in 1D colliding flows.}}
\bjtitle{\aap}
\bvolume{315},
\bfpage{265}--\blpage{283}
(\byear{1996})
\end{barticle}
\endbibitem

\bibitem[\protect\citeauthoryear{{Walder} and
  {Folini}}{1998}]{1998A&A...330L..21W}
\begin{barticle}
\bauthor{\binits{R.} \bsnm{{Walder}}}, \bauthor{\binits{D.} \bsnm{{Folini}}},
\batitle{{Knots, filaments, and turbulence in radiative shocks}}.
\bjtitle{\aap}
\bvolume{330},
\bfpage{21}--\blpage{24}
(\byear{1998})
\end{barticle}
\endbibitem

\bibitem[\protect\citeauthoryear{{Walder} and
  {Folini}}{2003}]{2003IAUS..212..139W}
\begin{botherref}
\oauthor{\binits{R.} \bsnm{{Walder}}}, \oauthor{\binits{D.} \bsnm{{Folini}}},
{3D-hydrodynamics of colliding winds in massive binaries},
in \textit{A Massive Star Odyssey: From Main Sequence to Supernova},
ed. by {K.~van der Hucht, A.~Herrero, \& C.~Esteban}.
IAU Symposium,
vol. 212,
2003,
p. 139
\end{botherref}
\endbibitem

\bibitem[\protect\citeauthoryear{{Williams} et~al.}{1997}]{1997MNRAS.289...10W}
\begin{barticle}
\bauthor{\binits{P.M.} \bsnm{{Williams}}}, \bauthor{\binits{S.M.}
  \bsnm{{Dougherty}}}, \bauthor{\binits{R.J.} \bsnm{{Davis}}},
  \bauthor{\binits{K.A.} \bsnm{{van der Hucht}}}, \bauthor{\binits{M.F.}
  \bsnm{{Bode}}}, \bauthor{\binits{D.Y.A.} \bsnm{{Setia Gunawan}}},
\batitle{{Radio and infrared structure of the colliding-wind Wolf-Rayet system
  WR147}}.
\bjtitle{\mnras}
\bvolume{289},
\bfpage{10}--\blpage{20}
(\byear{1997})
\end{barticle}
\endbibitem

\bibitem[\protect\citeauthoryear{{Woosley}}{1993}]{1993ApJ...405..273W}
\begin{barticle}
\bauthor{\binits{S.E.} \bsnm{{Woosley}}},
\batitle{{Gamma-ray bursts from stellar mass accretion disks around black
  holes}}.
\bjtitle{\apj}
\bvolume{405},
\bfpage{273}--\blpage{277}
(\byear{1993})
\end{barticle}
\endbibitem

\bibitem[\protect\citeauthoryear{{Woosley} and
  {Heger}}{2006}]{2006ApJ...637..914W}
\begin{barticle}
\bauthor{\binits{S.E.} \bsnm{{Woosley}}}, \bauthor{\binits{A.} \bsnm{{Heger}}},
\batitle{{The Progenitor Stars of Gamma-Ray Bursts}}.
\bjtitle{\apj}
\bvolume{637},
\bfpage{914}--\blpage{921}
(\byear{2006})
\end{barticle}
\endbibitem

\bibitem[\protect\citeauthoryear{{Wright}}{1973}]{1973MNRAS.162..339W}
\begin{barticle}
\bauthor{\binits{G.A.E.} \bsnm{{Wright}}},
\batitle{{Pinch instabilities in magnetic stars}}.
\bjtitle{\mnras}
\bvolume{162},
\bfpage{339}
(\byear{1973})
\end{barticle}
\endbibitem

\bibitem[\protect\citeauthoryear{{Yoon} and
  {Langer}}{2005}]{2005A&A...443..643Y}
\begin{barticle}
\bauthor{\binits{S.} \bsnm{{Yoon}}}, \bauthor{\binits{N.} \bsnm{{Langer}}},
\batitle{{Evolution of rapidly rotating metal-poor massive stars towards
  gamma-ray bursts}}.
\bjtitle{\aap}
\bvolume{443},
\bfpage{643}--\blpage{648}
(\byear{2005})
\end{barticle}
\endbibitem

\bibitem[\protect\citeauthoryear{{Zahn} et~al.}{2007}]{2007A&A...474..145Z}
\begin{barticle}
\bauthor{\binits{J.} \bsnm{{Zahn}}}, \bauthor{\binits{A.S.} \bsnm{{Brun}}},
  \bauthor{\binits{S.} \bsnm{{Mathis}}},
\batitle{{On magnetic instabilities and dynamo action in stellar radiation
  zones}}.
\bjtitle{\aap}
\bvolume{474},
\bfpage{145}--\blpage{154}
(\byear{2007})
\end{barticle}
\endbibitem

\bibitem[\protect\citeauthoryear{{Zhekov} and
  {Myasnikov}}{2000}]{2000ApJ...543L..53Z}
\begin{barticle}
\bauthor{\binits{S.A.} \bsnm{{Zhekov}}}, \bauthor{\binits{A.V.}
  \bsnm{{Myasnikov}}},
\batitle{{Colliding Stellar Winds: ``Asymmetric'' Thermal Conduction}}.
\bjtitle{\apjl}
\bvolume{543},
\bfpage{53}--\blpage{56}
(\byear{2000})
\end{barticle}
\endbibitem

\end{thebibliography}

\end{document}